\documentclass{aastex631}
\usepackage{ulem}

\usepackage{graphicx}

\usepackage{CJKutf8}
\usepackage{bm}
\usepackage{natbib}
\usepackage{amsmath}
\usepackage{booktabs}
\usepackage{soul}
\usepackage{xcolor}

\usepackage{threeparttable}

\begin{document}
\begin{CJK*}{UTF8}{gbsn}

\title{A Robust Launching Mechanism for Freely-Floating 
Planets from Host Stars with Close-in Planets}.

\author[0000-0002-7814-9185]{Xiaochen Zheng（郑晓晨)}
\affiliation{Beijing Planetarium,
Beijing, 100044, China}

\author[0009-0004-7622-3056]{Zhuoya Cao（曹卓雅）}
\affiliation{Department of Astronomy,
Westlake University, Hangzhou 310030, Zhejiang Province, China}
\affiliation{Department of Astronomy, 
Tsinghua University, Beijing, 100084, China}

\author{Shigeru Ida（井田茂）}
\affiliation{Earth and Life Science Institute, 
Institute of Science, Meguro, Tokyo, 152-8550, Japan}
\affiliation{Department of Astronomy,
Westlake University, Hangzhou 310030, Zhejiang Province, China}

\author[0000-0001-5466-4628]{Douglas N.C. Lin（林潮）}
\affiliation{Department of Astronomy and Astrophysics,
University of California, Santa Cruz, CA 95064, USA}
\affiliation{Department of Astronomy,
Westlake University, Hangzhou 310030, Zhejiang Province, China}

\author[0000-0001-8317-2788]{Shude Mao (毛淑德)}
\affiliation{Department of Astronomy,
Westlake University, Hangzhou 310030, Zhejiang Province, China}

\correspondingauthor{Douglas N. C. Lin, Shigeru Ida, Zhuoya Cao, Shude Mao}
\email{
lin@ucolick.org, 
ida@elsi.jp,
zhuoya\_cao@qq.com,
shude.mao@westlake.edu.cn}



\begin{abstract}

Secular perturbations from binary stars and distant massive planets can drive cold planets onto nearly parabolic orbits with pericenter passages extremely close to their host stars. Meanwhile, short-period super-Earths are frequently observed around nearby stars. Gravitational scattering between these two distinct populations can lead to substantial orbital energy exchange, liberating some intruders from the gravitational confinement of their host systems. This process offers a robust formation channel for a subset of the abundant freely floating planet population. It may also significantly perturb the original orbits of close-in planets, induce collisional trajectories between close-in planets and their host stars, and disrupt the dynamical evolution of cold planets toward close stellar encounters.
\end{abstract}



\section{Introduction} \label{sec:intro}

Over the past three decades, significant advancements in exoplanet detection techniques and methods have led to the confirmation of over 6,000 exoplanets (NASA Exoplanet Archive), with numerous additional candidates under investigation. This extensive sample has greatly enhanced our understanding of exoplanet demographics and planetary system architectures. 


The demographic census of exoplanets is conducted through various detection methods, each subject to observational selection biases \citep{perryman2018, zhudong2021}. Among these populations, 
close-in planets are considerably more abundant. This category includes all planets with $P < 400$ days \citep{zhu2021}, encompassing close-in Jupiters, super-Earths, and terrestrial planets. Most reside in multi-planet systems \citep{fabrycky2014}, typically exhibiting low eccentricities and nearly coplanar orbits, with some of them in or near mean-motion resonances \citep{dai2024}.


Occurrence rates show substantial variation across studies. Early HARPS data suggested $30–50\%$ of GK-type stars host planets with minimum masses of $3–30 M_{\oplus}$ and orbital periods $P < 50$ days \citep{mayor2009,mayor2010}, while \cite{howard2010} found a rate of $\sim 15\%$ for similar parameters and predicted $23\%$ of stars harbor smaller planets ($0.5–2 M_{\oplus}$). Subsequent work \citep{howard2012} reported occurrence rates of $\sim 0.13$, $\sim 0.023$, and $\sim 0.013$ planets per star for radius ranges of $2–4~R_{\oplus}$, $4–8~R_{\oplus}$, and $8–32~R_{\oplus}$, respectively, for $P < 50$ days.


Beyond this period, occurrence rates increase with orbital distance. \cite{dong2013} found rates of $\sim 28\%$ (Earth-size), $\sim 25\%$ (super-Earths), $\sim 7\%$ (Neptune-size), and $\sim 3\%$ (Jupiter-size) within 250 days. The overall occurrence rate for giant planets within 400 days is $\sim 4.6\%$ \citep{santerne2016}, with Jovian planets ($> 0.3 M_{\rm J}$) beyond 100 days occurring at $\sim 6.7\%$ \citep{wittenmyer2020}. These results collectively indicate a rising occurrence rate with orbital distance out to 1 a u, consistent with \textit{Kepler} statistics.

Long-period planets are primarily detected through direct imaging of young, luminous planets \citep{galicher2016} or inference from microlensing events \citep{gould2010}. Due to the sensitivity limitations of these techniques, most known long-period planets are relatively massive. These distant giants are expected to play a disproportionately important role in shaping the dynamical evolution and kinematic architecture of planetary systems. This hypothesis is supported by observational evidence showing that systems containing long-period planets exhibit significantly more diverse orbital eccentricities compared to their compact counterparts.


In addition to bound planets, a vast population of free-floating planets (FFPs) has been revealed through microlensing surveys. The initial evidence emerged from short-duration events (1–2 days), which \cite{sumi2011} attributed to a substantial population of unbound or wide-orbit Jupiter-mass planets, although such a population was found difficult to reproduce from population synthesis models \citep{MMI2016}. \cite{mroz_no_2017} later quantified this population, reporting an occurrence rate of 0.25 such planets per main-sequence star and identifying even shorter events suggestive of Earth- to super-Earth-mass FFPs. The first terrestrial-mass FFP candidate was confirmed by \cite{mroz2020}. Although \cite{Gould2022} estimated that FFPs outnumber bound planets by an order of magnitude, the first quantitative measurement was provided by \cite{sumi2023}. Using nine years of MOA-II data, they constructed the mass function of FFPs down to Earth mass, finding them to be $19_{-13}^{+23}$ times more abundant than bound planets beyond the snow line 
, with a total mass of $\sim 171 M_{\oplus}$ per star, suggesting a common formation origin in planetary systems. This abundant population of FFPs has been further supported by the James Webb Space Telescope (JWST) near-infrared survey in the Orion Nebula, which uncovered numerous planetary-mass candidates \citep{Tamura1998, Oasa1999, Bejar1999, Lucas2000, Barrado2002, pearson2023}.

These observational data suggest that freely floating planets 
may have been 1) the dominant byproducts of planet formation and
2) major agitators of planetary systems during their dissociation
from there host stars.  
Various formation pathways of FFPs 
originating from bound planetary systems have been proposed 
to account for the substantial FFP population observed. 

Planet–planet scattering during formation can generate unbound or wide-orbit planetary-mass objects \citep{zhoulin2007, ida2013}, while dynamical instabilities in mature, packed systems can excite eccentricities and disrupt planetary architectures \citep{rasio1996, weidenschilling1996, linida1997, zhoulinsun2007}. The likelihood of ejection increases for lower-mass planets, 
which require more massive perturbers to be scattered \citep{veras2005}. Additional mechanisms include stellar flybys \citep{spurzem2009, malmberg_effects_2010, zheng2015} and 
post-main-sequence evolution \citep{adams2013}. For example, circumbinary systems like 
Kepler-16 and Kepler-34 may eject numerous sub-Neptune-mass planets \citep{Coleman2023}. 
However, the abundance of ejected Earth-to-Neptune-mass planets remains poorly constrained due to 
limited studies and uncertainties in the wide-orbit population.
Observational studies of solar-type stars reveal that approximately $60\%$ of stars between 0.75 and $1.25 M_{\odot}$ exist in binary systems \citep{Offner2023}. Of these binary systems, about $23\%$ are classified as close binaries (with a separation ($a\leq 3$ au). Consequently, close binaries are present in roughly $14\%$ of all such stellar systems.
This population is significant in the context of planetary ejection. Recent work by \citet{Coleman2025} finds that dynamical interactions with the central binary in close circumbinary systems are likely the dominant progenitor for FFPs more massive than Earth. 
Conversely, planet–planet scattering in single or wide binary systems dominates the production of lower-mass planets at Mars-mass and below. 
However, in the study by \cite{Guo2025}, which examined FFPs produced by planet-planet scattering in single-star systems, a clear trend emerged: low-mass, close-in planets tend to remain bound to their host stars, while Neptune-like planets on wider orbits dominate the ejected population. This finding is broadly consistent with current observational data.

An alternative origin involves direct gravitational collapse of molecular cloud cores \citep{luhman2012}, similar to star formation. However, this requires unusually high densities significantly exceeding those in typical star-forming regions to fragment into planetary-mass objects. While this mechanism may contribute, its importance relative to ejection remains uncertain.

In this work, we explore the dynamical consequences of the assumption that most FFPs form near their host stars and are subsequently liberated. We propose that their progenitors likely originate from cold, less-bound orbits, where perturbations from distant massive bodies (e.g., stellar companions or giant planets) excite them onto high-eccentricity paths, a mechanism analogous to high-eccentricity migration invoked for hot Jupiters \citep{wu2003}. Motivated by the prevalence of compact inner systems, we investigate a previously unexplored channel involving strong scattering events between highly eccentric, massive planets and close-in, lower-mass planets. We demonstrate that such interactions can efficiently transfer orbital energy, unbinding the outer planet and producing FFP population.


We briefly recapitulate, in \S\ref{sec:excitation}, various eccentricity excitation mechanisms 
associated with high-e migration.  We derive, in
\S \ref{sec:analytic}, the physical basis for this ejection mechanism and determine an 
analytical criteria for successful ejection.  We discuss the 
role of tidal effects in \S\ref{sec:analytic_tide} based on an equilibrium-tide prescription
in \S\ref{sec:appendix}. We validate, in \S\ref{sec:simulations}, this process 
numerically using high-precision \textit{N}-body (\S\ref{sec:method}) simulations of 
systems hosting both close-in planets and a cold outer giant. 
In \S\ref{sec:tidaldamping}, we examine how tidal dissipation inside the planets and their host
stars may modify the dynamics and in \S\ref{sec:retained}, we highlight the kinematic changes
among the surviving planets.  
We summarize our findings and discuss their broad implications in \S\ref{sec:summary}.

\section{Relevant physical processes}

\begin{figure*}[ht!]
\centering
\includegraphics[width=1\columnwidth]{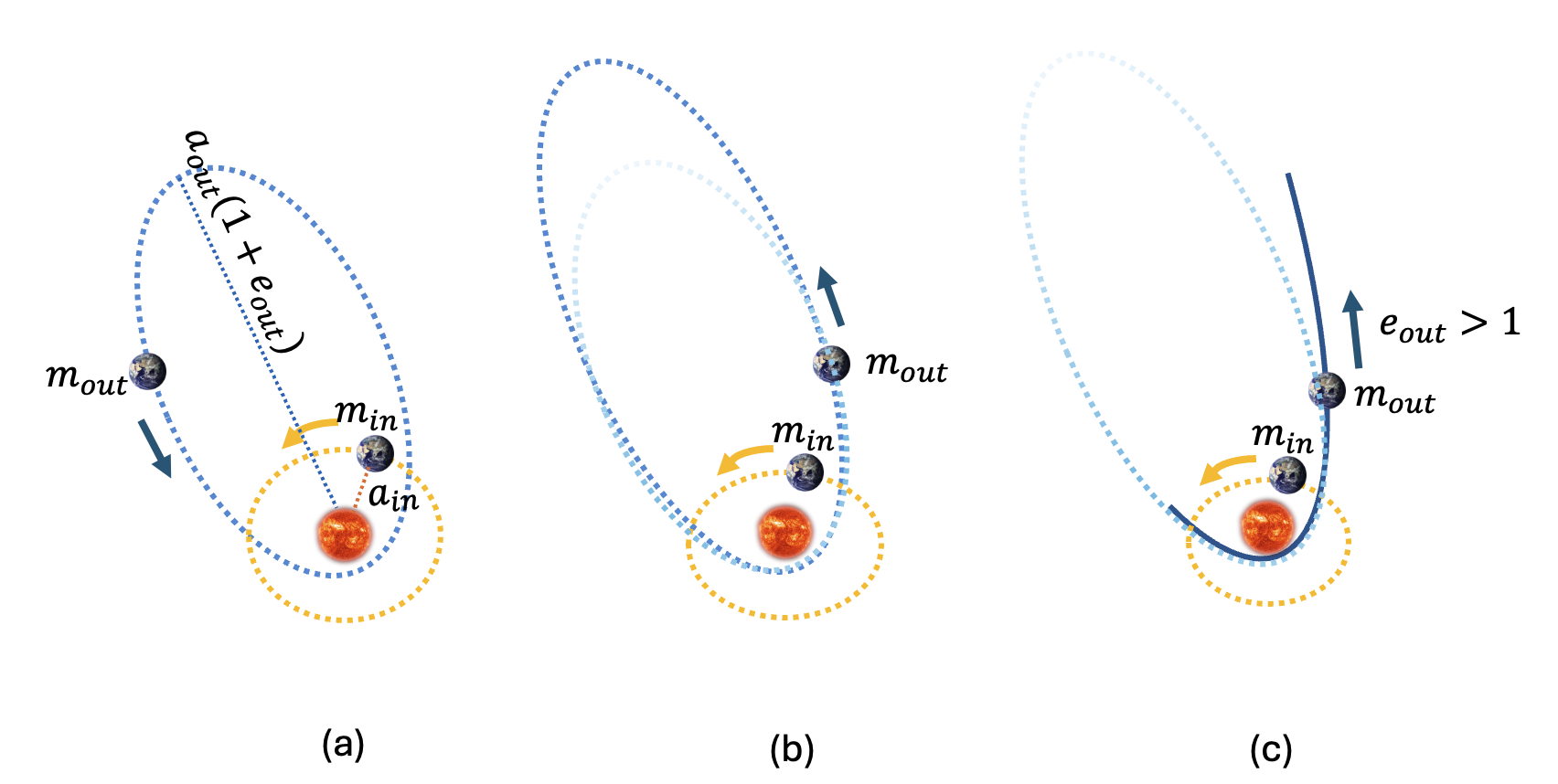}
\caption{The basic physical mechanism for producing free-floating planets through two-body scattering, depicted across three panels. Panel (a) shows an intruding cold planet ($m_{\rm in}$), characterized by an extremely eccentric orbit ($a_{\rm out}$, $e_{\rm out}$), as it intercepts a short-period super Earth ($m_{\rm in}$) with a nearly circular orbit ($a_{\rm in}$, $e_{\rm in}$). Panel (b) captures the outcome of their close gravitational encounter, where the intruding planet ($m_{\rm out}$) may gain sufficient energy to transition into a more loosely bound orbit. Panel (c) demonstrates the culmination of this process, where, after several orbital crossings, the intruding planet ($m_{\rm out}$) undergoes ejection from the system due to the cumulative energy changes from repeated interactions, ultimately becoming a free-floating object.
}
\label{fig:scenario}
\end{figure*}

In Figure~\ref{fig:scenario}, we schematically illustrate the physical mechanism for producing free-floating planets (FFPs) via gravitational scattering between close-in planets and cold, eccentric, outer intruders from an outer planetary population. For clarity, the parameters of the close-in planets, including planetary mass ($m$), semi-major axis ($a$), eccentricity ($e$), and inclination ($i$), are denoted throughout this work as $m_{\rm in}$, $a_{\rm in}$, $e_{\rm in}$ and $i_{\rm in}$, respectively. Analogously, the outer intruders are labeled $m_{\rm out}$, $a_{\rm out}$, $e_{\rm out}$ and $i_{\rm out}$.
We consider the systems with $m_{\rm out} \ge m_{\rm in}$, $a_{\rm out} \gg a_{\rm in}$, and $(1-e_{\rm out}^2) \ll 1$ such that $m_{\rm out}/2a_{\rm out}< m_{\rm in}/2a_{\rm in}$ and $m_{\rm out} \sqrt{a_{\rm out}(1-e_{\rm out}^2)} < m_{\rm in}\sqrt{a_{\rm in}}$.

As depicted in the left panel of Figure~\ref{fig:scenario}, an intruding cold planet on a highly eccentric orbit undergoes orbital crossings with the inner planetary population. During successive close encounters, the outer intruder may gain sufficient energy to attain an even more eccentric trajectory (middle panel). Ultimately, it can escape the host star’s gravitational influence entirely, as shown in the right panel of Figure~\ref{fig:scenario}.

\subsection{Excitation of Cold-planets' Nearly Parabolic Orbits}
\label{sec:excitation}

The rich populations of close-in super-Earths and distant cold Jupiters are well-established. 
We briefly recapitulate two mechanisms that can lead to high-eccentricity excitation for the
cold planets and their intrusion to the proximity of their host stars.
The underlying mechanism involves secular perturbations induced by a distant perturber. Within the context of this paper, cold perturbers encompass both unseen planets and distant stellar binary companions. 

High-resolution observations reveal that complex substructures in protoplanetary disks (PPDs) are 
ubiquitous around young stars \citep[e.g.,][]{vanBoekel2017, Avenhaus2018, Andrews2018, Long2019, Garufi2020}. These features are widely interpreted as signatures of planetary-mass perturbers 
interacting with their natal disks \citep{zheng2017b, Asensio-torres2021, Wang2022}. Stellar 
binaries are also prevalent in the universe. Over half of all star systems host one or more 
stellar companions, exhibiting diverse orbital configurations. 

The ubiquity of such distant companions, including massive planets or long-period 
stellar-mass companions profoundly influence the architectures of the inner 
planetary systems. Cold planetary populations, being nearer to these perturbers' gravitational 
sphere of influence and farther from their host stars, are particularly susceptible. 
In the following sections (\S \ref{sec:vZLK} and \S \ref{sec:SSR}), we detail 
two key mechanisms through which cold companions generate near-parabolic intruders: 
von Zeipel-Lidov-Kozai (vZLK) oscillation and sweeping secular resonance (SSR).  
Whereas vZLK is effective, requiring a modest to large inclination between the orbits
of the companion and planet, the SSR is a robust mechanism during the final depletion
of the planets' natal disk.  In \S \ref{chaos}, we also assess the dynamical feasibility 
of systems hosting both a close-in planet and an eccentric outer intruder.

\subsubsection{von Zeipel-Lidov-Kozai cycle} \label{sec:vZLK}

\begin{figure}[h]
    \centering
    \includegraphics[width=0.9\linewidth]{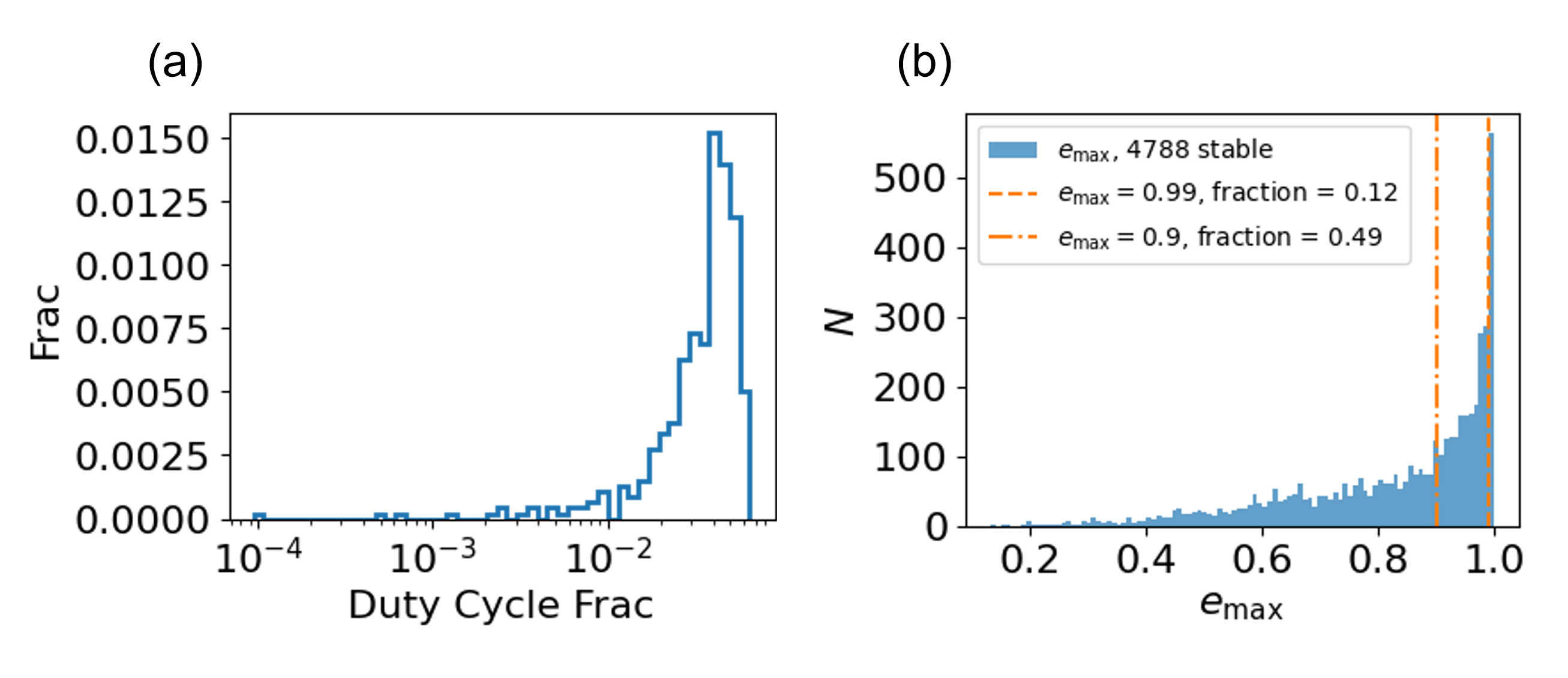}
    \caption{(a) Duty cycle fraction (the cumulative duration of extreme eccentricity excitation ($e > 0.99$) fraction) driven by the von Zeipel-Lidov-Kozai mechanism was quantified in a hierarchical triple system containing a Sun-like primary, a $M_\odot$ stellar companion ($a_{\rm pert} = 10^3$ au, $e_{\rm pert} = 0.5$), and a Jupiter-mass planet ($a_{\rm out} = 10$ au, $e_{\rm out}$ is sampled from a Rayleigh distribution, characterized by the probability density function: $f(e; \sigma) = e/\sigma^2 \exp\left(-e^2/(2\sigma^2)\right)$, where the scale parameter $\sigma \approx 0.3$. Numerical integrations over $3 \times 10^8$ years (neglecting tidal effects) demonstrate repeated eccentricity surges reaching $e_{\rm out} > 0.99$.  (b) Distribution of the theoretical maximum $e_{\rm out}$ attainable by the systems. 12\% of the systems can reach $e_{\rm out} = 0.99$, and 49\% can reach $e_{\rm out} = 0.9$. }
    \label{fig:kozai}
\end{figure}

The von Zeipel-Lidov-Kozai (vZLK) mechanism, driven by a largely inclined distant perturber 
induces significant orbital evolution through cumulative angular momentum transport \citep{vonZeipel1910, lidov1962, kozai1962, wu2003, naoz2016, bhaskar2021}. This process causes the cold planet to undergo large-amplitude inclination-eccentricity
($i-e$) modulation via resonant coupling. The characteristic vZLK oscillation timescale is 
\begin{equation}
\tau_{\rm vZLK} \simeq (M_{\star}/M_{\rm pert}) (a_{\rm pert}/a_{\rm out})^3 (1 - e_{\rm pert}^2)^{3/2} P_{\rm out} \gg P_{\rm out}
\label{eq:kozai}
\end{equation}
at $a_{\rm out} < a_{\rm pert}$, where $a_{\rm out}$ and $P_{\rm out}$ denote the semi-major axis 
and orbital period of the cold planet, while $M_{\rm pert}$,$a_{\rm pert}$, $e_{\rm pert}$ represent 
the mass and orbital parameters of the distant perturbing body. For the cold planet in the proximity 
of the perturbed, significant eccentricity excitation proceeds efficiently. We present illustrative 
numerical simulations to demonstrate that eccentricities of cold Jupiters (with $a \gtrsim 10$ au) 
can be driven to near-parabolic values by a distant perturber (at $\sim 10^3$ au) within a simulation
timescale ($\tau_{\rm vZLK} \sim 10^7 {\rm yr} \leq t_{\rm sim} \sim 3 \times 10^8$yr). 
These models establish this mechanism as a viable pathway for generating highly eccentric 
orbits for cold planet population. 

In order to quantify the cumulative duty cycle of extreme eccentricity excitation ($e > 0.99$) 
induced by the vZLK mechanism, we conducted numerical simulations using {\bf KozaiPy} 
, which is an open-source used for hierarchical 
three-body integration package (https://github.com/djmunoz/kozaipy). Our system configuration consists of a Sun-like 
primary, a solar mass stellar companion with $a_{\rm pert} = 10^3$ au and $e_{\rm pert} 
= 0.5$, and a Jupiter-mass cold planet at $a_{\rm out} = 10$ au with initial $e_{\rm out}$ 
chosen from a Rayleigh distribution with $\sigma \approx 0.3$, and a uniform inclination
distribution $-1 \leq {\rm cos}~i \leq 1$.  


First, we neglect tidal dissipation within both the planet and the host star. As shown in Figure \ref{fig:kozai}(a), the vZLK-driven high-eccentricity phase occupies approximately $3\%$ of the total simulated time—persisting over $10^6$–$10^7$ yr within a $3\times10^8$ yr integration. This duration substantially exceeds the characteristic ejection timescales of $10^3-10^6$ yr found for Jupiter-mass systems in our parameter study (see details from the later discussion on the ejection timescale). Such a temporal hierarchy confirms that vZLK-driven extreme eccentricity can indeed serve as a precursor to planetary ejection.
We simulated 5000 systems, among which 4788 remained stable. Figure \ref{fig:kozai}(b) indicates that 12\% of the systems can reach $e_{\rm out} = 0.99$, and 49\% can attain $e_{\rm out} = 0.9$. However, if the initial condition is set with $\cos(i)$ uniformly sampled from $[-0.1, 0.1]$, the percentage of systems reaching $e_{\rm out} = 0.99$ increases to 100\%. 
Therefore, the von Zeipel-Lidov-Kozai (vZLK) effect is one of the potential mechanisms for exciting an intruding planet's nearly parabolic orbit. This process is due to the 
secular perturbation induced by a more distant companion planet or star on mutually inclined planes, which leads to periodic modulation in the cold planet's eccentricity ($e_{\rm out}$) and inclination ($i_{\rm out}$).

\subsubsection{Sweeping secular resonance} \label{sec:SSR}

Besides the traditional vZLK mechanism, the sweeping secular resonance (SSR) induced by a depleting gas disk represents a promising alternative pathway for generating highly eccentric planetary orbits \citep{nagasawa2000a,nagasawa2000b,nagasawa2003, nagasawalin2005, zheng2020, zheng2021}. This mechanism operates during the lifetime of the protoplanetary disk through resonant interactions between a cold planet (initially in a low-eccentricity orbit) and a distant perturbing body. The critical physical scenario involves secular resonance, where the precession frequency of the distant perturber matches that induced in the cold planet via the disk's gravitational potential. At resonance, angular momentum is 
transferred from the cold planet to the distant perturber without any energy exchanges. This interaction leads to the monotonic excitation of the planet's eccentricity while its semi-major axis remains unchanged.  
As the disk depletes, the secular resonance location sweeps inward, traversing an extended radial range. This resonance sweeping process systematically drives eccentricity growth across multiple orbital locations, creating a population of eccentric outer intruders.
The asymptotic eccentricity is determined by the sweeping rate of the secular resonance, i.e.
the disk depletion timescale.  This mechanism shares similarities with secular resonance sweeping effects observed in our solar system dynamics, such as Jupiter's influence on asteroids' orbital evolution \citep{nagasawa2000a,zheng2017}.

\subsubsection{Secular chaos} \label{chaos}

Previous studies suggest that highly excited outer planets can facilitate marginally stable orbital configurations for close-in planetary populations \citep{nagasawa2008, nagasawa2011}. In their numerical simulations, \citet{Matsumura2013} explored the early dynamical evolution of multiple giant planets and their impact on the orbital stability of inner terrestrial worlds. Their results demonstrate that even in scenarios where orbital crossing occurs between an eccentric outer giant and a close-in terrestrial planet, such that the outer giant’s minimum pericenter distance falls within the semi-major axis of the inner planet, a significant fraction of terrestrial worlds still survives within 10 Myr. Notably, this survival rate is enhanced for planets situated close in proximity to their host stars. These findings underscore the resilience of inner planetary orbits under disruptive gravitational perturbations from eccentric, massive companions.


\subsection{Energy Exchange during Close Encounters}
\label{sec:analytic}

We consider an outer intruding planet on a nearly parabolic orbit with 
a semi-major axis $a_{\rm out} (\gg R_\star$, where $R_\star$ is the radius
of the host star) and an eccentricity $e_{\rm out} (=1-\epsilon$, where 
$\epsilon \ll 1$). This planet's total energy and angular momentum are
\begin{equation}
    E_{\rm out} = - m_{\rm out} {G M_\star \over 2 a_{\rm out}} \ \ \ \ \ \  {\rm and} \ \ \ \ \ \ 
    h_{\rm out} = m_{\rm out} \sqrt{G M_\star a_{\rm out} (1-e_{\rm out} ^2)}.
\end{equation}
when its orbit intercepts that of an inner planet with a semi-major axis 
$\sim a_{\rm in} (\ll a_{\rm out})$ and small eccentricity $e_{\rm in}$. 
At $\epsilon a_{\rm out} \ll a_{\rm in} \ll a_{\rm out}$, the outer intruder's radial
velocity ($V_{\rm out}$) and kinetic energy ($K_{\rm out}$) are 
\begin{equation}
    V_{\rm out} \simeq (2 G M_\star / a_{\rm in})^{1/2} \ \ \ \ \ \  {\rm and}
    \ \ \ \ \ \ K_{\rm out} \simeq m_{\rm out} G M_\star / a_{\rm in}.
\end{equation}

After these two planets undergo a close encounter with an impact 
parameter $b$, the impinging planets undergo a change both in velocity ($\Delta V$) and energy ($\Delta E$).
With a relative velocity $V$, the relative momentum  perpendicular to the incident relative velocity acquired by the encounter is given with the impulse approximation by a typical 
gravitational force between the two planets multiplied by a typical passing timescale:
\begin{equation}
   \frac{m_{\rm out} m_{\rm in}}{(m_{\rm out} + m_{\rm in})} \Delta V \sim {G m_{\rm out} m_{\rm in} \over b^2} {2 b \over V}.
   \label{eq:deltav1}
\end{equation}
The above impulsive approximation is adequate for close
encounters with impact parameters satisfy $(m_{\rm out} + m_{\rm in})/V^2 \ll b \lesssim R_{\rm Hill}$, 
where $R_{\rm Hill}=((m_{\rm in}+m_{\rm out})/3 M_\star)^{1/3} a_{\rm in}$ is planets mutual Hill's radius. 
Although $\Delta V$ is significantly reduced for
encounters with $b \gtrsim R_{\rm Hill}$, the outer intruder continues to cross the inner planet's
orbit.

The relative velocity $V$ between the close-in planet and the cold outer intruder for an encounter at $r_{\rm out} \sim a_{\rm in}$ is mostly given by a combination of primarily radial velocity $V_{\rm out}$ of the outer planet at 
$r_{\rm out} \sim a_{\rm in}$, and the Keplerian velocity of inner planet, $(GM_*/a_{\rm in})^{1/2}$, 
such that $V \sim V_{\rm out}$.  Since
\begin{equation}
\Delta V_{\rm out} = \frac{m_{\rm in}\Delta V}{(m_{\rm out} + m_{\rm in})},
\label{eq:deltav0}
\end{equation}
we obtain in the limit $m_{\rm in} \ll m_{\rm out}$,   
\begin{equation}
    \frac{\Delta V_{\rm out}}{V_{\rm out}}\sim {2 G m_{\rm in} \over b V_{\rm out}^2}.
    \label{eq:deltav2}
\end{equation}
In this case, while $\Delta V \propto m_{\rm out}$ 
(Equation~\ref{eq:deltav1}), $\Delta V_{\rm out} \propto \Delta V/m_{\rm out}$ 
(Equation~\ref{eq:deltav0}) so that the $m_{\rm out}$-dependence drops out in
$\Delta V_{\rm out}$ in Equation~(\ref{eq:deltav2}).
The energy change of the outer planet is 
$\Delta E \simeq m_{\rm out} V_{\rm out} \Delta V_{\rm out}$ such that 
\begin{equation}
    {\Delta E \over K_{\rm out}} \sim {\Delta V_{\rm out} \over V_{\rm out}}.
    \label{eq:deltav3}
\end{equation}
Since $K_{\rm out} \gg |E_{\rm out}|$, a small fractional change in $K_{\rm out}$ at $r_{\rm out} \sim a_{\rm in}$ can result in ejection of the outer planet. Moreover, 
$\Delta E/ K_{\rm out}$ is independent of $m_{\rm out}$ and
$\Delta E$ is also independent of $a_{\rm in}$.  

Combining these equations, we find 
\begin{equation}
    \left| {\Delta E \over E_{\rm out}} \right| \simeq {m_{\rm in} a_{\rm out} \over M_\star b} = 0.07 {m_{\rm in} \over M_\oplus}
{M_\odot \over M_\star} {a_{\rm out} \over 1~{\rm au}} {R_\oplus \over b}.
    \label{eq:deltaeovere}
\end{equation}
The above equation implies an encounter-ejection criteria, $\Delta E/E_{\rm out} \gtrsim 1$ with
\begin{equation}
    {b \over R_\oplus} \lesssim  0.07 {m_{\rm in} \over M_\oplus}
    {M_\odot \over M_\star} {a_{\rm out} \over 1~{\rm au}}.
    \label{eq:ejection_condition}
\end{equation}
Encounters with smaller impact parameters would lead to $\Delta E \gtrsim E_{\rm out}$
and hyperbolic encounters, albeit they occur less frequently (see below).  
The independence of $\vert {\Delta E / E_{\rm out}} \vert$ and $b/R_\oplus$ 
of $a_{\rm in}$ arises only for planet pairs with crossing orbits.   


Around G and M dwarf stars (with $M_\star \lesssim 1 M_\odot$), collision-free (dissipationless) encounters between a super-Earth with $m_{\rm in} \simeq 10 M_\oplus$ (regardless of $a_{\rm in}$) and an outer intruder with $m_{\rm out} \gtrsim 10 M_\oplus$ at $a_{\rm out} \gtrsim 10$au can lead to $\vert \Delta E/E_{\rm out} \vert \gtrsim 1$ (Equation \ref{eq:deltaeovere}), with either positive or negative sign, provided the sum of their radii $R_{\rm out}$ and $R_{\rm in}$ is less than $7R_\oplus$.
Such large fractional energy change introduces the ejection probability for the intruding planet.
Note that these conditions are independent of $m_{\rm out}$.
Close encounters without physical collision can also lead to the ejection of Jupiter-mass gas giants on more extended orbits ($a_{\rm out} \gtrsim 20$
au) by a super-Earth with a $m_{\rm in} \gtrsim 10 M_\oplus$ provided $R_{\rm out} + R_{\rm in}$ does not exceed 14 $R_\oplus$.

These conditions are also independent of $a_{\rm in}$ where the the planets' 
orbits cross each other.  But their occurrence frequency is 
\begin{equation}
    \omega_c \simeq (b/a_{\rm in})^2 \Omega_{\rm out} \ \ \ \ \ \ {\rm where} \ \ \ \ \ \ 
    \Omega_{\rm out} = (G M_\star /a_{\rm out}^3) ^{1/2} .
\end{equation}
If these planets' orbits persistently cross each other, 
the corresponding occurrence timescale is
\begin{equation}
    \tau_c \simeq {1 \over \omega_c} \simeq {R_\oplus^2 \over b^2} 
    {a_{\rm in}^2 \over R_\odot^2} {a_{\rm out} ^{3/2} \over {\rm au}^{3/2}} 
    10^4 {\rm yr}  \simeq 
    2 {M_\oplus^2 \over m_{\rm in}^2}
    {M_\star^2 \over M_\odot^2} {{\rm au}^{1 / 2} \over a_{\rm out}^{1/2}}
    {a_{\rm in}^2 \over R_\odot^2} 
    {\rm Myr}.
    \label{eq:tc}
\end{equation}
This associated chaotic lifetime is set by the combined rates of ejection and merger events. 
For encounters between a nearly parabolic cold planet (with $a_{\rm out} \lesssim 10^2$ au) and a
typical super-Earth (with $m_{\rm in} \gtrsim 10 M_\oplus$ and $a_{\rm in} \lesssim 0.1$ au), $\tau_c$ 
is generally much shorter than the multi-Gyr lifespan of G and M dwarfs.  
 If during a fraction ($f$) of the vZLK cycle the outer planet's peri-center distance $a_{\rm out} (1 - e_{\rm out}) < a_{\rm in}$, the occurrence timescale for an ejection or a merger event would be lengthened to $\tau_c^\prime = \tau_c/f$.

\subsection{Tidal Dissipation}
\label{sec:analytic_tide}

Tidal dissipation also leads to planet's eccentricity evolution
\citep{fabrycky2007, nagasawa2008, nagasawa2011}.  
In principle, the tidal torque is a vector which depends on the 
spin vectors of the planet ($\bm{\Omega}_{\rm p}$) and its star's
(${\bf \Omega}_\star$) as well its orbital angular momentum vector 
($\bm{A}_{\rm p}$) \citep{eggleton1998}.  It leads to changes in the 
planet's Runge-Lenz vector $\dot{\bm{e}}$ \citep{murray1999}. 
For simplicity, we consider the limiting case that all of these 
vectors are aligned.

In the equation of motion for individual planets, contribution due to the equilibrium-tidal dissipation can be incorporated into the total force \citep{1981A&A....99..126H} 
\begin{equation}
    \bm{F}=-G\frac{M m}{r^2}\left\{\bm{e}_r + 3q \left(\frac{R}{r}\right)^5 k_2
    \left[(1+3\frac{\dot{r}}{r}\tau_{\rm lag})\bm{e}_r -(\Omega - \dot{\theta})\tau_{\rm lag}
    \bm{e}_\theta \right]\right\} ,
\label{eq:tidalforce}
\end{equation}
where $M$ and $m$ represent the masses of the tidally deformed body and the perturbing companion (they can be either the intruding planet and its host star or the other way around)
respectively; $q = m/M$; $r$ is the separation distance between their centers of mass respectively;  
$R$ corresponds to the radius of the tidally affected body; $k_2$ is the perturbed body’s tidal Love
number; $\tau_{\rm lag}$ signifies a constant small time lag accounting for tidal amplitude 
and directional lag (\S\ref{sec:tidaldamping}); $\Omega$ and $\dot{\theta}$ denote the 
rotational angular velocity of the deformed body and the orbital angular velocity of the 
companion (true anomaly rate), respectively; while $\bm{e}_r$ and $\bm{e}_{\theta}$ are radial and 
tangential unit vectors in polar coordinates.  The conventional quality factor is
$Q = 1/ \omega_{\rm tide} \tau_{\rm lag}$ where $\omega_{\rm tide} = \Omega - \dot{\theta}$
is the tidal forcing frequency.

For the numerical simulations in \S\ref{sec:simulations}, we adopt $\bm{F}$ in Equation 
(\ref{eq:tidalforce}) for both the host star and the planets. For illustration purpose,
we briefly adopt, here, some simplifying approximations to highlight the main effects of
the star's tidal torque on the orbit of the intruding planet. Since the interaction between a star and its planet is symmetric, 
we can first consider only the case where the planet remains rigid. 
For deformable bodies such as gas giants, 
the calculations can be applied inversely, and the effects from both scenarios summed together. So changes in the star-planet's 
orbital angular momentum is caused by the combined tidal torque of the deformed planet (or star) on the star (or planet)  
\begin{equation}
    {\bf \Gamma}_{\rm tot}= {\bf \Gamma_\star}+{\bf \Gamma_{\rm p}} , 
    \ \ \ \ \ \ {\rm where}
\end{equation}
\begin{equation}
    {\bf \Gamma_{\rm \star, p}} = \bm{r} \times \bm{F}_{\rm \star, p}
    = 3 G\frac{M_{\star, \rm p}^2}{r}\left(\frac{R_{\rm p, \star}}
    {r}\right)^5 k_{2} ^{\rm p, \star} \left[(\Omega_{\rm p, \star} - 
    \dot{\theta}_{\rm p, \star})\tau_{\rm lag}^{\rm p, \star}
    \bm{e}_z\right] ,
\end{equation}
respectively, with $M=M_{\rm p, \star}$, $m=M_{\star, \rm p}$, $q=M_{\star, \rm p} / 
M_{\rm p, \star}$, and $R=R_{\rm p, \star}$ in Eq. \ref{eq:tidalforce}. For the 
discussion on tidal interaction, $M_{\rm p}$ and $M_\star$ refers to the mass of the 
planet and its host star.  Similarly,  $k_2 ^{\rm p, \star}$ and  $\tau_{\rm lag} 
^{\rm p, \star}$ refer to the Love number and lag time of the planet and star, respectively.

For a nearly parabolic orbit, tidal interaction occurs predominantly during peri-center passage
with $\Delta T \sim 2 \pi/{\dot \theta} (r_{\rm peri}) \sim \epsilon^{3/2} P$ where $\epsilon=1-e$, 
$r_{\rm peri}=\epsilon a$ is the perigee distance, $P= 2 \pi /\Omega_{\rm out}$ and $a$ are 
the period and semi-major axis respectively. 
Eccentricity damping is mostly induced by tidal dissipation in the planet during its peri-center
passage.  Averaged over many orbits (\S\ref{sec:appendix}), the eccentricity (as well as $a$ and $P$) evolves on a timescale 
\begin{equation}
\tau_e = {e / {\dot e}} \propto \mathcal{O} ( {\epsilon P / q^{2/3} k 
{\tilde \tau}_n {\tilde R}^5} )
\label{eq:tauemain}
\end{equation}
where $k=k_2 ^{\rm p}$, the normalized ${\tilde \tau}_{\rm n}=\tau_{\rm lag} ^{\rm p} / \Delta T$ 
and ${\tilde R}=R_{\rm p}/R_{\rm R}$ with $R_{\rm R} = q^{1/3} (1-e) a_p$ is the Roche 
radius at perigee (Eq. \ref{eq:taue}).
Adopting $R_p = R_{\rm J}$, $r_{\rm peri} = 0.1$ au, $M_{\rm p}= M_{\rm J}$,
$M_\star = 1 M_\odot$, $e_{\rm p}=0.99$, $a_{\rm p} = 10$ au (Table \ref{tab:model_f}),
we find $P \simeq 30$ yr, $r_{\rm peri}=0.1$ au, ${\tilde R} \simeq 0.05$, $\Delta T \simeq 10^6$s,
with $g_{12} \simeq 20.08$, $g_{22} \simeq 88.85$, $g_{32} \simeq 307.01$, and $g_{42} \simeq 26.87$
in Eqs. (\ref{eq:g12}) and (\ref{eq:g22g42}).
We take Jupiter's $\Omega = 2 \pi / (10~{\rm hr}) = 1.7 \times 10^{-4}~{\rm s}^{-1}$,
$k \simeq 0.3$, and $\tau^{\rm p} _{\rm lag} \simeq 10^{-7}$ yr (\S\ref{sec:tidaldamping}),
we infer from Eq. \ref{eq:tauemain},  ${\tilde \tau}_n \simeq 3.2\times 10^{-6}$, 
${\tilde \tau}_\Omega \simeq 8.9 \times 10^{-5}$. 
Applying these estimates to Eq. (\ref{eq:taue}), we find $\tau_e \gtrsim 10^8$ yr.

In our settings, the characteristic vZLK oscillation timescale $\tau_{\rm vzLK} \simeq 2 \times 10^7 {\rm yr}$ (Eq. \ref{eq:kozai}).  
The results in Fig.~\ref{fig:kozai}
show that typical duty cycle for a planet to attain $r_{\rm peri} \lesssim 0.1$ au is
$\sim 0.03 \tau_{\rm vZLK} \sim 6\times10^5$ yr.  Since this time scale is much smaller than $\tau_{\rm e}$,
we do not anticipate the tidal damping of eccentricity to significantly modify the outer planet's probability
of acquiring sufficiently high $e_{\rm out}$, through the vZLK effect, to cross the orbit
of the inner planet.  But in the limit $e_{\rm out} \gtrsim 
0.999$, $r_{\rm peri}$ is a factor of two larger than the stellar radius $R_\star$.
As ${\tilde R} \rightarrow 1$ and ${\tilde \tau}_{\rm n}$ increases, $\tau_{\rm e}$
becomes $\lesssim \tau_{\rm vZLK}$.  Tidal dissipation enhances   
orbital decay, inflation, and planet-star merger rate.   
These assessments are in agreement with the results of numerical simulations 
in \S\ref{sec:tidaldamping}.


\section{Numerical simulations}
\label{sec:simulations}

\begin{figure*}
\centering
    \includegraphics[width=0.5\columnwidth]{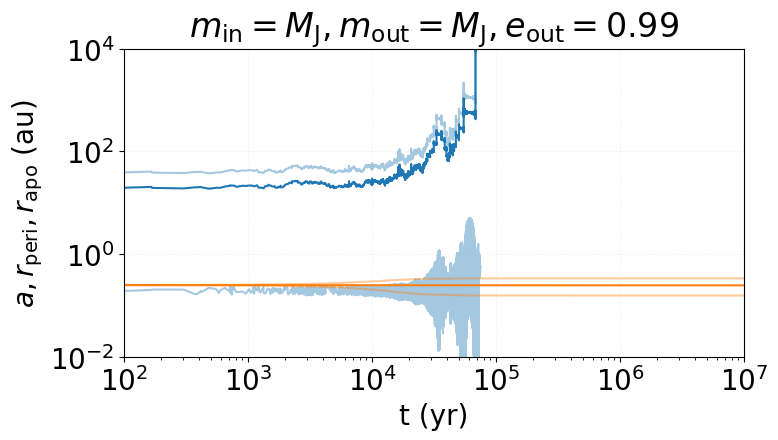}{(a)}
    \includegraphics[width=0.5\columnwidth]{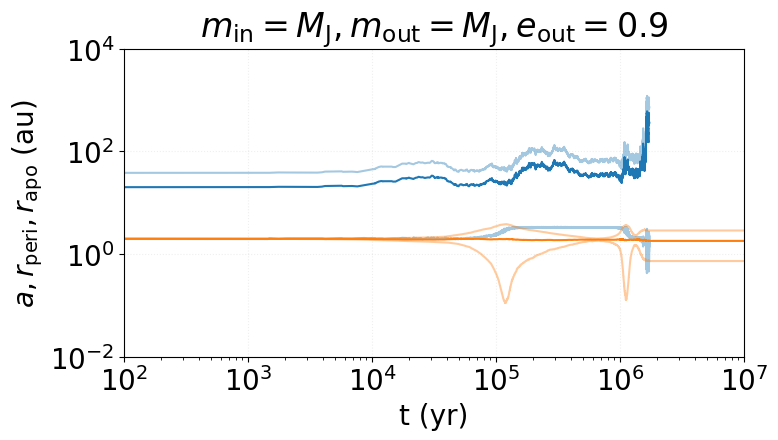} {(b)}
    \includegraphics[width=0.5\columnwidth]{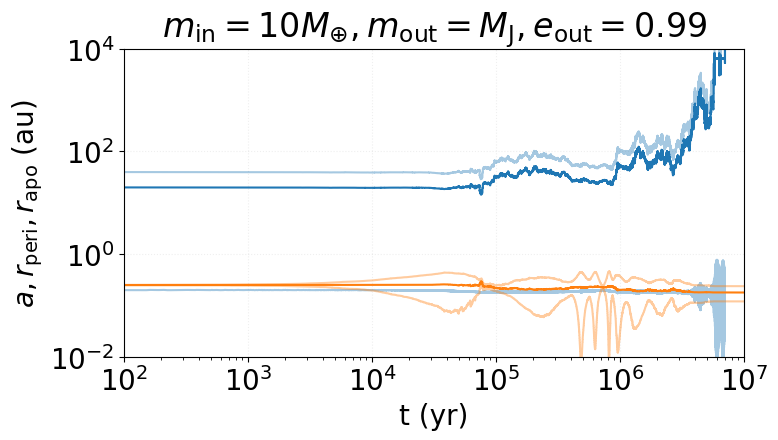}{(c)}
    \caption{Evolution of the orbital elements $r_{\rm peri}=a(1-e)$, $a$, and $r_{\rm apo}=a(1+e)$ for a 
    representative two-planet system over $10^7$ years. 
    The outer planet is represented by blue lines (dark line: semi-major axes $a$; light lines: pericenter $q$ and apocenter $Q$), while the close-in planet is shown in orange. Parameters in (a) are $m_{\rm in} = m_{\rm out} = m_{\rm J}$, $R_{\rm out} = R_{\rm in} = R_{\rm J}$, $e_{\rm out} = 0.99$, $e_{\rm in} = 0$, $a_{\rm out} = 20$ au, $a_{\rm in} = 0.25$ au. Parameters in (b) are $m_{\rm in} = m_{\rm out} = m_{\rm J}$, $R_{\rm out} = R_{\rm in} = R_{\rm J}$, $e_{\rm out} = 0.9$, $e_{\rm out} = 0.99$, $a_{\rm out} = 20$ au, $a_{\rm in} = 2$ au. Parameters in (c) are $m_{\rm in} = 10 M_{\oplus}$, $R_{\rm in} = R_{\oplus}(m_{\rm in}/M_{\oplus})^{1/3}$, $m_{\rm out} = M_{\rm J}$, $R_{\rm out} = R_{\rm J}$, $e_{\rm out} = 0.99$, $e_{\rm out} = 0.99$, $a_{\rm out} = 20$ au, $a_{\rm in} = 0.25$ au.}
\label{fig:a_t}
\end{figure*}

\subsection{Computational Method and Model Parameters}
\label{sec:method}

To explore the orbital dynamics of eccentric outer planets influenced by close encounters with the inner planetary population, we employ the open-source \textit{N}-body code {\bf REBOUND} \citep{rein_rebound:_2011}, enhanced with the {\bf REBOUNDx} library \citep{2020MNRAS.491.2885T}. This extension allows incorporation of complex additional forces into simulations, including 
velocity-dependent effects such as tidal interactions or relativistic corrections. 

For numerical integration, we utilize the IAS15 algorithm, a 15th-order, adaptive-timestep integrator \citep{everhart1985, rein2015}. As a non-symplectic method, IAS15 is designed for high-precision simulations involving close encounters or arbitrary forces. Its adaptive stepping mechanism dynamically adjusts the integration step size based on error estimates, ensuring accuracy during planetary close approaches. This robustness makes IAS15 particularly well-suited for studying orbital evolution in systems with dynamically unstable or hierarchical configurations. And all bodies have a finite size and merge directly if there are physical collisions (the distance between two particles is closer than the sum of their radii). Furthermore, the planets with positive orbital energy ($E > 0$) and located at large distances ($r > 1000$ au) from their host star are assumed to have undergone gravitational ejection from their original system.
For all the simulations presented in this paper, we run simulations up to 10 Myr, unless stated
otherwise.

To investigate the feasibility of inducing a nearly parabolic trajectory through gravitational interactions, we model a three-body dynamical system comprising a solar-like host star ($M_{\star} = M_{\odot}$, $R_{\star} = R_{\odot}$) and two planets. The intruding planet (also a designated outer planet) is initialized with a semi-major axis ($a_{\rm out}$) of either 10 au or 20 au and an eccentricity ($e_{\rm out}$) of 0.999 or 0.99 or even 0.9, mimicking highly elongated initial conditions. We consider three archetypal planet types for the outer planet: a super-Earth planet with $m_{\rm out} = 10 M_{\oplus}$ and $R_{\rm out} = R_{\oplus}(m_{\rm out}/M_{\oplus})^{1/3}$ or a Jupiter-mass giant planet with $m_{\rm out} = M_{\rm J}$ and $R_{\rm out} = R_{\rm J}$ or a Earth-like planet with $m_{\rm out} = M_{\oplus}$ and $R_{\rm out} = R_{\oplus}$. The close-in planet (also designated inner planet) is assumed to have a circular orbit ($e_{\rm in}=0$) initially throughout the work. The mass and semi-major axis of the close-in planets vary across a broad range. Taking the density divergence between rocky-like super-Earth and gaseous giant into consideration, the physical radii of the close-in planet are modeled using a piecewise function:
\begin{equation}
    R_{\rm in} = 
    \begin{cases}
R_{\oplus}(m_{\rm in}/M_{\oplus})^{1/3}  & m_{\rm in} \leq 10 M_{\oplus},\\
R_{\oplus}(10)^{1/3} + R_{\rm J}[(m_{\rm in}-10M_{\oplus})/M_{\rm J}]^{1/3} & 10 M_{\oplus} < m_{\rm in} < M_{\rm J}, \\
R_{\rm J} & m_{\rm in} = M_{\rm J} .
    \end{cases}
\end{equation}
These settings ensure a smooth physical size transition between the Jupiter-like gas giant and super-
Earth planets. For a more general stellar mass $M_\star$, the radius of solar-type stars 
$R_\star \simeq R_\odot (M_\star/M_\odot)^{0.6}$ where $R_\odot$ is the solar radius.

\subsection{Planetary ejection and retention}

Figure \ref{fig:a_t} highlights distinct dynamical behaviors of three representative 
successful-ejection events, including orbital instability and close encounters, as evidenced by abrupt changes in semi-major axes and eccentricity variations. In all cases, the outer intruder planet is initially set with a semi-major axis of $20$ au and a Jovian mass. The results indicate that a massive Jupiter-mass close-in planet is more likely to eject outer intruders effectively. 

However, this ejection mechanism exhibits inherent randomness, as the outer intruder does not monotonically gain energy during each close encounter with the close-in planet. This 
stochastic nature is reflected in the orbital evolution of the outer planet, where its semi-major axis is temporarily chaotic after some weak scattering events, and ejection only occurs following a threshold number of encounters.

According to the corresponding occurrence timescale for the ejection event as estimated in Equation~(\ref{eq:tc}), the outer intruder can be ejected by a Jupiter-mass inner planet (with $a_{\rm in} = 0.25$ au) within $\sim 10^5$ yrs. If the outer intruder is less eccentric with an initial eccentricity $e_{\rm out} = 0.9$, in order to have a close encounter with such an outer intruder, the close-in planet must be positioned at the pericenter distance (to the host star) of the outer planet as $a_{\rm in} = a_{\rm out}(1-e_{\rm out}) = 2$ au, the estimated ejection timescale increases to $\sim 10^6$ yrs. Most interestingly, even a super-Earth inner planet (with $m_{\rm in} = 10 M_{\oplus}$) is capable of kicking a Jovian planet out of the system within $10$ Myr. This inference may potentially account for the large number of unbounded Jovian mass free-floating planets predicted 
 based on the observed distribution of Einstein radius crossing timescale \citep{sumi2011}. 

 \begin{figure*}[ht!]
\centering
\includegraphics[width=0.9\columnwidth]{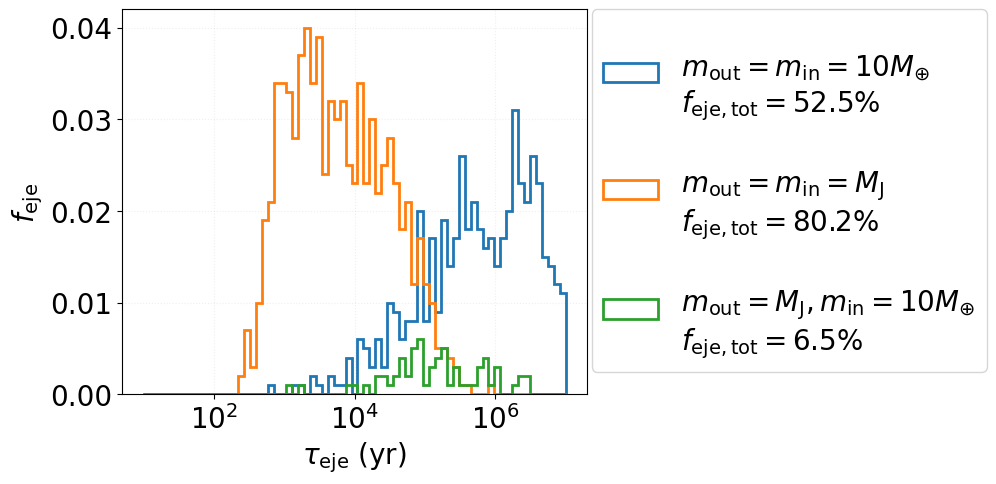}
\caption{
The distribution of ejection timescale ($t_{\rm eje}$) of the outer planet.
The fiducial settings are: $ a_{\rm out} = 10$ au, $e_{\rm out} = 0.99$, $a_{\rm in} = 0.1$ au, $e_{\rm in} = 0$. The ejection fraction is defined as $f_{\rm eje} = dN_{\rm eje}/ d\log_{10}(\tau_{\rm eje}/{\rm yr}) / N_{\rm tot}$, where $N_{\rm tot} = 1000$  is the total number of simulated systems (identical for all three cases).
The masses of the close-in planet and the outer planet are compared with various cases. We simply test three mass configurations: $m_{\rm out} = m_{\rm in} = 10 M_{\oplus}$ (blue steps), $m_{\rm out} = m_{\rm in} = M_{\rm J}$ (orange steps), and $m_{\rm out} = M_{\rm J}$, $m_{\rm in} = 10 M_{\oplus}$ (green steps). }
\label{fig:tesc_f_m}
\end{figure*}

\subsection{Ejection timescale}
\label{sec:tau_dep}

In Figure \ref{fig:tesc_f_m}, we present a statistical analysis of the ejection timescales for three distinct mass ratios between the intruding planet and the close-in planet, aiming to investigate the dependence on mass parameters. 
The blue steps correspond to the case where $m_{\rm out} = m_{\rm in} = 10 M_{\oplus}$, the orange steps represent $m_{\rm out} = m_{\rm in} = M_{\rm J}$, and the green steps indicate a scenario where the outer intruder planet is significantly more massive than the close-in planet, specifically $m_{\rm in} = 10 M_{\oplus}$ and $m_{\rm out} = M_{\rm J}$. 
The orbital parameters for the outer planet are configured with an extremely eccentric orbit ($e_{\rm out} = 0.99 $) and a semi-major axis of 10 au, while the inner planet is positioned at 0.1 au to facilitate orbital crossing with the outer planet. Each model configuration is simulated across 1000 systems to ensure statistical robustness.

Our analysis reveals that the ejection timescale of the intruding planet exhibits minimal dependence on its mass, aligning well with the analytical estimation (Eq. \ref{eq:tc}) that collisional 
conditions are independent of $m_{\rm out}$. However, the total ejection frequency demonstrates a clear 
correlation with the outer intruder's mass, as a more massive outer intruder is naturally more 
resistant to being ejected from the system. Consequently, the ejection frequency within 10 Myr 
is $\sim 50\%$ for the $m_{\rm out} = m_{\rm in} = 10 M_{\oplus}$ case and  
$\sim 10\%$ in the $m_{\rm out} = M_{\rm J}, m_{\rm in} = 10 M_{\oplus}$ model. 
In the context of intuitive planetary dynamics, a high-mass, close-in Jupiter exhibits a remarkable capacity to efficiently eject Jupiter-mass intruders from the outer system. This is exemplified by the case where both planets are Jupiter-mass ($m_{\rm out} = m_{\rm in} = M_{\rm J}$), which yields a robust ejection frequency of nearly $\sim 80\%$ within 10 Myr. These systems also exhibit short ejection timescales, with a significant fraction of outer planets removed within 0.01 Myr. This rapid timescale aligns with the estimation  $\tau_{eje} \propto 1/m_{\rm in}^2$ from Equation \ref{eq:tc}. Consequently, a Jupiter-mass inner planet leads to prompt ejections (orange step), whereas a super-Earth requires a much longer time to eject a Jupiter-mass outer planet (blue step).

This trend is further emphasized by the unstable timescale estimated in Equation (\ref{eq:tc}), which predicts that an Earth-mass inner planet would require several million years to eject the outer planet. These findings collectively underscore the critical role of the inner planet's mass in determining both the frequency and timescale of planetary ejections.

\begin{figure*}[ht!]
\centering
\includegraphics[width=1\columnwidth]{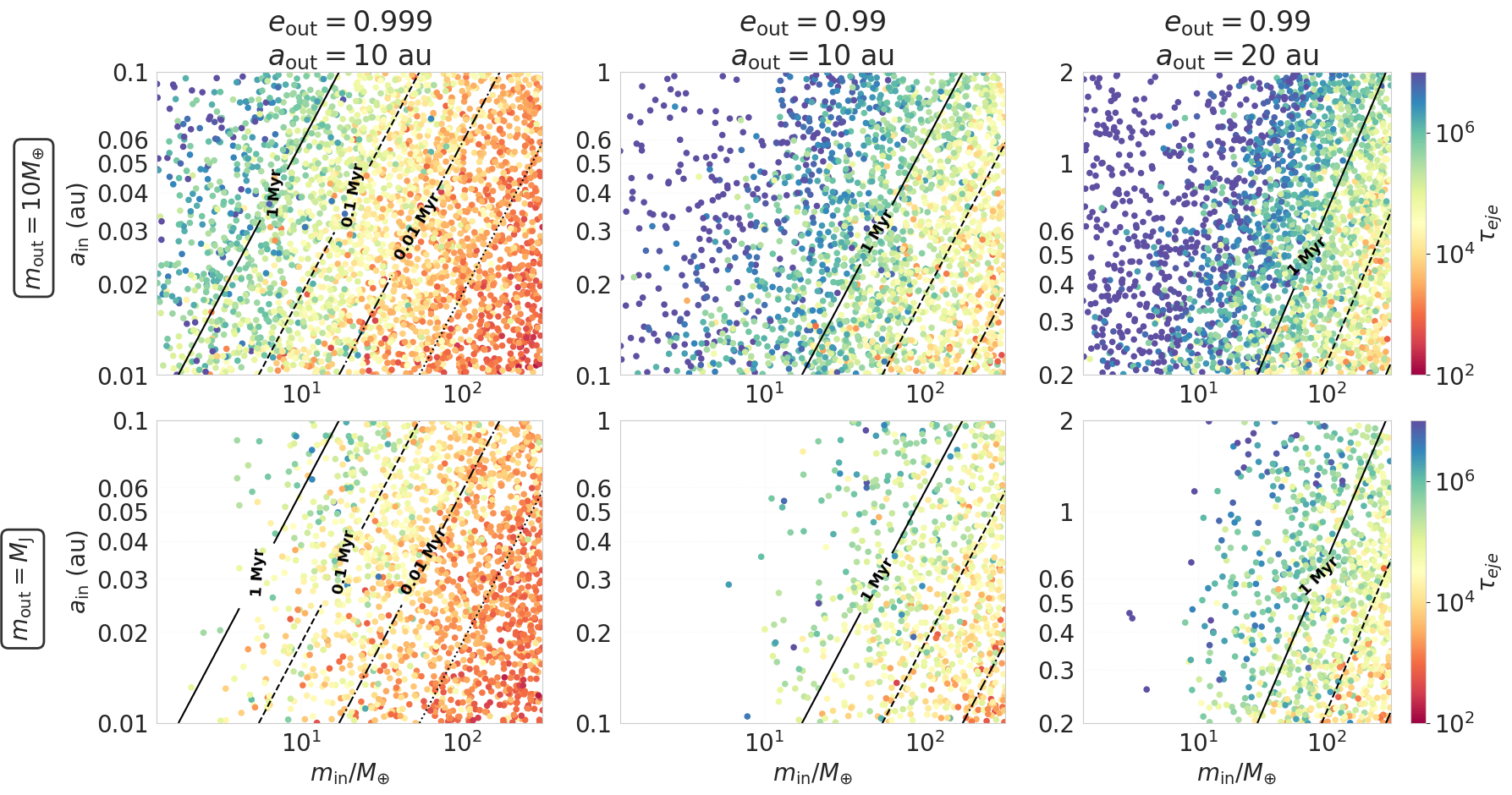}
\caption{The image illustrates how the escape timescale of the outer planet varies with the mass ($m_{\rm in}$) and initial semi-major axis ($a_{\rm in}$) of the inner planet. 
Both quantities are generated with a uniform logarithmic distribution.  
In the top panels, the outer planet is assigned a fixed mass of 10 Earth masses ($10 M_{\oplus}$), whereas in the bottom panels, it assumes a Jupiter mass ($M_{\rm J}$). Across the three panels, distinct orbital parameters of the outer planet are employed. Left panels: $a_{\rm out} = 10$ au, $e_{\rm out} = 0.999$; Middle panels: $a_{\rm out} = 10$ au, $e_{\rm out} = 0.99$; 
Right panels: $a_{\rm out} = 20$ au, $e_{\rm out} = 0.99$. 
The color bar indicates the escape timescale of the outer planet. Additionally, contours of the escaping occurrence timescale, based on Equation (\ref{eq:tc}), are plotted for comparison.
}
\label{fig:a2_e2_random}
\end{figure*}

In Section~\ref{sec:analytic} we derived the encounter–ejection criterion presented in Equation~\eqref{eq:ejection_condition}.  It reveals that an outer planet can be ejected regardless of its own mass, $m_{\rm out}$, and irrespective of the inner planet’s semi-major axis, $a_{\rm in}$. By contrast, the timescale on which instability manifests, whether through mergers or ejections, sensitive with $a_{\rm in}$  (Equation~\ref{eq:tc}). Both the instability threshold and its associated timescale do, however, depend on the outer planet’s semi-major axis, $a_{\rm out}$. This dependence is physically intuitive, the wider the orbit, the smaller the binding energy, and the more readily the planet is dislodged.  

\subsection{Stellar collision and planetary merger}

The results in Figure \ref{fig:tesc_f_m} also indicate a substantial fraction of the
system avoided ejection.  Most of these cases are associated with collisional mergers of 
the inner planets with their host stars 
(Fig. \ref{fig:a2_fesc_fmerg}). 
Angular momentum transfer from the inner to the outer planets during the close encounters
leads to the excitation of $e_{\rm in}$ with a fractional change in $a_{\rm in}$.  
We assume planets are consumed by their host stars when their $r_{\rm peri} \leq R_\star$.
In principle, the inner and outer planets can also merge when their impact parameter is less than
the sum of their radii.
Only a smaller fraction involve planet-planet collisions, with merger frequencies of  $3.3\%$, $1.1\%$ and $3.7\%$ for $m_{\rm in} = m_{\rm out} = 10 M_{\oplus}$, $m_{\rm in} = m_{\rm out} = M_{\rm J}$ and $m_{\rm in} = 10M_{\oplus}, m_{\rm out} = M_{\rm J}$ configurations, respectively.
This difference is due to planet-planet mergers requiring smaller impact parameters than that for small-angle deflections which lead to planet-star collision courses.
In the very small remaining fraction of ejection-free systems, close encounters between 
the planets can lead to significant eccentricity excitation of the inner planet and 
changes in the semi-major axis of the outer planets.
Further discussions of these outcomes are presented in \S\ref{sec:retained}.


\subsection{Ejection probability}
To validate the robustness of our analytical framework, we conducted extensive numerical simulations 
to determine the ejection probability and time scale across a broad range of $a_{\rm in}$ and 
$m_{\rm in}$ values for a fixed outer planet configuration in Figure \ref{fig:a2_e2_random}.
Specifically, we examined two distinct scenarios: one where the outer planet is a classical 
super-Earth with a mass of 10 Earth masses (illustrated in the top panels of Figure 
\ref{fig:a2_e2_random}), and another where the outer planet is a Jupiter-mass planet 
(depicted in the bottom panels of Figure \ref{fig:a2_e2_random}).

To model a more physically representative ejection scenario, we identify cold planets that remain marginally bound to the system at the simulation endpoint (10 Myr) but exhibit severely weakened gravitational binding. These systems satisfy $a_{\rm out, t=10~Myr} \gg a_{\rm out, t=0}$ and $e_{\rm out, t=10~Myr} > e_{\rm out, t=0}$, indicating extreme orbital expansion and eccentricity growth. Given their dynamical state, such planets are highly susceptible to eventual ejection. For classification 
simplicity, we assign all such cases an ejection timescale of $\tau_{\rm eje} = 100$ Myr. 

Three distinct sets of orbital elements were employed to characterize the outer planet: $a_{\rm out} = 
10$au with $e_{\rm out} = 0.999$, $a_{\rm out} = 10$au with $e_{\rm out} = 0.99$, and $a_{\rm out} = 
20$au with $e_{\rm out} = 0.99$. These correspond to the left, middle, and right panels of Figure 
\ref{fig:a2_e2_random}, respectively. The close-in planet was assigned masses ($m_{\rm in}$) with a 
uniform logarithmic distribution, masses ranging from $M_{\oplus}$ to $M_{\rm J}$. 
The semi-major axis of the inner planet's $a_{\rm in}$ was determined based on the perigee of the outer planet to ensure close encounters between the two planets during simulations. Consequently, for the outer planet with $a_{\rm out} = 10$au and $e_{\rm out} = 0.999$, the inner planet's $a_{\rm in}$
was specified with a uniform logarithmic distribution 
between 0.01au and 0.1au. For the outer planet with $a_{\rm out} = 10$au and $e_{\rm out} = 0.99$, the close-in planet's $a_{\rm in}$ was distributed between 0.1au and 1au. Finally, for the outer planet with $a_{\rm out} = 20$au and $e_{\rm out} = 0.99$, the close-in planet' $a_{\rm in}$ was distributed between 0.2au and 2au. 

Figure \ref{fig:a2_e2_random} represents a statistical analysis of the ejection timescale of the outer planet as a function of $a_{\rm in}$ and $m_{\rm in}$. The color bar indicates the ejection timescale derived from our simulations. Additionally, analytical estimations of the instability timescale (Equation \ref{eq:tc}) are overlaid as contour plots. The contours representing 1 Myr, 0.1 Myr, 0.01 Myr, and 1000 yr are depicted with solid, dashed, dash-dotted, and dotted lines, respectively. Each model comprises 4000 simulations.

As discussed in Section \ref{sec:analytic}, for a fixed outer planet, the inner planet can efficiently eject the intruder on a shorter timescale when it possesses a larger mass and a smaller semi-major axis ($\tau_c \propto a_{\rm in}^2/m_{\rm in}^2$). However, the minimum semi-major axis of an inner planet must satisfy the condition $a_{\rm in} \geq a_{\rm out}(1-e_{\rm out})$. Notably, a close-in gas giant planet can eject an extremely eccentric outer planet within 0.01 Myr, regardless of whether the outer intruder is a super-Earth or a Jupiter-mass gas giant. 
For warm super-Earths (with $m_{\rm in} \sim \text{few}~M_{\oplus}$ and $a_{\rm in} \sim 1 \ \text{au}$), it is still feasible to eject either a super-Earth or even a Jupiter from the system, albeit on a timescale extended to several Myr.

\begin{figure*}
    \centering
    \includegraphics[width=1\columnwidth]{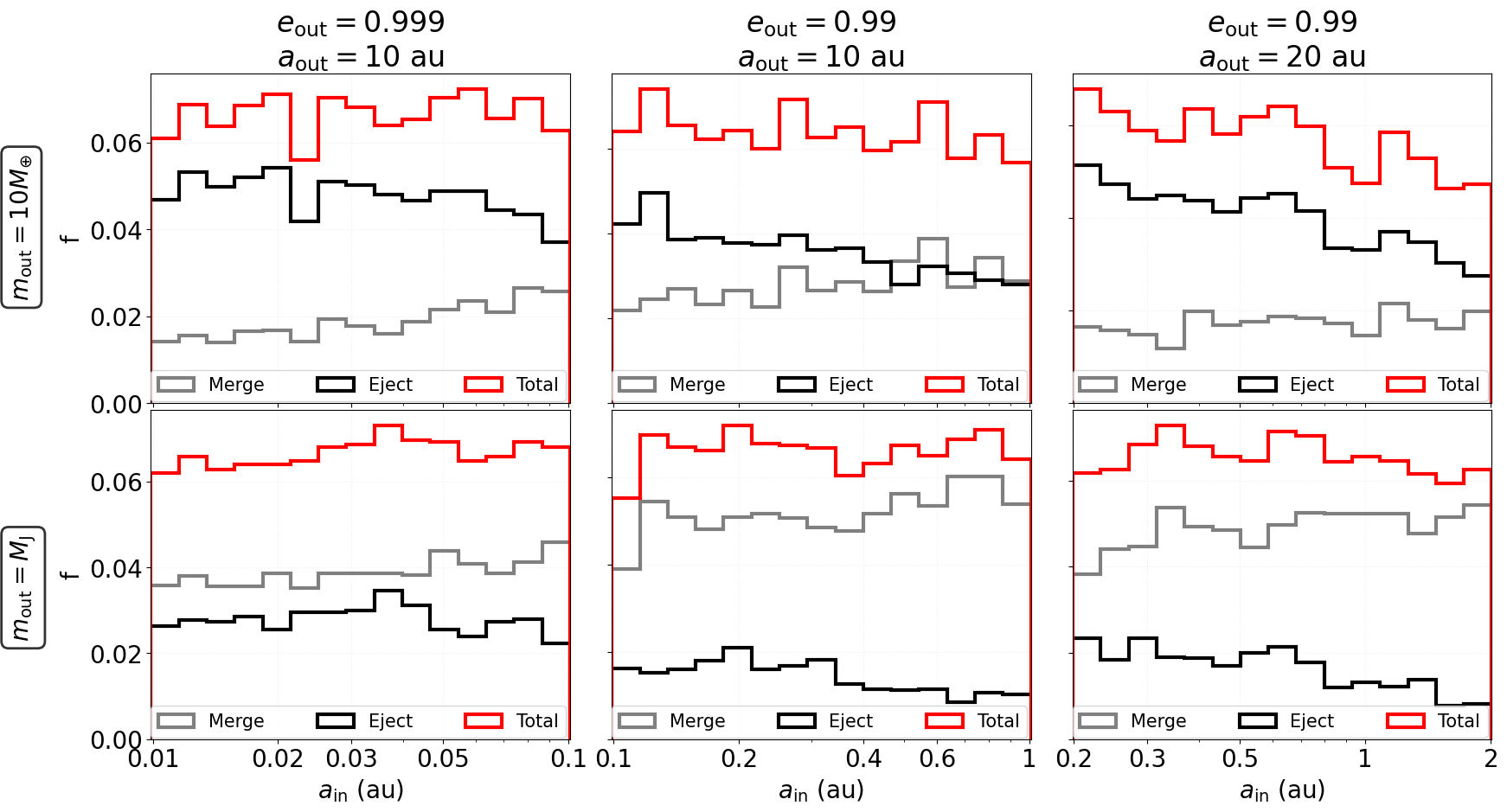}
    \caption{Fraction of planetary systems experiencing mergers or ejections. 
    These quantities are integrated over $m_{\rm in}$'s uniform logarithmic distribution 
    shown in Fig. \ref{fig:a2_e2_random}.
    The evolution of systems with an outer intruder planet is shown as a function of the initial close-in planet semi-major axis ($a_{\rm in}$). The black step function shows the fraction of systems where the intruder planet was ejected. The grey step function shows the fraction where planets merged  (mostly) with their host stars. The red step function shows the total unstable fraction (mergers + ejections). Top row: Outer intruder mass = $10 M_{\oplus}$. Bottom row: Outer intruder mass = $M_{\rm J}$. Columns (left to right):  $a_{\rm out} = 10$ au, $e_{\rm out} = 0.999$; $a_{\rm out} = 10$ au, $e_{\rm out} = 0.99$; $a_{\rm out} = 20$ au, $e_{\rm out} = 0.99$.}
    \label{fig:a2_fesc_fmerg}
\end{figure*}

In \S \ref{sec:analytic}, we demonstrated that the encounter-ejection criterion is independent of $m_{\rm out}$ and $a_{\rm in}$, though the instability timescale depends on $a_{\rm in}$. However, Figure \ref{fig:a2_e2_random} exhibits an apparent degeneracy with this conclusion, as the ejection fraction varies with both $a_{\rm in}$ and $m_{\rm out}$. 
Our analytic encounter-ejection criterion (§\ref{sec:analytic}) requires close encounters with sufficiently small impact parameters for significant energy exchange. These encounters 
include mergers (occurring when the impact parameter is smaller than the sum of the planet+star
or planet+planet radii), which compete with the ejection events. 
To resolve this discrepancy, we plot the combined fractions of ejections and mergers in 
Figure \ref{fig:a2_fesc_fmerg}. 
This quantity reveals that the total unstable fraction (mergers + ejections, red steps) is 
largely independent of $a_{\rm in}$ and $m_{\rm out}$, remaining near $\sim 6\%$ across all models.
The observed independence of the total unstable fraction from $a_{\rm in}$ and $m_{\rm out}$ aligns with Equation (\ref{eq:ejection_condition}), confirming that while the branching ratio (merger vs. ejection) depends on parameters, the onset of instability does not.

The lower panels of Figure~\ref{fig:a2_e2_random} (for $m_{\rm out}=M_{\rm J}$) show
a scarcity of ejection for systems with relatively small $m_{\rm in}$. This can be understood from energy conservation. 
The total (binding) energy of the system is $E= -GM_\star (m_{\rm in}/2 a_{\rm in})
- (m_{\rm out}/2 a_{\rm out})$, and to eject the outer planet, 
the semi-major axis of the inner planet must shrink to a value satisfying
\begin{equation}
    {a^\prime _{\rm in} \over a_{\rm in}} \leq {m_{\rm in} a_{\rm out} \over (m_{\rm in} a_{\rm out} 
    + m_{\rm out} a_{\rm in})}.
    \label{eq:aprime}
\end{equation}
This energy constraint leads to markedly different outcomes for different planetary masses.
In systems with a lower-mass outer planet ($m_{\rm out} = 10 M_{\oplus}$ at $a_{\rm out}=10~$au), 
ejecting the outer planet requires only a minor reduction in $a_{\rm in}$. 
Furthermore, the ejection condition (Eq. \ref{eq:ejection_condition}) can be satisfied with a relatively large impact parameter ($b \geq R_{\rm in} + R_{\rm out}$). As a result, the integrated ejection probability exceeds the merger probability (upper panel of Fig. \ref{fig:a2_fesc_fmerg}). This ejection probability is even higher for outer planets at $a_{\rm out}=20$ au, as they are less tightly bound.   

In contrast, ejecting a Jupiter-mass planet ($m_{\rm out}= M_{\rm J}$) requires a much more significant energy exchange. For inner planets with $m_{\rm in} \lesssim 3 M_\oplus$ and initial $a_{\rm in} \lesssim 0.1$ au, Equation (\ref{eq:aprime}) implies that $a^\prime_{\rm in}$ must decrease so drastically that the inner planet's pericenter distance falls below either the stellar radius $R_\star$ or the tidal disruption distance $d_{\rm TDE}$ (Eq. \ref{eq:ddte}). Consequently, for these systems, it is marginally more likely for the inner planet to be eliminated through a stellar merger than for the Jupiter-mass planet to be ejected as an FFP. This is consistent with the encounter-ejection criteria (Eq. \ref{eq:ejection_condition}), which for these compact systems requires a small impact parameter ($b \lesssim R_{\rm in} + R_{\rm out}$), making a merger more probable than an ejection (§\ref{sec:analytic}). The same trend is reflected in the Safronov-number perspective \citep{Safronov1972}. The gravitational focusing factor for the inner planet, $\Theta = v_{\rm esc}^2/v_{\rm Kep}^2 = 2(m_{\rm in}/M_{\star})(a_{\rm in}/R_{\rm in}) << 1$,  is very small for low-mass ($\lesssim 3 M_{\oplus}$), close-in planets. Such a low $\Theta$ indicates that physical mergers dominate over pure gravitational scattering, consistent with the outcome that stellar ingestion is more probable than the ejection of a Jupiter-mass companion.


\section{Limited tidal influence}
\label{sec:tidaldamping}



To assess the influence of tidal effects on dynamical ejections, we conducted numerical simulations using the REBOUNDx framework \citep{2020MNRAS.491.2885T} and its constant time lag tidal model
\citep{baronett_stellar_2022}. We initialized a two-gas-giant and a two-super-Earth systems, 
around a Solar-mass star. The inner and outer planets had semi-major axes of $a_{\rm in} = 0.1\ \mathrm{au}$ and $a_{\rm out} = 10\ \mathrm{au}$, respectively. The inner planet began on a circular orbit ($e_{\rm in} = 0$), while the outer planet was placed on a highly eccentric orbit ($e_{\rm out} = 0.99$). The outer planet's orbital orientation was randomized: its inclination followed an isotropic distribution ($\cos i_{\rm out} \sim \mathcal{U}[-1,1]$), and its argument of periastron ($\omega_{\rm out}$), longitude of the ascending node ($\Omega_{\rm out}$), and true anomaly ($f_{\rm out}$) were drawn uniformly from $\mathcal{U}[0^{\circ}, 360^{\circ}]$.

For tidal dissipation, we used a stellar Love number of $k_2^{\star} = 0.035$ and a planetary Love number of $k_2^{\rm p} = 0.3$ \citep{annurev:/content/journals/10.1146/annurev-astro-081913-035941}. The constant time lag for the star was fixed at $\tau_{\star} = 10^{-5}$ yr. We tested three planetary time lags ($\tau_{\rm p} = 10^{-5}, 10^{-7}, 10^{-8}$ yr) and found the ejection timescale to be largely insensitive to this parameter. Consequently, we adopted an intermediate value of $\tau_{\rm p} = 10^{-7}$ yr for all subsequent simulations for Jupiter and $\tau_{\rm p} = 10^{-5}$ yr for Super earth.

\subsection{Tidal disruption}
Planets disintegrate when their periastron is reduced below their tidal disruption distance
\begin{equation}
    d_{\rm TDE} = 1.05 (M_\star/\rho_p)^{1/3} = 1.7 R_\star (\rho_\star/\rho_{\rm p})^{1/3} ,
\label{eq:ddte}
\end{equation}
where $\rho_\star$ and $\rho_{\rm p}$ are the average density of the star and planet, respectively \citep{sridhar1992}.
For a Sun-like star, $\rho_\star \simeq \rho_{\rm p}$  for a Jupiter-mass planet and is about 0.25 times that of Earth.  These density ratios imply $d_{\rm TDE} \simeq 1.7 R_\odot$ and $R_\odot$ for Jupiter and Earth respectively. We classify gas giants with $R_\star < r_{\rm peri} \lesssim d_{\rm TDE}$ as planet-star mergers. Due to severe tidal perturbations, gas giants with $r_{\rm peri}$ 
slightly larger than $d_{\rm TDE}$ may still be disrupted after their first survivable stellar encounters 
\citep{liu2013}. We note that our model does not include tidal inflation of gas giants during circularization \citep{gu2003}, which could increase their radius, reduce density, enlarge $d_{\rm TDE}$, and further enhance the merger rate.





\subsection{Gas giant systems}
\begin{figure}
    \centering
    \includegraphics[width=1\linewidth]{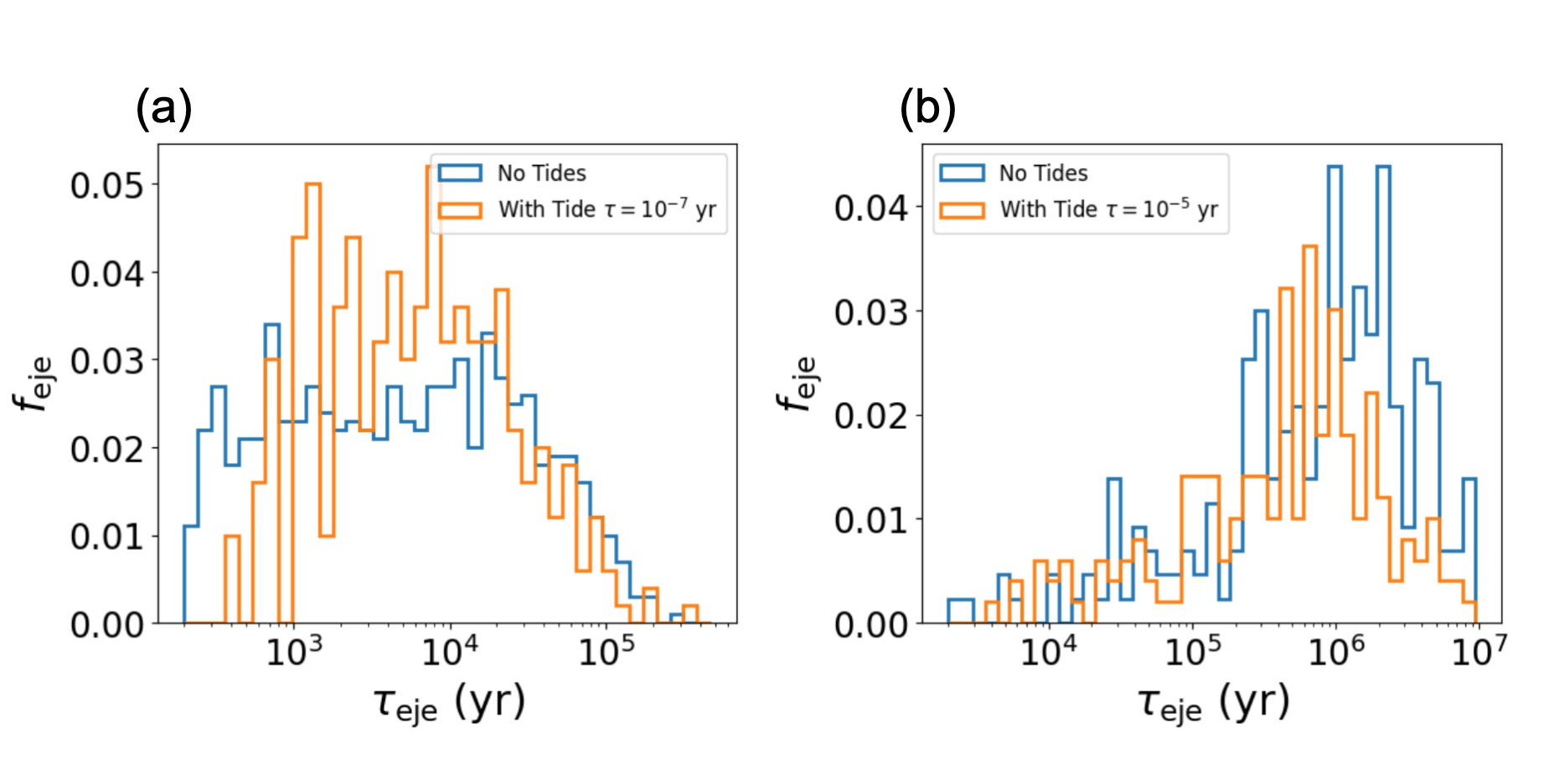}
    \caption{The ejection fraction analysis for Jupiter ($m_{\rm out} = m_{\rm in} = M_{\rm J}$
    on the left panel) or super-Earth systems ($m_{\rm out} = m_{\rm in} = 10 M_\oplus$ on the right
    panel). Here, $e_{\rm in} = 0$, $a_{\rm out} = 10$ au, $e_{\rm out}=0.99$ and $a_{\rm in}=0.1$ au. The influence of tides on the ejection probability for Jupiter remains very limited.}
    \label{fig:tide_tau}
\end{figure}

Our REBOUNDx simulations spanning 1 Myr reveal tidal influences when comparing dissipative and non-dissipative regimes. 
For the $m_{\rm in}= m_{\rm out} = M_{\rm J}$ gas giant model 
(left panel Fig. \ref{fig:tide_tau}), 
gravitational ejections dominate the outcomes, occurring in 80\% of cases for outer 
planets, while inner planet ejections remain exceptionally rare 
(Table \ref{tab:model_f}).   
This large difference is due to the outer planet has much smaller initial semi-major axis and total energy per unit mass than the inner planets. Collisions occur in $\sim 20\%$ of the systems, with more than two-thirds of these collision cases involving coalescence with the Sun (Fig.~\ref{fig:a2_fesc_fmerg}). Among these, collisions between the inner planet and the Sun constitute the majority. 

In principle,  tidal interactions modify these evolutionary pathways. When the tidal effects are 
included,  planets' rate of direct stellar impact is reduced by $\sim 5\%$ while their frequency of intrusion into the host stars' $d_{\rm TDE}$ is increased by $\sim 10\%$.  
The outer planet ejection probabilities remain nearly the same. 
Tidal dissipation inside the planets and their host stars also induces orbital decay, steering the
outer planets toward stellar collisions or tidal disruption. But at least for 
this case, there are still a significant number of ejection cases that shows 
the feasibility of producing a free-floating Jupiter through this mechanism 
is not significantly suppressed by the tidal dissipation. 

\subsection{Super-Earth systems}

For the super-Earth systems ($m_{\rm out} = m_{\rm in} = 10 M_\oplus$) with the same 
orbital elements, the tide-free ejection fraction decreases 
greatly (Table \ref{tab:model_f}), indicating that high ejection probability (53\%) remains
though slightly lower than that for Jupiter-mass systems.  
The ejection probability is slightly reduced ($\sim 10\%$) in models which include the tidal 
effects.  Since $d_{\rm TDE} \simeq R_\star$ for these high-$\rho_{\rm p}$ planets, 
the stellar collision is not enhanced by the tidal disruption process.  Moreover,
orbital evolution due to tidal dissipation in these planets and their host stars 
remains weak.  

We also consider a systems with $e_{\rm out}=0.999$ and the initial pericenter distance
of the outer planets $r_{\mathrm{peri, out}} = 0.01$au= $2 R_\star$). The super-Earth 
ejection probability from tide-free scattering remains high ($\sim 52\%$).  However, 
with the tidal effect, the ejection probability decreases by a factor of $\gtrsim 2$ 
(Table \ref{tab:model_f}) despite $d_{\rm TDE} = R_\star$. This reduction is 
accompanied by a large increase in the planet-star merger rate.  With $\tau_{\rm e}$ being 
a steeply increasing function  of either $r_{\rm peri, in}$ or $r_{\rm peri, out}$ 
(Eq. \ref{eq:tauemain}), tidal influence on their semi major axis and eccentricity evolution 
intensifies (Eqs. \ref{eq:adottot} and \ref{eq:edottot}) for close-in planets and their host
stars.  Planets with such small periastron distances   
undergo orbital decay on a timescale shorter than the von Zeipel-Lidov-Kozai (vZLK) timescale 
($\tau_{\rm e} \lesssim \tau_{\rm vZLK}$, \S\ref{sec:analytic_tide}), leading to prolific mergers
with their host stars.


\begin{table}[h]
\centering
\caption{Statistical frequencies of planetary ejection, merging, and retention across various models within 10 Myr evolution. Each model presents results for both tidal ($+$) and non-tidal ($-$) circumstances. }
\begin{threeparttable}
\setlength{\tabcolsep}{8pt}
\begin{tabular}{@{}cccccccc@{}}
\toprule
$m_{\rm out}$ & $m_{\rm in}$ & $a_{\rm out}$ & $a_{\rm in}$ & $e_{\rm out}$ & Tide & $f_{\rm eje}$\tnote{1} & $f_{\rm eje, in}$\tnote{2}\\
& & (au) & (au) & & & (\%) & (\%) \\
\midrule
$10 M_{\oplus}$ & $10 M_{\oplus}$ & 10 & 0.1 & 0.99 & $-$ & 52.5  & 0.0  \\
& & & & & $+$ & 36.3  & 0.0   \\
\addlinespace
$10 M_{\oplus}$ & $10 M_{\oplus}$ & 10 & 0.1 & 0.999 & $-$ & 51.5  & 0.0  \\
& & & & & $+$ & 27.4  & 0.0  \\
\addlinespace
$M_{\rm J}$ & $M_{\rm J}$ & 10 & 0.1 & 0.99 & $-$ & 80.2  & 0.0 \\
& & & & & $+$ & 77.4  & 0.02 \\
\addlinespace
$M_{\rm J}$ & $10 M_{\oplus}$ & 10 & 0.1 & 0.99 & $-$ & 6.5  & 0.0  \\
& & & & & $+$ & 6.8  & 0.02  \\
\bottomrule
\end{tabular}

\begin{tablenotes}
\item[1] Ejection rate of the intruder planet. 
\item[2] Ejection rate of the close-in planet.
\end{tablenotes}
\end{threeparttable}
\label{tab:model_f}
\end{table}

\section{Orbits of the Retained Population}
\label{sec:retained}

We now examine the changes in orbital elements for planets that survive the dynamical instability. The outcomes are presented separately for systems where the outer planet is ejected and for those where a planet-star merger occurs.  Planet-planet mergers are rare compared to planet-star mergers. Furthermore, if the outer planet collides with the star, the inner planet is typically also consumed, clearing the system entirely. Therefore, we focus on the subset of systems where only the inner planet is consumed.
We analyze three model types: a) super-Earths only ($m_{\rm in}=m_{\rm out}=
10 M_\oplus$), b) Jupiters only ($m_{\rm in}=m_{\rm out}= M_{\rm J}$), and c) mixed systems ($m_{\rm in}=10 M_\oplus$, 
$m_{\rm out}= M_{\rm J}$). All models have $a_{\rm out}=10$ au,  $a_{\rm in}=0.1$ au, $e_{\rm out}=0.99$, $e_{\rm in}=0$, and neglect tidal effects.

\begin{figure}
    \centering
    \includegraphics[width=1\linewidth]{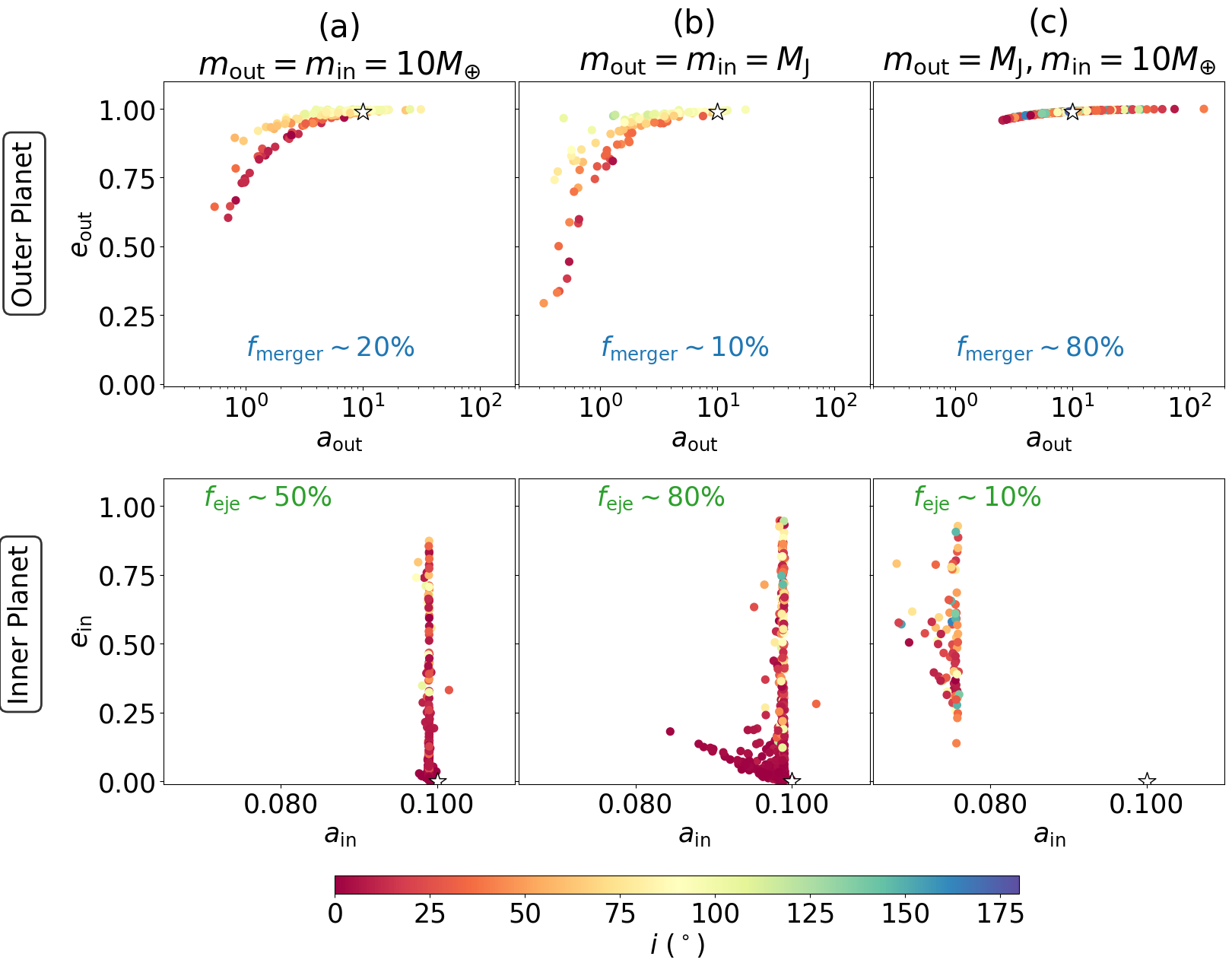}
    \caption{ {\bf Top Panels}: Final orbital elements of the outer planet remaining after the merger of its close-in companion with the host star. {\bf Bottom Panels:} Final orbital elements of the close-in planet remaining after the ejection of the outer planet. 
    All three configurations share the following initial conditions: an outer planet semi-major axis of $a_{\rm out} = 10$ au ($=100~a_{\rm in}$), an outer planet eccentricity of $e_{\rm out} = 0.99$, and an inner planet eccentricity of $e_{\rm in} = 0$. 
    The panels vary the planetary mass ratios, {\bf Left (a):} $m_{\rm out} = m_{\rm in} = 10 M_{\oplus}$; {\bf Center (b):} $m_{\rm out} = m_{\rm in} = M_{\rm J}$; {\bf Right (c):} $m_{\rm out} = M_{\rm J}, m_{\rm in} = 10 M_{\oplus}$.
    The top panels are annotated with the fraction ($f_{\rm merger}$) of simulations where 
    the inner planet merged with the host star. The bottom panels show the ejection fraction 
    ($f_{\rm eje}$) of the outer planet. 
    The initial conditions for each planet are marked by an open black star in its respective final-state panel. Color bars indicate the final inclination of the outer planet (top) and the inner planet (bottom). 
    }
    \label{fig:merge}
\end{figure}

The lower panels of Fig. \ref{fig:merge} show systems where the outer planet is ejected. Because the inner planet is more tightly bound, its ejection is far less likely than that of the outer planet ($f_{\rm eje, in} \ll f_{\rm eje}$, Table \ref{tab:model_f}).
The orbital energy conservation (Eq. \ref{eq:aprime}) dictates that the surviving inner planets undergo significant modification, characterized by a reduction in their semi-major axes, $a^\prime _{\rm in}$, and excitation of their eccentricities. While these planets begin on circular orbits, the outcome is a broad eccentricity distribution with a prominent peak near 0.1. The model with $m_{\rm out} = M_{\rm J}, m_{\rm in} = 10 M_{\oplus}$ (bottom-right panel of Fig. \ref{fig:merge})
is a notable exception. Here, the exceptionally large energy exchange leads to a severe alteration of the semi-major axis and drives the average eccentricity to roughly 0.5.
Inner planets with $a^\prime _{\rm in} 
(1-e^\prime _{\rm in}) \leq R_\star$ merge with the host stars. Subsequent tidal circularization would further reduce $a_{\rm in}$, potentially forming ultrashort-period planets.

The ejection process also randomizes orbital inclinations. While the inner planets begin in a coplanar disk aligned with the stellar spin, the gravitational interaction with the ejected outer planet leaves the survivors with a nearly uniform inclination distribution, including retrograde orbits (Fig. \ref{fig:merge}, yellow to blue dots). This implies the generation of a broad star-planet obliquity distribution. This mechanism is a variant of the conventional scenario for spin-orbit misalignment, which typically invokes the vZLK effect on single-planet systems \citep{wu2003}. The existence of planets on retrograde, nearly parabolic orbits with small inclinations could thus serve as dynamical fossils, indicating a past FFP ejection event.

We next consider changes to the outer planet's orbit following the merger of the inner planet with the host star. The upper panels of Fig. \ref{fig:merge} reveal distinct outcomes:
\begin{itemize}
\item 
In Jupiter-only systems (model b), the most common outcomes are the ejection of the outer planet or the consumption of both planets by the star. Very few single planets survive the merger of their partner with the host star.

\item 
A larger fraction of super-Earth systems (model a) produce single survivors after the inner planets merge with the host stars.

\item 
In mixed systems (model c), the mergers of the inner super-Earths with the stars significantly
alter the orbits of the outer gas giants in approximately half of the cases.
\end{itemize}

The semi-major axes of the surviving outer planets are dispersed widely from their initial values (open black stars in Fig. \ref{fig:merge}). As the inner planet loses angular momentum to collide with the star, the outer planet can either gain or lose orbital energy during its close encounters, leading to either an increase or decrease in $a_{\rm out}$. This angular momentum exchange also circularizes the outer planet's orbit, reducing its eccentricity from the initial value of 0.99, though the final eccentricities still cluster near unity. The initial inclination distribution of the survivors is largely preserved.

\section{Estimated Contribution to the Free-Floating Planet Population}

Of Sun‑like stars, about $\sim 30\%$ ($P_{\rm SE}$) host inner super‑Earths (SE) and roughly $\sim 10\%$ ($P_{\rm CJ}$) host a cold Jupiter (CJ). Studies indicate that nearly $\sim 90\%$ of cold Jupiters are expected to have interior small planets \citep{ZhuWu2018, Bryan2019}. Among these inner super‑Earths, approximately one‑third are hot super‑Earths (HSE, $R_{\rm in} < 4 R_{\oplus} $) with orbital periods $P_{\rm in} < 10$ days. Based on transit and radial‑velocity samples, it is estimated that about $40\%$ of Sun‑like stars with a cold Jupiter also host a hot super‑Earth (Liu, Zhu, et al. in prep), i.e. $P_{\rm HSE|CJ} \approx 0.4$.

\begin{figure}[h]
    \centering
    \includegraphics[width=0.9\linewidth]{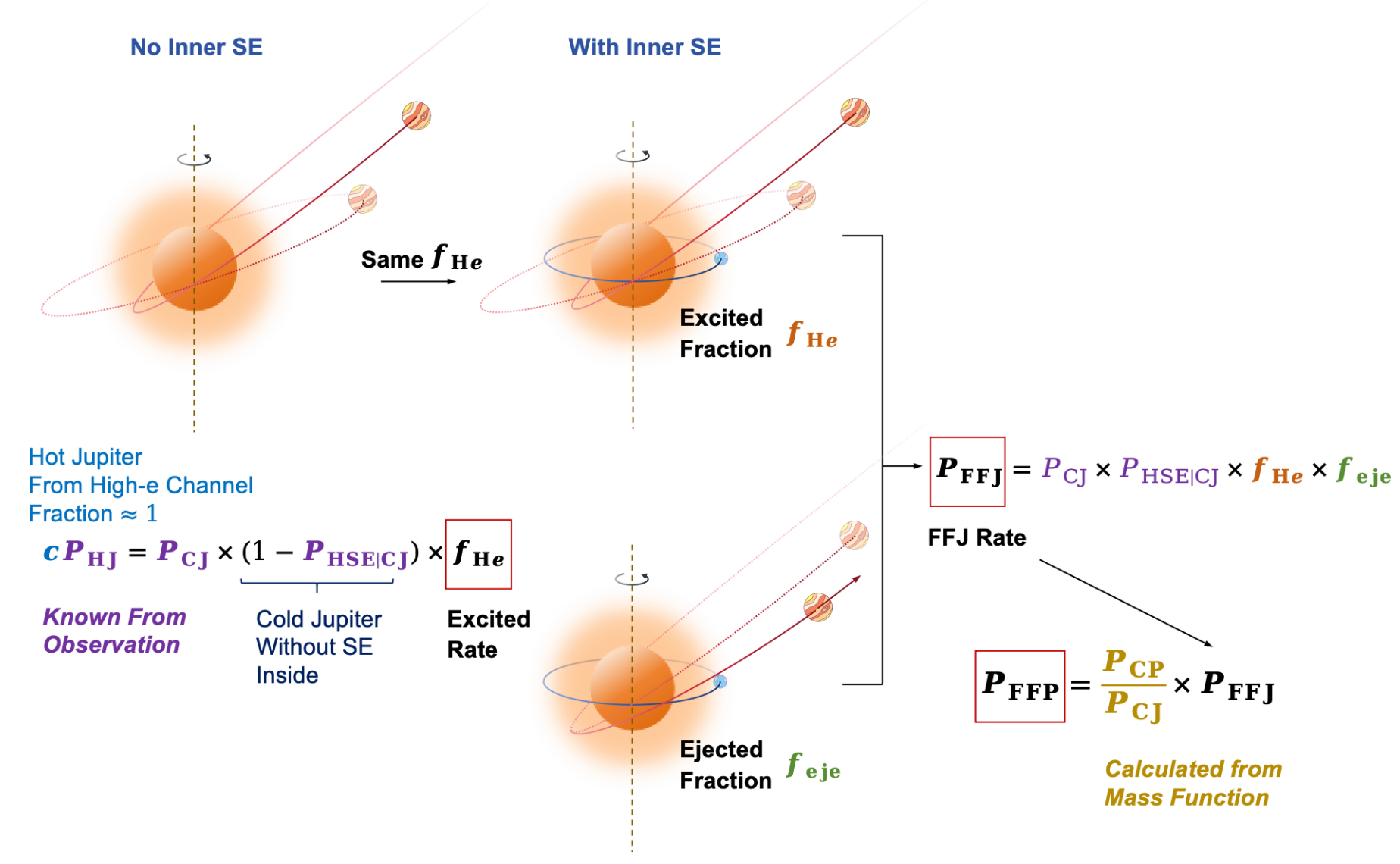}
    \caption{A flowchart illustrating the methodology for estimating the approximate production rate of FFPs (free-floating planets). First, the excitation frequency of cold Jupiters is estimated using systems containing a cold Jupiter but no inner super-Earth. This frequency is assumed to be identical for cold Jupiter systems with an inner super-Earth. Subsequently, the estimated FFJ (free-floating Jupiter) production rate is derived. Finally, by applying the mass function to obtain the ratio of cold planets to cold Jupiters, the overall FFP production rate is calculated.}
    \label{fig:flow_chart}
\end{figure}

The observed occurrence rate of hot Jupiters (HJ) is around $1\%$ ($P_{\rm HJ}$) \citep{howard2012, santerne2016}. However, studies such as \citep{Gan2023} report a lower rate of $\sim 0.2\%$ around M dwarfs. 
Hot Jupiters can form either via disk migration or high-eccentricity migration. We denote by $c$ the fraction of hot Jupiters that form through the high-eccentricity migration channel. 
If we assume that they form exclusively via the high‑eccentricity migration channel{\color{purple} ($c\approx1$)}, thus only those cold Jupiters that do not harbor an inner hot super‑Earth can be excited to high eccentricity by external perturbers—a process independent of the inner system—then the fraction of cold Jupiters that undergo such excitation, $ f_{\rm He}$, can be approximated by

\begin{equation}
 cP_{\rm HJ} = P_{\rm CJ} \times \bigl(1 - P_{\rm HSE|CJ}\bigr) \times f_{\rm He}.   
\end{equation}

Inserting the observational values yields $f_{\rm He} \approx 0.01 / [0.1 \times (1 - 0.4)] \approx 17\%$. 
This estimate is consistent with the simulation results obtained by \citet{Ochiai2014}. Their simulations show that, among cold Jupiters experiencing orbital instability, approximately $15-20\%$ undergo eccentricity excitation to $e \sim 1$. Consequently, if the fraction of cold Jupiters that undergo orbital instability is $\sim 50-80\%$, the resulting fraction $f_{\mathrm{He}}$ is estimated to be $\sim 7.5-16\%$. 

Using this excitation rate, we can then estimate the production rate of free‑floating Jupiters (FFJs) through close encounters between excited cold Jupiters and inner hot super‑Earths:

\begin{equation}
 P_{\rm FFJ} = P_{\rm CJ} \times P_{\rm HSE|CJ} \times f_{\rm He} \times f_{\rm eje},   
\end{equation}
where $f_{\rm eje}$ is the ejection fraction during such encounters. 
According to \citet{Ochiai2014}, the ejection rate of cold Jupiter without an inner super-Earth is 40\%-60\%. Based on our results, assuming the ejection frequency of cold Jupiter with a close-in super-Earth 
is nearly certain ($f_{\rm eje} \approx 1$) in this regime, the resulting FFP production rate is $\approx 0.7\%$.

However, this FFJ production rate applies only to objects of Jovian mass. To extrapolate this rate to the total population of free-floating planets, we must consider the complete mass distribution of cold planets derived from microlensing surveys \citep{zang2025}:

\begin{equation}
\frac{dN}{d~{\rm log}~q} = f({\rm log}~q), \quad {\rm where} \quad f({\rm log}~q) \approx 0.52 \times 10^{-0.73 \times ({\rm log}~q/7.4q_\oplus)^2} + 0.058 \times 10^{-1.8 \times ({\rm log}~q/770q_{\oplus} )^2}.
\end{equation}

Within this distribution, cold Jupiters (defined here as objects with masses between Saturn's mass and $13\,M_{\rm J}$) constitute only a small fraction of the total cold-planet population. This fraction, $P_{\rm CJ}/P_{\rm CP}$, is given by:

\begin{equation}
    \frac{P_{\rm CJ}}{P_{\rm CP}} = \frac{\int_{{\rm log}~(M_{\rm Sat}/M_{\odot})}^{{\rm log}~(13 M_{\rm J}/M_{\odot})} f({\rm log}~q) \, d{\rm log}~q}{{\int_{-6}^{-1.5}} f({\rm log}~q) \, d{\rm log}~q} \approx 0.085.
\end{equation}

The occurrence rate of cold Jupiters around M dwarfs is approximately one-fourth of the corresponding frequency inferred for such planets orbiting FGK stars (\citealt{Clanton2014, Montet2014}; see also \citealt{GanGuoMao2024}). Therefore, for an order-of-magnitude estimate, the occurrence rate for M dwarfs may serve as a reasonable approximation for that of Sun-like stars. Furthermore, population-synthesis studies utilizing a realistic IMF and microlensing constraints support the plausibility of high ejection efficiencies also in M-dwarf-dominated stellar populations \citep{Guo2025}.

\citet{Guo2025} concluded that the dominant population contributing to the free-floating planet population are planets of Neptune-like masses. 
The estimated total free-floating planet production rate arises from dynamical encounters between cold planets (across all masses) and close-in super-Earths. This rate can therefore be scaled from the free-floating Jupiter ($P_{\rm FFJ}$) rate as follows:

\begin{equation}
    P_{\rm FFP} = P_{\rm FFJ} \times \frac{P_{\rm CP}}{P_{\rm CJ}} \approx 8 \% .
\end{equation}

Finally, it is important to note that this estimate is conservative. It is derived specifically from interactions involving the observed populations of close-in super-Earths and cold planets. In reality, other populations—such as hot Neptunes and even hot Jupiters—can also eject cold planets, potentially contributing to a higher overall FFP production rate than calculated here.

\section{Summary and discussions}
\label{sec:summary}


Recent microlensing surveys indicate that freely floating planets (FFPs) may outnumber known bound planets. This work explores a viable formation channel for FFPs arising from gravitational encounters between distinct planetary populations within a planetary system. 

Our study focuses specifically on the post-gas-disk phase, when the gas disk is severely depleted or absent. At this stage, high-eccentricity tidal migration of cold Jupiters is a leading scenario for forming hot Jupiters. We adopt this established framework as our starting point, asking a subsequent question: if such a highly eccentric giant planet progenitor exists, what is the dynamical fate of such a planetary system that already hosts close-in planets?


Close-in planets are likely to have formed at larger orbital distances and subsequently migrated to their current locations via disk migration \citep{lin1996} or high-eccentricity migration \citep{wu2003}. High-eccentricity migration, in particular, proposes that these planets were injected toward the star by perturbations from distant companions, through mechanisms such as the von Zeipel–Lidov–Kozai (vZLK) effect, sweeping secular resonances, or secular chaos \citep{ida2004}. 


Our work investigates a logical outcome of such dynamical processes: when multiple planets are driven to small periastron distances, or when a long-period intruding planet enters the orbital region of existing close-in planets, their orbits may intersect. During close encounters, strong two-body scattering can take place. Since the intruding planet possesses significantly lower orbital energy per unit mass ($-GM_\star/2a$) than the close-in planets, even a modest fractional energy exchange can lead to the ejection of the intruder from the system, forming an FFP. We analyze this FFP-formation channel using both analytical estimates (\S \ref{sec:analytic}) and numerical simulations (\S \ref{sec:simulations}). Our results demonstrate that such interactions provide an efficient and robust mechanism for the production of free-floating planets.

On the technical side, this paper analyzes the energy exchange during close encounters between intruding cold planets and close-in planets. Using an impulse approximation, we derive an analytic expression for the critical impact parameter that leads to significant energy exchange and potential ejection. This quantity (Eq.~\ref{eq:ejection_condition}) depends on the inner planet's mass ($m_{\rm in}$) and the outer planet's semi-major axis ($a_{\rm out}$), but is independent of $a_{\rm in}$ and $m_{\rm out}$. Consequently, under certain conditions:
\begin{itemize}
    \item Close encounters with short-period super-Earths can eject more massive, long-period gas giants.
\end{itemize}


This analytical prediction is validated by numerical simulations (\S\ref{sec:simulations}), which demonstrate that both the ejection timescales and frequencies depend on the planetary mass ratio and the semi-major axis of the close-in planet (Fig.~\ref{fig:a2_e2_random}), consistent with the theoretical framework in \S\ref{sec:analytic}. We further find that:
\begin{itemize}
    \item As inner planets, hot Jupiters are able to eject nearly all intruding long-period planets (Table~\ref{tab:model_f}).
\end{itemize}



This result implies that the presence of hot Jupiters would prevent the implantation of nearby planets through high-eccentricity migration. Any closely-packed multiple planets in the vicinity of hot Jupiters, if they exist, must have acquired their present-day orbital configurations through disk migration.

Close-in ($a_{\rm in} \leq 0.1$ au) super-Earths are significantly more common than hot Jupiters. 
\begin{itemize}
    \item They also exhibit a high probability of ejecting intruding cold super-Earths (Fig.~\ref{fig:tesc_f_m}).
\end{itemize}


Comets greatly outnumber planets. In a subsequent study, we will demonstrate that their ejection via close encounters with close-in super-Earths provides a robust production mechanism for interstellar comets and asteroids \citep{zhang2020}. When their dynamical domain is encroached by cold gas giants,
\begin{itemize}
    \item close-in super-Earths become vulnerable to $e_{\rm in}$-excitation and are susceptible to stellar merger (Fig.~\ref{fig:a2_fesc_fmerg}).
\end{itemize}


Profuse stellar consumption of super-Earths can produce observable abundance dispersions among stars in young clusters \citep{shen2005ApJ} and metallicity differences in binary systems \citep{gonzales2013, liu2014, saffe2019}.
\begin{itemize}
    \item The substantial depletion of close-in super-Earths consequently reduces their efficiency in ejecting gas giants (Table~\ref{tab:model_f}).
\end{itemize}


The preferential ejection probability of intruding cold super-Earths over gas giants aligns with the mass distribution of free-floating planets (FFPs) inferred from microlensing surveys. In addition to stellar collisions, super-Earths exhibit the following behavior:
\begin{itemize}
    \item Close-in super-Earths have a small but non-negligible probability of undergoing mergers with intruding gas giants (Table~\ref{tab:model_f}).
\end{itemize}


Planetary mergers can contribute to the diversity in the internal structures of gas giant planets \citep{liu2015, liu2019}. Although tidal dissipation and disruption can increase the likelihood of mergers (\S\ref{sec:analytic_tide} and \S\ref{sec:tidaldamping}),
\begin{itemize}
    \item the ejection mechanism described here remains a viable formation channel for FFPs (Fig.~\ref{fig:tide_tau}).
\end{itemize}


Following the ejection of the outer planets, the orbital elements of the surviving inner planets exhibit significant modifications:
\begin{itemize}
    \item their eccentricities become markedly excited, in some cases reaching retrograde configurations; and
    \item their semi-major axes experience a fractional decrease (lower panel of Fig.~\ref{fig:merge}).
\end{itemize}
Both outcomes consequently induce subsequent tidal orbital evolution. A natural extension of this work is to investigate the stability and potential disintegration of multiple close-in super-Earth systems when perturbed by intruding super-Earths and gas giants.


In models where the inner planets experience collisions with their host star, the orbital parameters of the surviving outer planets undergo systematic modifications:
\begin{itemize}
    \item their semi-major axes spread out by approximately a factor of two; and
    \item their eccentricities are reduced by up to a factor of two (upper panel of Fig.~\ref{fig:merge}).
\end{itemize}
In a subsequent investigation, we intend to examine systems with long-period inner planets and explore whether their close encounters with long-period gas giants on nearly parabolic orbits could lead to the formation of eccentric warm Jupiters. 
Within this post-disk context, the presence of a close-in super-Earth—a common outcome of in-situ formation or disk migration—can efficiently halt high-eccentricity migration. Through strong gravitational scattering, it can eject the incoming giant planet, thereby protecting the inner terrestrial environment and contributing to the free-floating planet population. This result highlights a critical constraint on the high-eccentricity migration channel.
We explore implications for the observed orbital alignments of warm and hot Jupiters further in a follow-up study (Cao et al., in prep).









\begin{acknowledgments}
We thank Drs. Tianjun Gan, Wei Zhu, and Weicheng Zang for useful discussions.
The authors acknowledge the Tsinghua Astrophysics High-Performance Computing platform at Tsinghua University for providing computational and data storage resources that have contributed to the results in this paper. 
XCZ is supported by the National Natural Science Foundation of China (Grant No. 12203007) and the Mengya Program of Beijing Academy of Science and Technology (BGS202203). ZYC and SM are partly supported by the National Science Foundation of China (Grant No. 12133005).
\end{acknowledgments}

\software{REBOUNDx \citep{2020MNRAS.491.2885T}, REBOUND \citep{rein_rebound:_2011}, KozaiPy (https://github.com/djmunoz/kozaipy)}

\begin{appendix}
\section{Tidal dissipation theory analysis}
\label{sec:appendix}


The force responsible for tidal dissipation in the planet and star includes two parts: 
\begin{equation}
    \bm{F}_{\rm tidal}=F_{r}\hat{\bm{e}}_r+F_{\theta}\hat{\bm{e}}_\theta \ \ \ \ \ \ \ \ \ \ \ \ {\rm with}
\end{equation}
\begin{equation}
    F_{r}=-3G\frac{m^{2}}{r^{2}}\frac{R^5}{r^5}k\left(1+3\frac{\dot{r}}{r}\tau\right), \ \ \ \ \ \
    F_{\theta}=3G\frac{m^{2}}{r^{2}}\frac{R^5}{r^5}k\tau
    \left(\Omega-\dot{\theta}\right) ,
\end{equation}
where $\Omega$ is the spin rate of the deformed body. 
For a planet (or a star) as the deformed body and its stellar (or planetary companion) as 
the perturbing body, $M=M_{\rm p, \star}$, $m=M_{\star, \rm p}$, $q \equiv M/m =M_{\rm p, \star} / M_{\star, \rm p}$, $\Omega=\Omega_{\rm p, \star}$, $R=R_{\rm p, \star}$, 
$k=k_2 ^{\rm p, \star}$, and $\tau=\tau_{\rm lag} ^{\rm p, \star}$, respectively.
In the present context, we consider the nearly parabolic planet as the deformed body 
and its host star as the tidal perturbing body (with $m=M_{\rm p}$ and $M=M_\star$). 
Since the influence of each orbital tide constitutes a minor perturbation, the effects of tidal forces can be decomposed into the radial and azimuthal components for separate analysis, followed by linear superposition. The variations induced by $F_r$ are denoted by $\dot{a}_r$ and $\dot{e}_r$, while those caused by $F_{\theta}$ are represented as $\dot{a}_{\theta}$ and $\dot{e}_{\theta}$. 
According to \citet{1981A&A....99..126H}, both $a_r$ and $a_\theta$  as well as  $e_r$ and $e_\theta$ components contribute to the changing rate of eccentricity and semi-major axis, respectively:
\begin{equation}
{\dot a} = {\dot a}_r + {\dot a}_\theta \ \ \ \ \ \ {\rm and}
\ \ \ \ \ \ \dot e = \dot e_r + \dot e_{\theta}.
\end{equation}

We first consider the force in the radial direction.  At radial distance from the star and the radial velocity are 
\begin{equation}
r=a\frac{(1-e^{2})}{(1+e\cos\theta)}, \ \ \ \ \ \ 
\dot{r}= r \frac{e\sin\theta}{(1+e\cos\theta)} \dot{\theta} 
={\sqrt{G(M+m) \over a(1-e^2)}}e\sin\theta.
\end{equation}
The total angular momentum is given by
\begin{equation}
    h={{Mm}r^2\over (M+m)} \dot{\theta} , \ \ \ \ \ \ {\rm with} \ \ \ \ \ \ 
    h^2 = G\frac{M^2m^2}{(M+m)}a(1-e^2) .\\
    \label{eq:angularmomentum}
\end{equation}
Energy change is calculated as
\begin{equation}
\begin{aligned}
\Delta E_r & =\int_{\delta t}F_r dr 
=\int_{-\Delta\theta}^{\Delta\theta}F_r\frac{dr}{d\theta}d\theta 
 = \int_{-\Delta\theta}^{\Delta\theta}F_ra\frac{e\left(1-e^2\right)}{\left(1+e\cos\theta\right)^2}\sin\theta d\theta \\
 & =\int_{-\Delta\theta}^{\Delta\theta}(-9G)\frac{m^2}{r^2}\left(\frac{R}{r}\right)^5k\frac{\dot{r}}{r}\tau a\frac{e\left(1-e^2\right)}{\left(1+e\cos\theta\right)^2}\sin\theta d\theta \\
 &=-9Gm^2R^5k\tau ae\left(1-e^2\right)\int_{-\Delta\theta}^{\Delta\theta}\frac{\dot r \sin\theta}{r^8\left(1+e\cos\theta\right)^2}d\theta.
\end{aligned}
\end{equation}
Since the magnitude of $\Delta E$ decreases steeply with $r$, we adopt the integration interval 
$\Delta \theta = \pi/2$ for computational convenience.  For nearly parabolic orbit with
$e=1-\epsilon$ and $\epsilon < < 1$, 
$r (\theta=\pi/2) \simeq 2 r_{\rm peri}$ where $r_{\rm peri}$ is the peri-apsis distant 
$r_{\rm peri} = a (1-e) = \epsilon a$. Integration beyond this range of $\Delta \theta$ would 
modify the magnitude of $\Delta E$ by $\lesssim 2 \%$.  With this approximation,
\begin{equation}
\begin{aligned}
 \Delta E_r& \simeq -9G^{\frac{3}{2}}m^{2}R^{5}k\tau e^{2}\frac{(M+m)^{1/2}}{(1-e^{2})^{15/2}a^{15/2}}\int_{-\pi/2}^{\pi/2}\sin^{2}\theta(1+e\cos\theta)^{6}d\theta \\
&=-9f_1(e)G^{\frac{3}{2}}m^{2}R^{5}k\tau e^{2}\frac{(M+m)^{1/2}}{(1-e^{2})^{15/2}a^{15/2}} ,\\
{\rm where}  \quad \quad f_1(e) &= \frac{\pi}{2}+4e+\frac{15\pi}{8}e^2+\frac{16}{3}e^3+\frac{15\pi}{16}e^4+\frac{32}{35}e^5+\frac{5\pi}{128}e^6.
\\
\end{aligned}
\end{equation}




To calculate the rate of $E_r$ change, we use the average rate over the period, $P = 2\pi /n$
where $n=\sqrt{G (M+m)/a^3}$ is the mean motion,
\begin{equation}
\begin{aligned}
    \dot E_r &\simeq \frac{\Delta E_r}{P}
    = -\frac{9 f_1(e)}{2\pi} G^2m^{2}R^{5}k\tau e^{2}\frac{(M+m)}{(1-e^{2})^{15/2}a^{9}} .
\end{aligned}
\end{equation}
Based on the total energy $E  = -{GMm}/{2a}$, we can calculate the rate of change 
in semi-major axis $\dot a_r$ due to $F_r$ such that 
\begin{equation}
\begin{aligned}
\dot{a}_{r} & =\frac{2a^{2}\dot{E_r}}{GMm} 
   = -\frac{9 f_1(e)}{\pi} GR^{5}k\tau e^{2}\frac{m(M+m)}{M(1-e^{2})^{15/2}a^{7}} .
\end{aligned}
\end{equation}

Since $F_r$ does not lead to any angular momentum changes, we find from (\ref{eq:angularmomentum}):
\begin{equation}
\begin{aligned}
 2h \dot h&=G\frac{M^{2}m^{2}}{(M+m)}[\dot{a}_r(1-e^{2})-2ae\dot{e}_r]=0 \\
  \Rightarrow\dot{e}_{r}&=\frac{\dot{a}_r(1-e^{2})}{2ae}
= -\frac{9 f_1(e)}{2\pi} GR^{5}k\tau e\frac{m(M+m)}{M(1-e^{2})^{13/2}a^{8}} .
\end{aligned}
\label{eq:erdot}
\end{equation}

In sum, for $\hat r$ direction, replacing $e$ with $1-\epsilon$ and only considering the first order of $\epsilon$, we get

\begin{equation}
    \begin{aligned}
        \dot a_r & \simeq \frac{-9 g_{11}(\epsilon)}{128\sqrt{2}\pi} GR^{5}k\tau \frac{m(M+m)}{M\epsilon^{15/2}a^{7}} ,\\ 
        \dot e_r & \simeq -\frac{9 g_{12}(\epsilon)}{128\sqrt{2}\pi} GR^{5}k\tau \frac{m(M+m)}{M\epsilon^{13/2}a^{8}} ,
    \end{aligned}
\end{equation}

where
\begin{equation}
    \begin{aligned}
        g_{11}(\epsilon) &\simeq \frac{1076}{105}+\frac{429\pi}{128}-\left(\frac{676}{15}+\frac{231\pi}{16}\right)\epsilon , \\
        g_{12}(\epsilon) &\simeq \frac{1076}{105}+\frac{429\pi}{128}-\left(\frac{3656}{105}+\frac{1419\pi}{128}\right)\epsilon .
    \end{aligned}
\label{eq:g12}
\end{equation}

Next, we consider the $\hat \theta$ direction,

\begin{equation}
\begin{aligned}
 \Delta E_{\theta}&=\int_{-\Delta\theta}^{\Delta\theta}rF_{\theta}d\theta
 \simeq 3Gm^{2}R^{8}k\tau\int_{-\pi/2} ^{\pi/2}{(\Omega-\dot{\theta}) \over r^7} d\theta \\
 & =3\frac{k}{T}\frac{m^{2}}{M}{R^{8} \over a^{6}(1-e^{2})^{6}}\int_{-\pi/2}^{\pi/2}\left[\left(1+e\cos\theta\right)^{6}\Omega-{(1+e\cos\theta)^{8}
 \over \left(1-e^{2}\right)^{3/2}}n\right]d\theta \\
 & = 3\frac{k}{T}\frac{m^{2}}{M}{R^{8} \over a^{6}\left(1-e^{2}\right)^{6}}\left[f_2(e)\Omega-
 {f_3(e) \over (1-e^{2})^{3/2}}n\right]  ,
\end{aligned}
\end{equation}
where 
$T
={R^3}/{GM\tau}={((M+m)/ M n^2 \tau)}({R^3 / a^3})$,
\begin{equation}
\begin{aligned}
f_2(e) &= \pi+12e + \frac{15\pi}{2}e^2+\frac{80}{3}e^3+\frac{45\pi}{8}e^4+\frac{32}{5}e^5
+\frac{5\pi}{16}e^6 ,\\
f_3(e) &= \pi +16e+14\pi e^2+\frac{224}{3}e^3+\frac{105\pi}{4}e^4+\frac{896}{15}e^5+\frac{35\pi}{4}e^6+\frac{256}{35}e^7+\frac{35\pi}{128}e^8 .\\
\end{aligned}
\end{equation}

From the time-averaged rate of change in $E_\theta$, 
\begin{equation}
    {\dot E}_\theta = {{\Delta E}_\theta \over P} = \frac{3}{2\pi}\frac{nk}{T}\frac{m^{2}}{M}{R^{8} \over a^{6}\left(1-e^{2}\right)^{6}}\left[f_2(e)\Omega-
 {f_3(e) \over (1-e^{2})^{3/2}}n\right] ,
\end{equation}
we derive 
\begin{equation}
\begin{aligned}
{\dot a}_\theta &= {2 a^2 {\dot E}_\theta \over G M m}\\
&=\frac{3}{\pi}\frac{k}{nT}\frac{m(M+m)}{M^2}{R^{8} \over a^{7}\left(1-e^{2}\right)^{6}}\left[f_2(e)\Omega-
 {f_3(e) \over (1-e^{2})^{3/2}}n\right] .\\
\end{aligned}
\label{eq:adottheta}
\end{equation}

For nearly parabolic orbits (with $\epsilon < < 1$), the perigee distance $r_p = (1-e) a
\simeq \epsilon a$.  Its rate of fractional changes ${\dot r}_p/r_p = {\dot a} /a - 
{\dot e} /(1-e)$ is dominated by changes in $e$.  In order to derive an expression for ${\dot e}$, we first obtain from eq.(\ref{eq:angularmomentum}),
\begin{equation}
    \begin{aligned}
    \dot \theta &= \frac{\sqrt{aG(1-e^2)(M+m)}}{r^2}
     = n {(1+e \cos \theta)^2 \over (1-e^2)^{3/2}} ,
    \end{aligned}
\end{equation}

\begin{equation}
\begin{aligned}
 \Delta h_{\theta}&=\int_{-\Delta\theta}^{\Delta\theta}rF_{\theta}\frac{d\theta}{\dot \theta}
 =-{3 \over (1-e^2)^6}\frac{k}{T}\frac{m^{2}}{M}{R^{8} \over a^{6}}\int_{-\Delta\theta}^{\Delta\theta}\left[\left(1+e\cos\theta\right)^{6}-\left(1-e^{2}\right)^{3/2}\left(1+e\cos\theta\right)^{4}\frac{\Omega}{n}\right]d\theta \\
 & =-{3\over (1-e^2)^6}\frac{k}{T}\frac{m^{2}}{M}{R^{8}\over a^{6}}\left[f_2(e)-\left(1-e^{2}\right)^{3/2}f_4(e)\frac{\Omega}{n}\right] ,
\end{aligned}
\end{equation}
where
\begin{equation}
\begin{aligned}
f_4(e)&=\pi+8e+3\pi e^2+\frac{16}{3}e^3+\frac{3\pi}{8}e^4 .
\end{aligned}
\end{equation}

Averaging over the orbital period $P$,


\begin{equation}
\begin{aligned}
\dot h_{\theta} &= \frac{\Delta h_{\theta}}{P}\\
 &=-{3\over 2\pi(1-e^2)^6}\frac{nk}{T}\frac{m^{2}}{M}{R^{8}\over a^{6}}\left[f_2(e)-\left(1-e^{2}\right)^{3/2}f_4(e)\frac{\Omega}{n}\right] .
\end{aligned}
\end{equation}

Substitute $\dot a_{\theta}$ in Eq. (\ref{eq:adottheta}) into Eq. (\ref{eq:angularmomentum}), we find

\begin{equation}
\begin{aligned}
\dot e_{\theta}&=\frac{1}{2}\frac{(1-e^2)}{ae} \dot a_{\theta} -\frac{(M+m)h}{GM^2m^2ae} \dot h_\theta\\
&=\frac{3}{2\pi}\frac{k}{nT}\frac{m(M+m)}{M^2}{R^{8} \over a^{8}e\left(1-e^{2}\right)^{5}}\left[f_2(e)\Omega-
 {f_3(e) \over (1-e^{2})^{3/2}}n\right] \\
&~~ -{3\over 2\pi}\frac{k}{nT}\frac{m(M+m)}{M^2}{R^{8}\over a^{8}e(1-e^2)^{4}}\left[\frac{f_2(e)}{\left(1-e^{2}\right)^{3/2}}n-f_4(e)\Omega\right] .
\end{aligned}
\end{equation}

Replacing $e$ with $1-\epsilon$, we get

\begin{equation}
    \begin{aligned}
        \dot a_{\theta} &= \frac{3}{128\pi}\frac{k}{nT}\frac{m(M+m)}{M^2}{R^{8} \over a^{7}\epsilon^{6}}\left[g_{21}(\epsilon)\Omega-
 {g_{31}(\epsilon) \over 2\sqrt{2}\epsilon^{3/2}}n\right] ,\\
        \dot e_{\theta} &= \frac{3}{64\pi}\frac{k}{nT}\frac{m(M+m)}{M^2}{R^{8} \over a^{8}\epsilon^{5}}\left[g_{22}(\epsilon)\Omega-
 {g_{32}(\epsilon) \over 2\sqrt{2}\epsilon^{3/2}}n\right] \\
&~~ -{3\over 32\pi}\frac{k}{nT}\frac{m(M+m)}{M^2}{R^{8}\over a^{8}\epsilon^{4}}\left[\frac{g_{22}(\epsilon)}{2\sqrt{2}\epsilon^{3/2}}n-g_{42}(\epsilon)\Omega\right] ,
    \end{aligned}
\end{equation}
where 
\begin{equation}
    \begin{aligned}
g_{21}(\epsilon) &\simeq \frac{676}{15}+\frac{231\pi}{16}-(124+\frac{315\pi}{8})\epsilon ,\\
g_{31}(\epsilon)  &\simeq \frac{1104}{7}+\frac{6435\pi}{128}-(\frac{8848}{15}+\frac{3003\pi}{16})\epsilon ,\\
g_{22}(\epsilon)  &\simeq \frac{676}{15}+\frac{231\pi}{16}-(\frac{1184}{15}+\frac{399\pi}{16})\epsilon ,\\
g_{32}(\epsilon)  &\simeq \frac{1104}{7}+\frac{6435\pi}{128}-(\frac{45376}{105}+\frac{17589\pi}{128})\epsilon ,\\
g_{42}(\epsilon)  &\simeq \frac{40}{3}+\frac{35\pi}{8}-(\frac{32}{3}+\frac{25\pi}{8})\epsilon .
  \end{aligned}
  \label{eq:g22g42}
\end{equation}

The total rate of change

\begin{equation}
\begin{aligned}
    {\dot a}  & = {\dot a}_r + {\dot a}_\theta
    \simeq \frac{-9 g_{11}(\epsilon)}{128\sqrt{2}\pi} \frac{k}{T}\frac{m(M+m)}{M^2}{R^{8} \over a^{7}\epsilon^{15/2}} \\
    &~~~+\frac{3}{128\pi}\frac{k}{nT}\frac{m(M+m)}{M^2}{R^{8} \over a^{7}\epsilon^{6}}\left[g_{21}(\epsilon)\Omega-
 {g_{31}(\epsilon) \over 2\sqrt{2}\epsilon^{3/2}}n\right] \ \ \ \ \ \  {\rm and}
\end{aligned}
\label{eq:adottot}
\end{equation}

\begin{equation}
\begin{aligned}
\dot e &= \dot e_r + \dot e_{\theta}
\simeq-\frac{9 g_{12}(\epsilon)}{128\sqrt{2}\pi}\frac{k}{T}\frac{m(M+m)}{M^2}{R^{8} \over a^{8}\epsilon^{13/2}}\\
&~~+\frac{3}{64\pi}\frac{k}{nT}\frac{m(M+m)}{M^2}{R^{8} \over a^{8}\epsilon^{5}}\left[g_{22}(\epsilon)\Omega-
 {g_{32}(\epsilon) \over 2\sqrt{2}\epsilon^{3/2}}n\right] \\
&~~ -{3\over 32\pi}\frac{k}{nT}\frac{m(M+m)}{M^2}{R^{8}\over a^{8}\epsilon^{4}}\left[\frac{g_{22}(\epsilon)}{2\sqrt{2}\epsilon^{3/2}}n-g_{42}(\epsilon)\Omega\right] .
\end{aligned}
\label{eq:edottot}
\end{equation}


For physical interpretations, it is useful to recast the above expressions in terms of 
the dimensionless ratio ${\tilde R} \equiv R/R_{\rm R}$ where $R_{\rm R}=q^{1/3} r_{\rm peri}
=q^{1/3} a \epsilon$ is the planet's Roche radius at perigee, $q=M/m=M_{\rm p}/M_\star$ is the planet to star mass ratio. It is also informative to introduce dimensionless response time, 
${\tilde \tau}_\Omega = \Omega \tau$ where $\Omega$ is the planet's spin frequency 
and ${\tilde \tau}_n \equiv \tau/\Delta T$ where $\Delta T= P \epsilon^{3/2}$ is the time interval of perigee passage. These timescales are relevant to the inertial-mode
dynamical tides \citep{ogilvie2004} and visco-elastic equilibrium tides  \citep{2012A&A...541A.165R}.
The first, second, and third terms in Eq (\ref{eq:edottot}) become


\begin{equation}
\begin{aligned}
{\dot e} = & -{9 \pi q^{2/3} \over 32 \sqrt{2}} 
{k {\tilde \tau}_n \over P}  {\tilde R}^5  g_{12} (\epsilon)
+ \frac{3}{32} q^{2/3} {k {\tilde \tau}_\Omega \over P} {\tilde R}^5 
\left[g_{22}(\epsilon)-{\pi g_{32}(\epsilon) 
\over \sqrt{2}\Omega \Delta T}\right] \\
& -\frac{3}{16}\epsilon q^{2/3} {k {\tilde \tau}_\Omega \over P} {\tilde R}^5 
\left[g_{42}(\epsilon)-{\pi g_{22}(\epsilon) 
\over \sqrt{2}\Omega \Delta T}\right] \\
= & - q^{2/3}{k {\tilde \tau}_n \over P}  {\tilde R}^5 
\Bigg\{{9 \pi g_{12} \over 32 \sqrt{2}} - \frac{3}{32}
\left[g_{22}(\epsilon)-{\pi g_{32}(\epsilon) \over \sqrt{2}\Omega \Delta T}\right]
+\frac{3}{16}\epsilon \left[g_{42}(\epsilon)-{\pi g_{22}(\epsilon) 
\over \sqrt{2}\Omega \Delta T}\right]
\Bigg\}.
\end{aligned}
\end{equation}

In principle $\Omega \Delta T > 1$ implies fast spin and eccentricity excitation
(i.e. with ${\dot e}_\theta > 1$). In contrast, $\Omega \Delta T < 1$ implies slow spin and 
eccentricity damping (i.e. with ${\dot e}_\theta < 1$). The key point we wish to show is
that the eccentricity evolution timescale 
\begin{equation}
    \tau_e = {e \over {\dot e}} \propto \mathcal{O} \left( {\epsilon P \over q^{2/3} k 
    {\tilde \tau}_n {\tilde R}^5} \right) .
    \label{eq:taue}
\end{equation}
is longer than the von Zeipel-Kozai-Lidov cycle and the ejection time scale
(\S\ref{sec:analytic_tide} and \ref{sec:tidaldamping}).

\end{appendix}
%

\bibliography{ffp}

@ARTICLE{Gan2023,
       author = {{Gan}, Tianjun and {Wang}, Sharon X. and {Wang}, Songhu and {Mao}, Shude and {Huang}, Chelsea X. and {Collins}, Karen A. and {Stassun}, Keivan G. and {Shporer}, Avi and {Zhu}, Wei and {Ricker}, George R. and {Vanderspek}, Roland and {Latham}, David W. and {Seager}, Sara and {Winn}, Joshua N. and {Jenkins}, Jon M. and {Barkaoui}, Khalid and {Belinski}, Alexander A. and {Ciardi}, David R. and {Evans}, Phil and {Girardin}, Eric and {Maslennikova}, Nataliia A. and {Mazeh}, Tsevi and {Panahi}, Aviad and {Pozuelos}, Francisco J. and {Radford}, Don J. and {Schwarz}, Richard P. and {Twicken}, Joseph D. and {W{\"u}nsche}, Ana{\"e}l and {Zucker}, Shay},
        title = "{Occurrence Rate of Hot Jupiters Around Early-type M Dwarfs Based on Transiting Exoplanet Survey Satellite Data}",
      journal = {\aj},
         year = 2023,
        month = jan,
       volume = {165},
       number = {1},
          eid = {17},
        pages = {17},
          doi = {10.3847/1538-3881/ac9b12},
archivePrefix = {arXiv},
       eprint = {2210.08313},
 primaryClass = {astro-ph.EP},
       adsurl = {https://ui.adsabs.harvard.edu/abs/2023AJ....165...17G}
}

@INPROCEEDINGS{Offner2023,
       author = {{Offner}, S.~S.~R. and {Moe}, M. and {Kratter}, K.~M. and {Sadavoy}, S.~I. and {Jensen}, E.~L.~N. and {Tobin}, J.~J.},
        title = "{The Origin and Evolution of Multiple Star Systems}",
    booktitle = {Protostars and Planets VII},
         year = 2023,
       editor = {{Inutsuka}, S. and {Aikawa}, Y. and {Muto}, T. and {Tomida}, K. and {Tamura}, M.},
       series = {Astronomical Society of the Pacific Conference Series},
       volume = {534},
        month = jul,
        pages = {275},
          doi = {10.48550/arXiv.2203.10066},
archivePrefix = {arXiv},
       eprint = {2203.10066},
 primaryClass = {astro-ph.SR},
       adsurl = {https://ui.adsabs.harvard.edu/abs/2023ASPC..534..275O}
}

@ARTICLE{Coleman2025,
       author = {{Coleman}, Gavin A.~L. and {DeRocco}, William},
        title = "{Predicting the Galactic population of free-floating planets from realistic initial conditions}",
      journal = {\mnras},
         year = 2025,
        month = mar,
       volume = {537},
       number = {3},
        pages = {2303-2312},
          doi = {10.1093/mnras/staf138},
archivePrefix = {arXiv},
       eprint = {2407.05992},
 primaryClass = {astro-ph.EP},
       adsurl = {https://ui.adsabs.harvard.edu/abs/2025MNRAS.537.2303C}
}

@ARTICLE{Bryan2019,
       author = {{Bryan}, Marta L. and {Knutson}, Heather A. and {Lee}, Eve J. and {Fulton}, B.~J. and {Batygin}, Konstantin and {Ngo}, Henry and {Meshkat}, Tiffany},
        title = "{An Excess of Jupiter Analogs in Super-Earth Systems}",
      journal = {\aj},
         year = 2019,
        month = feb,
       volume = {157},
       number = {2},
          eid = {52},
        pages = {52},
          doi = {10.3847/1538-3881/aaf57f},
archivePrefix = {arXiv},
       eprint = {1806.08799},
 primaryClass = {astro-ph.EP},
       adsurl = {https://ui.adsabs.harvard.edu/abs/2019AJ....157...52B}
}

@ARTICLE{ZhuWu2018,
       author = {{Zhu}, Wei and {Wu}, Yanqin},
        title = "{The Super Earth-Cold Jupiter Relations}",
      journal = {\aj},
         year = 2018,
        month = sep,
       volume = {156},
       number = {3},
          eid = {92},
        pages = {92},
          doi = {10.3847/1538-3881/aad22a},
archivePrefix = {arXiv},
       eprint = {1805.02660},
 primaryClass = {astro-ph.EP},
       adsurl = {https://ui.adsabs.harvard.edu/abs/2018AJ....156...92Z}
}

@BOOK{Safronov1972,
       author = {{Safronov}, V.~S.},
        title = "{Evolution of the protoplanetary cloud and formation of the earth and planets.}",
         year = 1972,
       adsurl = {https://ui.adsabs.harvard.edu/abs/1972epcf.book.....S}
}

@ARTICLE{Oasa1999,
       author = {{Oasa}, Yumiko and {Tamura}, Motohide and {Sugitani}, Koji},
        title = "{A Deep Near-Infrared Survey of the Chamaeleon I Dark Cloud Core}",
      journal = {\apj},
         year = 1999,
        month = nov,
       volume = {526},
       number = {1},
        pages = {336-343},
          doi = {10.1086/307964},
       adsurl = {https://ui.adsabs.harvard.edu/abs/1999ApJ...526..336O}
}

@ARTICLE{Tamura1998,
       author = {{Tamura}, Motohide and {Itoh}, Yoichi and {Oasa}, Yumiko and {Nakajima}, Tadashi},
        title = "{Isolated and Companion Young Brown Dwarfs in the Taurus and Chamaeleon Molecular Clouds}",
      journal = {Science},
         year = 1998,
        month = nov,
       volume = {282},
        pages = {1095},
          doi = {10.1126/science.282.5391.1095},
       adsurl = {https://ui.adsabs.harvard.edu/abs/1998Sci...282.1095T}
}

@ARTICLE{Bejar1999,
       author = {{B{\'e}jar}, V.~J.~S. and {Zapatero Osorio}, M.~R. and {Rebolo}, R.},
        title = "{A Search for Very Low Mass Stars and Brown Dwarfs in the Young {\ensuremath{\sigma}} Orionis Cluster}",
      journal = {\apj},
         year = 1999,
        month = aug,
       volume = {521},
       number = {2},
        pages = {671-681},
          doi = {10.1086/307583},
archivePrefix = {arXiv},
       eprint = {astro-ph/9903217},
 primaryClass = {astro-ph},
       adsurl = {https://ui.adsabs.harvard.edu/abs/1999ApJ...521..671B}
}

@ARTICLE{Lucas2000,
       author = {{Lucas}, P.~W. and {Roche}, P.~F.},
        title = "{A population of very young brown dwarfs and free-floating planets in Orion}",
      journal = {\mnras},
         year = 2000,
        month = jun,
       volume = {314},
       number = {4},
        pages = {858-864},
          doi = {10.1046/j.1365-8711.2000.03515.x},
archivePrefix = {arXiv},
       eprint = {astro-ph/0003061},
 primaryClass = {astro-ph},
       adsurl = {https://ui.adsabs.harvard.edu/abs/2000MNRAS.314..858L}
}

@INPROCEEDINGS{Barrado2002,
       author = {{Barrado y Navascu{\'e}s}, David and {Zapatero Osorio}, Mar{\'\i}a Rosa and {B{\'e}jar}, V{\'\i}ctor and {Rebolo}, Rafael and {Mart{\'\i}n}, Eduardo L. and {Mundt}, Reinhard and {Bailer-Jones}, Coryn A.~L.},
        title = "{VLT/FORS Spectroscopy in {\ensuremath{\sigma}} Orionis: Isolated Planetary Mass Candidate Members}",
    booktitle = {The Origin of Stars and Planets: The VLT View},
         year = 2002,
       editor = {{Alves}, Jo{\~a}o F. and {McCaughrean}, Mark J.},
        month = jan,
        pages = {195},
          doi = {10.1007/10856518_24},
       adsurl = {https://ui.adsabs.harvard.edu/abs/2002osp..conf..195B}
}

@ARTICLE{nagasawa2000b,
       author = {{Nagasawa}, Makiko and {Ida}, Shigeru},
        title = "{Sweeping Secular Resonances in the Kuiper Belt Caused by Depletion of the Solar Nebula}",
      journal = {\aj},
         year = 2000,
        month = dec,
       volume = {120},
       number = {6},
        pages = {3311-3322},
          doi = {10.1086/316856},
       adsurl = {https://ui.adsabs.harvard.edu/abs/2000AJ....120.3311N}
}

@ARTICLE{nagasawa2000a,
       author = {{Nagasawa}, Makiko and {Tanaka}, Hidekazu and {Ida}, Shigeru},
        title = "{Orbital Evolution of Asteroids during Depletion of the Solar Nebula}",
      journal = {\aj},
         year = 2000,
        month = mar,
       volume = {119},
       number = {3},
        pages = {1480-1497},
          doi = {10.1086/301246},
       adsurl = {https://ui.adsabs.harvard.edu/abs/2000AJ....119.1480N}
}

@ARTICLE{wittenmyer2020,
       author = {{Wittenmyer}, Robert A. and {Wang}, Songhu and {Horner}, Jonathan and {Butler}, R.~P. and {Tinney}, C.~G. and {Carter}, B.~D. and {Wright}, D.~J. and {Jones}, H.~R.~A. and {Bailey}, J. and {O'Toole}, S.~J. and {Johns}, Daniel},
        title = "{Cool Jupiters greatly outnumber their toasty siblings: occurrence rates from the Anglo-Australian Planet Search}",
      journal = {\mnras},
         year = 2020,
        month = feb,
       volume = {492},
       number = {1},
        pages = {377-383},
          doi = {10.1093/mnras/stz3436},
archivePrefix = {arXiv},
       eprint = {1912.01821},
 primaryClass = {astro-ph.EP},
       adsurl = {https://ui.adsabs.harvard.edu/abs/2020MNRAS.492..377W}
}

@ARTICLE{Wang2022,
       author = {{Wang}, Shijie and {Kanagawa}, Kazuhiro D. and {Suto}, Yasushi},
        title = "{Architecture of Planetary Systems Predicted from Protoplanetary Disks Observed with ALMA. II. Evolution Outcomes and Dynamical Stability}",
      journal = {\apj},
         year = 2022,
        month = jun,
       volume = {932},
       number = {1},
          eid = {31},
        pages = {31},
          doi = {10.3847/1538-4357/ac68de},
archivePrefix = {arXiv},
       eprint = {2204.08826},
 primaryClass = {astro-ph.EP},
       adsurl = {https://ui.adsabs.harvard.edu/abs/2022ApJ...932...31W}
}

@ARTICLE{Asensio-Torres2021,
       author = {{Asensio-Torres}, R. and {Henning}, Th. and {Cantalloube}, F. and {Pinilla}, P. and {Mesa}, D. and {Garufi}, A. and {Jorquera}, S. and {Gratton}, R. and {Chauvin}, G. and {Szul{\'a}gyi}, J. and {van Boekel}, R. and {Dong}, R. and {Marleau}, G. -D. and {Benisty}, M. and {Villenave}, M. and {Bergez-Casalou}, C. and {Desgrange}, C. and {Janson}, M. and {Keppler}, M. and {Langlois}, M. and {M{\'e}nard}, F. and {Rickman}, E. and {Stolker}, T. and {Feldt}, M. and {Fusco}, T. and {Gluck}, L. and {Pavlov}, A. and {Ramos}, J.},
        title = "{Perturbers: SPHERE detection limits to planetary-mass companions in protoplanetary disks}",
      journal = {\aap},
         year = 2021,
        month = aug,
       volume = {652},
          eid = {A101},
        pages = {A101},
          doi = {10.1051/0004-6361/202140325},
archivePrefix = {arXiv},
       eprint = {2103.05377},
 primaryClass = {astro-ph.EP},
       adsurl = {https://ui.adsabs.harvard.edu/abs/2021A&A...652A.101A}
}

@ARTICLE{Garufi2020,
       author = {{Garufi}, A. and {Avenhaus}, H. and {P{\'e}rez}, S. and {Quanz}, S.~P. and {van Holstein}, R.~G. and {Bertrang}, G.~H. -M. and {Casassus}, S. and {Cieza}, L. and {Principe}, D.~A. and {van der Plas}, G. and {Zurlo}, A.},
        title = "{Disks Around T Tauri Stars with SPHERE (DARTTS-S). II. Twenty-one new polarimetric images of young stellar disks}",
      journal = {\aap},
         year = 2020,
        month = jan,
       volume = {633},
          eid = {A82},
        pages = {A82},
          doi = {10.1051/0004-6361/201936946},
archivePrefix = {arXiv},
       eprint = {1911.10853},
 primaryClass = {astro-ph.EP},
       adsurl = {https://ui.adsabs.harvard.edu/abs/2020A&A...633A..82G}
}

@ARTICLE{Long2019,
       author = {{Long}, Feng and {Herczeg}, Gregory J. and {Harsono}, Daniel and {Pinilla}, Paola and {Tazzari}, Marco and {Manara}, Carlo F. and {Pascucci}, Ilaria and {Cabrit}, Sylvie and {Nisini}, Brunella and {Johnstone}, Doug and {Edwards}, Suzan and {Salyk}, Colette and {Menard}, Francois and {Lodato}, Giuseppe and {Boehler}, Yann and {Mace}, Gregory N. and {Liu}, Yao and {Mulders}, Gijs D. and {Hendler}, Nathanial and {Ragusa}, Enrico and {Fischer}, William J. and {Banzatti}, Andrea and {Rigliaco}, Elisabetta and {van de Plas}, Gerrit and {Dipierro}, Giovanni and {Gully-Santiago}, Michael and {Lopez-Valdivia}, Ricardo},
        title = "{Compact Disks in a High-resolution ALMA Survey of Dust Structures in the Taurus Molecular Cloud}",
      journal = {\apj},
         year = 2019,
        month = sep,
       volume = {882},
       number = {1},
          eid = {49},
        pages = {49},
          doi = {10.3847/1538-4357/ab2d2d},
archivePrefix = {arXiv},
       eprint = {1906.10809},
 primaryClass = {astro-ph.SR},
       adsurl = {https://ui.adsabs.harvard.edu/abs/2019ApJ...882...49L}
}

@ARTICLE{Andrews2018,
       author = {{Andrews}, Sean M. and {Huang}, Jane and {P{\'e}rez}, Laura M. and {Isella}, Andrea and {Dullemond}, Cornelis P. and {Kurtovic}, Nicol{\'a}s T. and {Guzm{\'a}n}, Viviana V. and {Carpenter}, John M. and {Wilner}, David J. and {Zhang}, Shangjia and {Zhu}, Zhaohuan and {Birnstiel}, Tilman and {Bai}, Xue-Ning and {Benisty}, Myriam and {Hughes}, A. Meredith and {{\"O}berg}, Karin I. and {Ricci}, Luca},
        title = "{The Disk Substructures at High Angular Resolution Project (DSHARP). I. Motivation, Sample, Calibration, and Overview}",
      journal = {\apjl},
         year = 2018,
        month = dec,
       volume = {869},
       number = {2},
          eid = {L41},
        pages = {L41},
          doi = {10.3847/2041-8213/aaf741},
archivePrefix = {arXiv},
       eprint = {1812.04040},
 primaryClass = {astro-ph.SR},
       adsurl = {https://ui.adsabs.harvard.edu/abs/2018ApJ...869L..41A}
}

@ARTICLE{Avenhaus2018,
       author = {{Avenhaus}, Henning and {Quanz}, Sascha P. and {Garufi}, Antonio and {Perez}, Sebastian and {Casassus}, Simon and {Pinte}, Christophe and {Bertrang}, Gesa H. -M. and {Caceres}, Claudio and {Benisty}, Myriam and {Dominik}, Carsten},
        title = "{Disks around T Tauri Stars with SPHERE (DARTTS-S). I. SPHERE/IRDIS Polarimetric Imaging of Eight Prominent T Tauri Disks}",
      journal = {\apj},
         year = 2018,
        month = aug,
       volume = {863},
       number = {1},
          eid = {44},
        pages = {44},
          doi = {10.3847/1538-4357/aab846},
archivePrefix = {arXiv},
       eprint = {1803.10882},
 primaryClass = {astro-ph.SR},
       adsurl = {https://ui.adsabs.harvard.edu/abs/2018ApJ...863...44A}
}

@ARTICLE{vanBoekel2017,
       author = {{van Boekel}, R. and {Henning}, Th. and {Menu}, J. and {de Boer}, J. and {Langlois}, M. and {M{\"u}ller}, A. and {Avenhaus}, H. and {Boccaletti}, A. and {Schmid}, H.~M. and {Thalmann}, Ch. and {Benisty}, M. and {Dominik}, C. and {Ginski}, Ch. and {Girard}, J.~H. and {Gisler}, D. and {Lobo Gomes}, A. and {Menard}, F. and {Min}, M. and {Pavlov}, A. and {Pohl}, A. and {Quanz}, S.~P. and {Rabou}, P. and {Roelfsema}, R. and {Sauvage}, J. -F. and {Teague}, R. and {Wildi}, F. and {Zurlo}, A.},
        title = "{Three Radial Gaps in the Disk of TW Hydrae Imaged with SPHERE}",
      journal = {\apj},
         year = 2017,
        month = mar,
       volume = {837},
       number = {2},
          eid = {132},
        pages = {132},
          doi = {10.3847/1538-4357/aa5d68},
archivePrefix = {arXiv},
       eprint = {1610.08939},
 primaryClass = {astro-ph.EP},
       adsurl = {https://ui.adsabs.harvard.edu/abs/2017ApJ...837..132V}
}

@ARTICLE{santerne2016,
       author = {{Santerne}, A. and {Moutou}, C. and {Tsantaki}, M. and {Bouchy}, F. and {H{\'e}brard}, G. and {Adibekyan}, V. and {Almenara}, J. -M. and {Amard}, L. and {Barros}, S.~C.~C. and {Boisse}, I. and {Bonomo}, A.~S. and {Bruno}, G. and {Courcol}, B. and {Deleuil}, M. and {Demangeon}, O. and {D{\'\i}az}, R.~F. and {Guillot}, T. and {Havel}, M. and {Montagnier}, G. and {Rajpurohit}, A.~S. and {Rey}, J. and {Santos}, N.~C.},
        title = "{SOPHIE velocimetry of Kepler transit candidates. XVII. The physical properties of giant exoplanets within 400 days of period}",
      journal = {\aap},
         year = 2016,
        month = mar,
       volume = {587},
          eid = {A64},
        pages = {A64},
          doi = {10.1051/0004-6361/201527329},
archivePrefix = {arXiv},
       eprint = {1511.00643},
 primaryClass = {astro-ph.EP},
       adsurl = {https://ui.adsabs.harvard.edu/abs/2016A&A...587A..64S}
}

@ARTICLE{dong2013,
       author = {{Dong}, Subo and {Zhu}, Zhaohuan},
        title = "{Fast Rise of ``Neptune-size'' Planets (4-8 R $_{{\ensuremath{\oplus}}}$) from P \raisebox{-0.5ex}\textasciitilde 10 to \raisebox{-0.5ex}\textasciitilde250 Days{\textemdash}Statistics of Kepler Planet Candidates up to \raisebox{-0.5ex}\textasciitilde0.75 AU}",
      journal = {\apj},
         year = 2013,
        month = nov,
       volume = {778},
       number = {1},
          eid = {53},
        pages = {53},
          doi = {10.1088/0004-637X/778/1/53},
archivePrefix = {arXiv},
       eprint = {1212.4853},
 primaryClass = {astro-ph.EP},
       adsurl = {https://ui.adsabs.harvard.edu/abs/2013ApJ...778...53D}
}

@ARTICLE{Clanton2014,
       author = {{Clanton}, Christian and {Gaudi}, B. Scott},
        title = "{Synthesizing Exoplanet Demographics from Radial Velocity and Microlensing Surveys. II. The Frequency of Planets Orbiting M Dwarfs}",
      journal = {\apj},
         year = 2014,
        month = aug,
       volume = {791},
       number = {2},
          eid = {91},
        pages = {91},
          doi = {10.1088/0004-637X/791/2/91},
archivePrefix = {arXiv},
       eprint = {1404.7500},
 primaryClass = {astro-ph.EP},
       adsurl = {https://ui.adsabs.harvard.edu/abs/2014ApJ...791...91C}
}

@ARTICLE{howard2012,
       author = {{Howard}, Andrew W. and {Marcy}, Geoffrey W. and {Bryson}, Stephen T. and {Jenkins}, Jon M. and {Rowe}, Jason F. and {Batalha}, Natalie M. and {Borucki}, William J. and {Koch}, David G. and {Dunham}, Edward W. and {Gautier}, III, Thomas N. and {Van Cleve}, Jeffrey and {Cochran}, William D. and {Latham}, David W. and {Lissauer}, Jack J. and {Torres}, Guillermo and {Brown}, Timothy M. and {Gilliland}, Ronald L. and {Buchhave}, Lars A. and {Caldwell}, Douglas A. and {Christensen-Dalsgaard}, J{\o}rgen and {Ciardi}, David and {Fressin}, Francois and {Haas}, Michael R. and {Howell}, Steve B. and {Kjeldsen}, Hans and {Seager}, Sara and {Rogers}, Leslie and {Sasselov}, Dimitar D. and {Steffen}, Jason H. and {Basri}, Gibor S. and {Charbonneau}, David and {Christiansen}, Jessie and {Clarke}, Bruce and {Dupree}, Andrea and {Fabrycky}, Daniel C. and {Fischer}, Debra A. and {Ford}, Eric B. and {Fortney}, Jonathan J. and {Tarter}, Jill and {Girouard}, Forrest R. and {Holman}, Matthew J. and {Johnson}, John Asher and {Klaus}, Todd C. and {Machalek}, Pavel and {Moorhead}, Althea V. and {Morehead}, Robert C. and {Ragozzine}, Darin and {Tenenbaum}, Peter and {Twicken}, Joseph D. and {Quinn}, Samuel N. and {Isaacson}, Howard and {Shporer}, Avi and {Lucas}, Philip W. and {Walkowicz}, Lucianne M. and {Welsh}, William F. and {Boss}, Alan and {Devore}, Edna and {Gould}, Alan and {Smith}, Jeffrey C. and {Morris}, Robert L. and {Prsa}, Andrej and {Morton}, Timothy D. and {Still}, Martin and {Thompson}, Susan E. and {Mullally}, Fergal and {Endl}, Michael and {MacQueen}, Phillip J.},
        title = "{Planet Occurrence within 0.25 AU of Solar-type Stars from Kepler}",
      journal = {\apjs},
         year = 2012,
        month = aug,
       volume = {201},
       number = {2},
          eid = {15},
        pages = {15},
          doi = {10.1088/0067-0049/201/2/15},
archivePrefix = {arXiv},
       eprint = {1103.2541},
 primaryClass = {astro-ph.EP},
       adsurl = {https://ui.adsabs.harvard.edu/abs/2012ApJS..201...15H}
}

@ARTICLE{zhu2021,
       author = {{Zhu}, Wei and {Dong}, Subo},
        title = "{Exoplanet Statistics and Theoretical Implications}",
      journal = {\araa},
         year = 2021,
        month = sep,
       volume = {59},
        pages = {291-336},
          doi = {10.1146/annurev-astro-112420-020055},
archivePrefix = {arXiv},
       eprint = {2103.02127},
 primaryClass = {astro-ph.EP},
       adsurl = {https://ui.adsabs.harvard.edu/abs/2021ARA&A..59..291Z}
}

@ARTICLE{howard2010,
       author = {{Howard}, Andrew W. and {Marcy}, Geoffrey W. and {Johnson}, John Asher and {Fischer}, Debra A. and {Wright}, Jason T. and {Isaacson}, Howard and {Valenti}, Jeff A. and {Anderson}, Jay and {Lin}, Doug N.~C. and {Ida}, Shigeru},
        title = "{The Occurrence and Mass Distribution of Close-in Super-Earths, Neptunes, and Jupiters}",
      journal = {Science},
         year = 2010,
        month = oct,
       volume = {330},
       number = {6004},
        pages = {653},
          doi = {10.1126/science.1194854},
archivePrefix = {arXiv},
       eprint = {1011.0143},
 primaryClass = {astro-ph.EP},
       adsurl = {https://ui.adsabs.harvard.edu/abs/2010Sci...330..653H}
}

@INPROCEEDINGS{mayor2010,
       author = {{Mayor}, Michel},
        title = "{Overview of Radial Velocity Surveys and Planet Discoveries}",
    booktitle = {Exoplanets Rising: Astronomy and Planetary Science at the Crossroads},
         year = 2010,
        month = mar,
          eid = {4},
        pages = {4},
       adsurl = {https://ui.adsabs.harvard.edu/abs/2010erap.confE...4M}
}

@ARTICLE{mayor2009,
       author = {{Mayor}, M. and {Udry}, S. and {Lovis}, C. and {Pepe}, F. and {Queloz}, D. and {Benz}, W. and {Bertaux}, J. -L. and {Bouchy}, F. and {Mordasini}, C. and {Segransan}, D.},
        title = "{The HARPS search for southern extra-solar planets. XIII. A planetary system with 3 super-Earths (4.2, 6.9, and 9.2 M$_{{\ensuremath{\oplus}}}$)}",
      journal = {\aap},
         year = 2009,
        month = jan,
       volume = {493},
       number = {2},
        pages = {639-644},
          doi = {10.1051/0004-6361:200810451},
archivePrefix = {arXiv},
       eprint = {0806.4587},
 primaryClass = {astro-ph},
       adsurl = {https://ui.adsabs.harvard.edu/abs/2009A&A...493..639M}
}

@ARTICLE{bhaskar2021,
       author = {{Bhaskar}, Hareesh and {Li}, Gongjie and {Hadden}, Sam and {Payne}, Matthew J. and {Holman}, Matthew J.},
        title = "{Mildly Hierarchical Triple Dynamics and Applications to the Outer Solar System}",
      journal = {\aj},
         year = 2021,
        month = jan,
       volume = {161},
       number = {1},
          eid = {48},
        pages = {48},
          doi = {10.3847/1538-3881/abcbfc},
archivePrefix = {arXiv},
       eprint = {2008.04335},
 primaryClass = {astro-ph.EP},
       adsurl = {https://ui.adsabs.harvard.edu/abs/2021AJ....161...48B}
}

@ARTICLE{naoz2016,
       author = {{Naoz}, Smadar},
        title = "{The Eccentric Kozai-Lidov Effect and Its Applications}",
      journal = {\araa},
         year = 2016,
        month = sep,
       volume = {54},
        pages = {441-489},
          doi = {10.1146/annurev-astro-081915-023315},
archivePrefix = {arXiv},
       eprint = {1601.07175},
 primaryClass = {astro-ph.EP},
       adsurl = {https://ui.adsabs.harvard.edu/abs/2016ARA&A..54..441N}
}

@ARTICLE{kozai1962,
       author = {{Kozai}, Yoshihide},
        title = "{Secular perturbations of asteroids with high inclination and eccentricity}",
      journal = {\aj},
         year = 1962,
        month = nov,
       volume = {67},
        pages = {591-598},
          doi = {10.1086/108790},
       adsurl = {https://ui.adsabs.harvard.edu/abs/1962AJ.....67..591K}
}

@ARTICLE{lidov1962,
       author = {{Lidov}, M.~L.},
        title = "{The evolution of orbits of artificial satellites of planets under the action of gravitational perturbations of external bodies}",
      journal = {\planss},
         year = 1962,
        month = oct,
       volume = {9},
       number = {10},
        pages = {719-759},
          doi = {10.1016/0032-0633(62)90129-0},
       adsurl = {https://ui.adsabs.harvard.edu/abs/1962P&SS....9..719L}
}

@ARTICLE{vonZeipel1910,
       author = {{von Zeipel}, H.},
        title = "{Sur l'application des s{\'e}ries de M. Lindstedt {\`a} l'{\'e}tude du mouvement des com{\`e}tes p{\'e}riodiques}",
      journal = {Astronomische Nachrichten},
         year = 1910,
        month = mar,
       volume = {183},
       number = {22},
        pages = {345},
          doi = {10.1002/asna.19091832202},
       adsurl = {https://ui.adsabs.harvard.edu/abs/1910AN....183..345V}
}

@ARTICLE{nagasawalin2005,
       author = {{Nagasawa}, M. and {Lin}, D.~N.~C.},
        title = "{The Dynamical Evolution of the Short-Period Extrasolar Planet around {\ensuremath{\upsilon}} Andromedae in the Pre-Main-Sequence Stage}",
      journal = {\apj},
         year = 2005,
        month = oct,
       volume = {632},
       number = {2},
        pages = {1140-1156},
          doi = {10.1086/433162},
       adsurl = {https://ui.adsabs.harvard.edu/abs/2005ApJ...632.1140N}
}

@ARTICLE{luhman2012,
       author = {{Luhman}, Kevin L.},
        title = "{The Formation and Early Evolution of Low-Mass Stars and Brown Dwarfs}",
      journal = {\araa},
         year = 2012,
        month = sep,
       volume = {50},
        pages = {65-106},
          doi = {10.1146/annurev-astro-081811-125528},
archivePrefix = {arXiv},
       eprint = {1208.5800},
 primaryClass = {astro-ph.GA},
       adsurl = {https://ui.adsabs.harvard.edu/abs/2012ARA&A..50...65L}
}

@ARTICLE{adams2013,
       author = {{Adams}, Fred C. and {Anderson}, Kassandra R. and {Bloch}, Anthony M.},
        title = "{Evolution of planetary systems with time-dependent stellar mass-loss}",
      journal = {\mnras},
         year = 2013,
        month = jun,
       volume = {432},
       number = {1},
        pages = {438-454},
          doi = {10.1093/mnras/stt479},
archivePrefix = {arXiv},
       eprint = {1303.3841},
 primaryClass = {astro-ph.SR},
       adsurl = {https://ui.adsabs.harvard.edu/abs/2013MNRAS.432..438A}
}

@ARTICLE{linida1997,
       author = {{Lin}, D.~N.~C. and {Ida}, Shigeru},
        title = "{On the Origin of Massive Eccentric Planets}",
      journal = {\apj},
         year = 1997,
        month = mar,
       volume = {477},
       number = {2},
        pages = {781-791},
          doi = {10.1086/303738},
       adsurl = {https://ui.adsabs.harvard.edu/abs/1997ApJ...477..781L}
}

@ARTICLE{weidenschilling1996,
       author = {{Weidenschilling}, Stuart J. and {Marzari}, Francesco},
        title = "{Gravitational scattering as a possible origin for giant planets at small stellar distances}",
      journal = {\nat},
         year = 1996,
        month = dec,
       volume = {384},
       number = {6610},
        pages = {619-621},
          doi = {10.1038/384619a0},
       adsurl = {https://ui.adsabs.harvard.edu/abs/1996Natur.384..619W}
}

@ARTICLE{rasio1996,
       author = {{Rasio}, Frederic A. and {Ford}, Eric B.},
        title = "{Dynamical instabilities and the formation of extrasolar planetary systems}",
      journal = {Science},
         year = 1996,
        month = nov,
       volume = {274},
        pages = {954-956},
          doi = {10.1126/science.274.5289.954},
       adsurl = {https://ui.adsabs.harvard.edu/abs/1996Sci...274..954R}
}

@ARTICLE{sumi2023,
       author = {{Sumi}, Takahiro and {Koshimoto}, Naoki and {Bennett}, David P. and {Rattenbury}, Nicholas J. and {Abe}, Fumio and {Barry}, Richard and {Bhattacharya}, Aparna and {Bond}, Ian A. and {Fujii}, Hirosane and {Fukui}, Akihiko and {Hamada}, Ryusei and {Hirao}, Yuki and {Silva}, Stela Ishitani and {Itow}, Yoshitaka and {Kirikawa}, Rintaro and {Kondo}, Iona and {Matsubara}, Yutaka and {Miyazaki}, Shota and {Muraki}, Yasushi and {Olmschenk}, Greg and {Ranc}, Cl{\'e}ment and {Satoh}, Yuki and {Suzuki}, Daisuke and {Tomoyoshi}, Mio and {Tristram}, Paul. J. and {Vandorou}, Aikaterini and {Yama}, Hibiki and {Yamashita}, Kansuke},
        title = "{Free-floating Planet Mass Function from MOA-II 9 yr Survey toward the Galactic Bulge}",
      journal = {\aj},
         year = 2023,
        month = sep,
       volume = {166},
       number = {3},
          eid = {108},
        pages = {108},
          doi = {10.3847/1538-3881/ace688},
archivePrefix = {arXiv},
       eprint = {2303.08280},
 primaryClass = {astro-ph.EP},
       adsurl = {https://ui.adsabs.harvard.edu/abs/2023AJ....166..108S}
}

@ARTICLE{Mroz2020,
       author = {{Mr{\'o}z}, Przemek and {Poleski}, Rados{\l}aw and {Gould}, Andrew and {Udalski}, Andrzej and {Sumi}, Takahiro and {Szyma{\'n}ski}, Micha{\l} K. and {Soszy{\'n}ski}, Igor and {Pietrukowicz}, Pawe{\l} and {Koz{\l}owski}, Szymon and {Skowron}, Jan and {Ulaczyk}, Krzysztof and {OGLE Collaboration} and {Albrow}, Michael D. and {Chung}, Sun-Ju and {Han}, Cheongho and {Hwang}, Kyu-Ha and {Jung}, Youn Kil and {Kim}, Hyoun-Woo and {Ryu}, Yoon-Hyun and {Shin}, In-Gu and {Shvartzvald}, Yossi and {Yee}, Jennifer C. and {Zang}, Weicheng and {Cha}, Sang-Mok and {Kim}, Dong-Jin and {Kim}, Seung-Lee and {Lee}, Chung-Uk and {Lee}, Dong-Joo and {Lee}, Yongseok and {Park}, Byeong-Gon and {Pogge}, Richard W. and {KMT Collaboration}},
        title = "{A Terrestrial-mass Rogue Planet Candidate Detected in the Shortest-timescale Microlensing Event}",
      journal = {\apjl},
         year = 2020,
        month = nov,
       volume = {903},
       number = {1},
          eid = {L11},
        pages = {L11},
          doi = {10.3847/2041-8213/abbfad},
archivePrefix = {arXiv},
       eprint = {2009.12377},
 primaryClass = {astro-ph.EP},
       adsurl = {https://ui.adsabs.harvard.edu/abs/2020ApJ...903L..11M}
}

@ARTICLE{Coleman2023,
       author = {{Coleman}, Gavin A.~L. and {Nelson}, Richard P. and {Triaud}, Amaury H.~M.~J.},
        title = "{Global N-body simulations of circumbinary planet formation around Kepler-16 and -34 analogues I: Exploring the pebble accretion scenario}",
      journal = {\mnras},
         year = 2023,
        month = jul,
       volume = {522},
       number = {3},
        pages = {4352-4373},
          doi = {10.1093/mnras/stad833},
archivePrefix = {arXiv},
       eprint = {2303.09899},
 primaryClass = {astro-ph.EP},
       adsurl = {https://ui.adsabs.harvard.edu/abs/2023MNRAS.522.4352C}
}

@ARTICLE{pearson2023,
       author = {{Pearson}, Samuel G and {McCaughrean}, Mark J},
        title = "{Jupiter Mass Binary Objects in the Trapezium Cluster}",
      journal = {arXiv e-prints},
         year = 2023,
        month = oct,
          eid = {arXiv:2310.01231},
        pages = {arXiv:2310.01231},
          doi = {10.48550/arXiv.2310.01231},
archivePrefix = {arXiv},
       eprint = {2310.01231},
 primaryClass = {astro-ph.EP},
       adsurl = {https://ui.adsabs.harvard.edu/abs/2023arXiv231001231P}
}

@ARTICLE{gould2022,
       author = {{Gould}, Andrew and {Jung}, Youn Kil and {Hwang}, Kyu-Ha and {Dong}, Subo and {Albrow}, Michael D. and {Chung}, Sun-Ju and {Han}, Cheongho and {Ryu}, Yoon-Hyun and {Shin}, In-Gu and {Shvartzvald}, Yossi and {Yang}, Hongjing and {Yee}, Jennifer C. and {Zang}, Weicheng and {Cha}, Sang-Mok and {Kim}, Dong-Jin and {Kim}, Seung-Lee and {Lee}, Chung-Uk and {Lee}, Dong-Joo and {Lee}, Yongseok and {Park}, Byeong-Gon and {Pogge}, Richard W.},
        title = "{Free-Floating Planets, the Einstein Desert, and 'OUMUAMUA}",
      journal = {Journal of Korean Astronomical Society},
         year = 2022,
        month = oct,
       volume = {55},
        pages = {173-194},
          doi = {10.5303/JKAS.2022.55.5.173},
archivePrefix = {arXiv},
       eprint = {2204.03269},
 primaryClass = {astro-ph.EP},
       adsurl = {https://ui.adsabs.harvard.edu/abs/2022JKAS...55..173G}
}

@ARTICLE{zheng2021,
       author = {{Zheng}, Xiaochen and {Lin}, Douglas N.~C. and {Mao}, Shude},
        title = "{The Influence of the Secular Perturbation of an Intermediate-mass Companion. II. Ejection of Hypervelocity Stars from the Galactic Center}",
      journal = {\apj},
         year = 2021,
        month = jun,
       volume = {914},
       number = {1},
          eid = {33},
        pages = {33},
          doi = {10.3847/1538-4357/abf5de},
archivePrefix = {arXiv},
       eprint = {2104.02989},
 primaryClass = {astro-ph.GA},
       adsurl = {https://ui.adsabs.harvard.edu/abs/2021ApJ...914...33Z}
}

@ARTICLE{zheng2020,
       author = {{Zheng}, Xiaochen and {Lin}, Douglas N.~C. and {Mao}, Shude},
        title = "{The Influence of the Secular Perturbation of an Intermediate-mass Companion. I. Eccentricity Excitation of Disk Stars at the Galactic Center}",
      journal = {\apj},
         year = 2020,
        month = dec,
       volume = {905},
       number = {2},
          eid = {169},
        pages = {169},
          doi = {10.3847/1538-4357/abc8e5},
archivePrefix = {arXiv},
       eprint = {2011.04653},
 primaryClass = {astro-ph.GA},
       adsurl = {https://ui.adsabs.harvard.edu/abs/2020ApJ...905..169Z}
}

@ARTICLE{zheng2017b,
       author = {{Zheng}, Xiaochen and {Lin}, Douglas N.~C. and {Kouwenhoven}, M.~B.~N. and {Mao}, Shude and {Zhang}, Xiaojia},
        title = "{Clearing Residual Planetesimals by Sweeping Secular Resonances in Transitional Disks: A Lone-planet Scenario for the Wide Gaps in Debris Disks around Vega and Fomalhaut}",
      journal = {\apj},
         year = 2017,
        month = nov,
       volume = {849},
       number = {2},
          eid = {98},
        pages = {98},
          doi = {10.3847/1538-4357/aa8ef3},
archivePrefix = {arXiv},
       eprint = {1709.07382},
 primaryClass = {astro-ph.EP},
       adsurl = {https://ui.adsabs.harvard.edu/abs/2017ApJ...849...98Z}
}

@ARTICLE{zheng2017,
       author = {{Zheng}, Xiaochen and {Lin}, Douglas N.~C. and {Kouwenhoven}, M.~B.~N.},
        title = "{Planetesimal Clearing and Size-dependent Asteroid Retention by Secular Resonance Sweeping during the Depletion of the Solar Nebula}",
      journal = {\apj},
         year = 2017,
        month = feb,
       volume = {836},
       number = {2},
          eid = {207},
        pages = {207},
          doi = {10.3847/1538-4357/836/2/207},
archivePrefix = {arXiv},
       eprint = {1610.09670},
 primaryClass = {astro-ph.EP},
       adsurl = {https://ui.adsabs.harvard.edu/abs/2017ApJ...836..207Z}
}

@ARTICLE{Matsumura2013,
       author = {{Matsumura}, Soko and {Ida}, Shigeru and {Nagasawa}, Makiko},
        title = "{Effects of Dynamical Evolution of Giant Planets on Survival of Terrestrial Planets}",
      journal = {\apj},
         year = 2013,
        month = apr,
       volume = {767},
       number = {2},
          eid = {129},
        pages = {129},
          doi = {10.1088/0004-637X/767/2/129},
archivePrefix = {arXiv},
       eprint = {1209.1320},
 primaryClass = {astro-ph.EP},
       adsurl = {https://ui.adsabs.harvard.edu/abs/2013ApJ...767..129M}
}

@ARTICLE{sumi2011,
       author = {{Sumi}, T. and {Kamiya}, K. and {Bennett}, D.~P. and {Bond}, I.~A. and {Abe}, F. and {Botzler}, C.~S. and {Fukui}, A. and {Furusawa}, K. and {Hearnshaw}, J.~B. and {Itow}, Y. and {Kilmartin}, P.~M. and {Korpela}, A. and {Lin}, W. and {Ling}, C.~H. and {Masuda}, K. and {Matsubara}, Y. and {Miyake}, N. and {Motomura}, M. and {Muraki}, Y. and {Nagaya}, M. and {Nakamura}, S. and {Ohnishi}, K. and {Okumura}, T. and {Perrott}, Y.~C. and {Rattenbury}, N. and {Saito}, To. and {Sako}, T. and {Sullivan}, D.~J. and {Sweatman}, W.~L. and {Tristram}, P.~J. and {Udalski}, A. and {Szyma{\'n}ski}, M.~K. and {Kubiak}, M. and {Pietrzy{\'n}ski}, G. and {Poleski}, R. and {Soszy{\'n}ski}, I. and {Wyrzykowski}, {\L}. and {Ulaczyk}, K. and {Microlensing Observations in Astrophysics (MOA) Collaboration}},
        title = "{Unbound or distant planetary mass population detected by gravitational microlensing}",
      journal = {\nat},
         year = 2011,
        month = may,
       volume = {473},
       number = {7347},
        pages = {349-352},
          doi = {10.1038/nature10092},
archivePrefix = {arXiv},
       eprint = {1105.3544},
 primaryClass = {astro-ph.EP},
       adsurl = {https://ui.adsabs.harvard.edu/abs/2011Natur.473..349S}
}

@ARTICLE{rein2015,
       author = {{Rein}, Hanno and {Spiegel}, David S.},
        title = "{IAS15: a fast, adaptive, high-order integrator for gravitational dynamics, accurate to machine precision over a billion orbits}",
      journal = {\mnras},
         year = 2015,
        month = jan,
       volume = {446},
       number = {2},
        pages = {1424-1437},
          doi = {10.1093/mnras/stu2164},
archivePrefix = {arXiv},
       eprint = {1409.4779},
 primaryClass = {astro-ph.EP},
       adsurl = {https://ui.adsabs.harvard.edu/abs/2015MNRAS.446.1424R}
}

@INPROCEEDINGS{everhart1985,
       author = {{Everhart}, E.},
        title = "{An efficient integrator that uses Gauss-Radau spacings}",
    booktitle = {IAU Colloq. 83: Dynamics of Comets: Their Origin and Evolution},
         year = 1985,
       editor = {{Carusi}, A. and {Valsecchi}, G.~B.},
       series = {Astrophysics and Space Science Library},
       volume = {115},
        month = jan,
        pages = {185},
          doi = {10.1007/978-94-009-5400-7_17},
       adsurl = {https://ui.adsabs.harvard.edu/abs/1985ASSL..115..185E}
}

@ARTICLE{eggleton1998,
       author = {{Eggleton}, Peter P. and {Kiseleva}, Ludmila G. and {Hut}, Piet},
        title = "{The Equilibrium Tide Model for Tidal Friction}",
      journal = {\apj},
         year = 1998,
        month = may,
       volume = {499},
       number = {2},
        pages = {853-870},
          doi = {10.1086/305670},
archivePrefix = {arXiv},
       eprint = {astro-ph/9801246},
 primaryClass = {astro-ph},
       adsurl = {https://ui.adsabs.harvard.edu/abs/1998ApJ...499..853E}
}

@BOOK{murray1999,
       author = {{Murray}, Carl D. and {Dermott}, Stanley F.},
        title = "{Solar System Dynamics}",
         year = 1999,
          doi = {10.1017/CBO9781139174817},
       adsurl = {https://ui.adsabs.harvard.edu/abs/1999ssd..book.....M}
}

@ARTICLE{nagasawa2003,
       author = {{Nagasawa}, M. and {Lin}, D.~N.~C. and {Ida}, S.},
        title = "{Eccentricity Evolution of Extrasolar Multiple Planetary Systems Due to the Depletion of Nascent Protostellar Disks}",
      journal = {\apj},
         year = 2003,
        month = apr,
       volume = {586},
       number = {2},
        pages = {1374-1393},
          doi = {10.1086/367884},
archivePrefix = {arXiv},
       eprint = {astro-ph/0205104},
 primaryClass = {astro-ph},
       adsurl = {https://ui.adsabs.harvard.edu/abs/2003ApJ...586.1374N}
}

@ARTICLE{2020MNRAS.491.2885T,
       author = {{Tamayo}, Daniel and {Rein}, Hanno and {Shi}, Pengshuai and {Hernandez}, David M.},
        title = "{REBOUNDx: a library for adding conservative and dissipative forces to otherwise symplectic N-body integrations}",
      journal = {\mnras},
         year = 2020,
        month = jan,
       volume = {491},
       number = {2},
        pages = {2885-2901},
          doi = {10.1093/mnras/stz2870},
archivePrefix = {arXiv},
       eprint = {1908.05634},
 primaryClass = {astro-ph.EP},
       adsurl = {https://ui.adsabs.harvard.edu/abs/2020MNRAS.491.2885T}
}

@article{annurev:/content/journals/10.1146/annurev-astro-081913-035941,
   author = "Ogilvie, Gordon I.",
   title = "Tidal Dissipation in Stars and Giant Planets", 
   journal= "Annual Review of Astronomy and Astrophysics",
   year = "2014",
   volume = "52",
   number = "Volume 52, 2014",
   pages = "171-210",
   doi = "https://doi.org/10.1146/annurev-astro-081913-035941",
   url = "https://www.annualreviews.org/content/journals/10.1146/annurev-astro-081913-035941",
   publisher = "Annual Reviews",
   issn = "1545-4282",
   type = "Journal Article",
  }

@article{rein_rebound:_2011,
    title = {{REBOUND}: {An} open-source multi-purpose {N}-body code for collisional dynamics},
    volume = {128},
    issn = {0004-6361},
    url = {http://arxiv.org/abs/1110.4876%0Ahttp://dx.doi.org/10.1051/0004-6361/201118085},
    doi = {10.1051/0004-6361/201118085},
    author = {Rein, Hanno and Liu, Shang-Fei},
    year = {2011},
    note = {arXiv: 1110.4876},
    pages = {1--10},
}

@article{baronett_stellar_2022,
	title = {Stellar evolution and tidal dissipation in {REBOUNDx}},
	volume = {510},
	copyright = {https://academic.oup.com/journals/pages/open\_access/funder\_policies/chorus/standard\_publication\_model},
	issn = {0035-8711, 1365-2966},
	url = {https://academic.oup.com/mnras/article/510/4/6001/6502369},
	doi = {10.1093/mnras/stac043},
	language = {en},
	number = {4},
	urldate = {2025-05-05},
	journal = {Monthly Notices of the Royal Astronomical Society},
	author = {Baronett, Stanley A and Ferich, Noah and Tamayo, Daniel and Steffen, Jason H},
	month = jan,
	year = {2022},
	pages = {6001--6009},
}

@article{mroz_no_2017,
    title = {No large population of unbound or wide-orbit {Jupiter}-mass planets},
    volume = {548},
    issn = {14764687},
    doi = {10.1038/nature23276},
    number = {7666},
    journal = {Nature},
    author = {Mróz, Przemek and Udalski, Andrzej and Skowron, Jan and Poleski, Radosław and Kozłowski, Szymon and Szymański, Michał K. and Soszyński, Igor and Wyrzykowski, Łukasz and Pietrukowicz, Paweł and Ulaczyk, Krzysztof and Skowron, Dorota and Pawlak, Michał},
    year = {2017},
    pmid = {28738410},
    note = {arXiv: 1707.07634},
    pages = {183--186},
}

@article{malmberg_effects_2010,
    title = {The effects of fly-bys on planetary systems},
    volume = {22},
    number = {September},
    journal = {Astronomy},
    author = {Malmberg, Daniel and Davies, Melvyn B and Heggie, Douglas C},
    year = {2010},
    note = {arXiv: 1009.4196v1},
    pages = {1--22},
}

@ARTICLE{ida2004,
       author = {{Ida}, S. and {Lin}, D.~N.~C.},
        title = "{Toward a Deterministic Model of Planetary Formation. I. A Desert in the Mass and Semimajor Axis Distributions of Extrasolar Planets}",
      journal = {\apj},
         year = 2004,
        month = mar,
       volume = {604},
       number = {1},
        pages = {388-413},
          doi = {10.1086/381724},
archivePrefix = {arXiv},
       eprint = {astro-ph/0312144},
 primaryClass = {astro-ph},
       adsurl = {https://ui.adsabs.harvard.edu/abs/2004ApJ...604..388I}
}

@ARTICLE{zhudong2021,
       author = {{Zhu}, Wei and {Dong}, Subo},
        title = "{Exoplanet Statistics and Theoretical Implications}",
      journal = {\araa},
         year = 2021,
        month = sep,
       volume = {59},
        pages = {291-336},
          doi = {10.1146/annurev-astro-112420-020055},
archivePrefix = {arXiv},
       eprint = {2103.02127},
 primaryClass = {astro-ph.EP},
       adsurl = {https://ui.adsabs.harvard.edu/abs/2021ARA&A..59..291Z}
}

@ARTICLE{dai2024,
       author = {{Dai}, Fei and {Goldberg}, Max and {Batygin}, Konstantin and {van Saders}, Jennifer and {Chiang}, Eugene and {Choksi}, Nick and {Li}, Rixin and {Petigura}, Erik A. and {Gilbert}, Gregory J. and {Millholland}, Sarah C. and {Dai}, Yuan-Zhe and {Bouma}, Luke and {Weiss}, Lauren M. and {Winn}, Joshua N.},
        title = "{The Prevalence of Resonance Among Young, Close-in Planets}",
      journal = {\aj},
         year = 2024,
        month = dec,
       volume = {168},
       number = {6},
          eid = {239},
        pages = {239},
          doi = {10.3847/1538-3881/ad83a6},
archivePrefix = {arXiv},
       eprint = {2406.06885},
 primaryClass = {astro-ph.EP},
       adsurl = {https://ui.adsabs.harvard.edu/abs/2024AJ....168..239D}
}

@ARTICLE{fabrycky2014,
       author = {{Fabrycky}, Daniel C. and {Lissauer}, Jack J. and {Ragozzine}, Darin and {Rowe}, Jason F. and {Steffen}, Jason H. and {Agol}, Eric and {Barclay}, Thomas and {Batalha}, Natalie and {Borucki}, William and {Ciardi}, David R. and {Ford}, Eric B. and {Gautier}, Thomas N. and {Geary}, John C. and {Holman}, Matthew J. and {Jenkins}, Jon M. and {Li}, Jie and {Morehead}, Robert C. and {Morris}, Robert L. and {Shporer}, Avi and {Smith}, Jeffrey C. and {Still}, Martin and {Van Cleve}, Jeffrey},
        title = "{Architecture of Kepler's Multi-transiting Systems. II. New Investigations with Twice as Many Candidates}",
      journal = {\apj},
         year = 2014,
        month = aug,
       volume = {790},
       number = {2},
          eid = {146},
        pages = {146},
          doi = {10.1088/0004-637X/790/2/146},
archivePrefix = {arXiv},
       eprint = {1202.6328},
 primaryClass = {astro-ph.EP},
       adsurl = {https://ui.adsabs.harvard.edu/abs/2014ApJ...790..146F}
}

@ARTICLE{galicher2016,
       author = {{Galicher}, R. and {Marois}, C. and {Macintosh}, B. and {Zuckerman}, B. and {Barman}, T. and {Konopacky}, Q. and {Song}, I. and {Patience}, J. and {Lafreni{\`e}re}, D. and {Doyon}, R. and {Nielsen}, E.~L.},
        title = "{The International Deep Planet Survey. II. The frequency of directly imaged giant exoplanets with stellar mass}",
      journal = {\aap},
         year = 2016,
        month = oct,
       volume = {594},
          eid = {A63},
        pages = {A63},
          doi = {10.1051/0004-6361/201527828},
archivePrefix = {arXiv},
       eprint = {1607.08239},
 primaryClass = {astro-ph.EP},
       adsurl = {https://ui.adsabs.harvard.edu/abs/2016A&A...594A..63G}
}

@ARTICLE{zang2025,
       author = {{Zang}, Weicheng and {Jung}, Youn Kil and {Yee}, Jennifer C. and {Hwang}, Kyu-Ha and {Yang}, Hongjing and {Udalski}, Andrzej and {Sumi}, Takahiro and {Gould}, Andrew and {Mao}, Shude and {Albrow}, Michael D. and {Chung}, Sun-Ju and {Han}, Cheongho and {Ryu}, Yoon-Hyun and {Shin}, In-Gu and {Shvartzvald}, Yossi and {Cha}, Sang-Mok and {Kim}, Dong-Jin and {Kim}, Hyoun-Woo and {Kim}, Seung-Lee and {Lee}, Chung-Uk and {Lee}, Dong-Joo and {Lee}, Yongseok and {Park}, Byeong-Gon and {Pogge}, Richard W. and {Zhang}, Xiangyu and {Kuang}, Renkun and {Wang}, Hanyue and {Zhang}, Jiyuan and {Hu}, Zhecheng and {Zhu}, Wei and {Mr{\'o}z}, Przemek and {Skowron}, Jan and {Poleski}, Rados{\l}aw and {Szyma{\'n}ski}, Micha{\l} K. and {Soszy{\'n}ski}, Igor and {Pietrukowicz}, Pawe{\l} and {Koz{\l}owski}, Szymon and {Ulaczyk}, Krzysztof and {Rybicki}, Krzysztof A. and {Iwanek}, Patryk and {Wrona}, Marcin and {Gromadzki}, Mariusz and {Abe}, Fumio and {Barry}, Richard and {Bennett}, David P. and {Bhattacharya}, Aparna and {Bond}, Ian A. and {Fujii}, Hirosane and {Fukui}, Akihiko and {Hamada}, Ryusei and {Hirao}, Yuki and {Silva}, Stela Ishitani and {Itow}, Yoshitaka and {Kirikawa}, Rintaro and {Koshimoto}, Naoki and {Matsubara}, Yutaka and {Miyazaki}, Shota and {Muraki}, Yasushi and {Olmschenk}, Greg and {Ranc}, Cl{\'e}ment and {Rattenbury}, Nicholas J. and {Satoh}, Yuki and {Suzuki}, Daisuke and {Tomoyoshi}, Mio and {Tristram}, Paul J. and {Vandorou}, Aikaterini and {Yama}, Hibiki and {Yamashita}, Kansuke},
        title = "{Microlensing events indicate that super-Earth exoplanets are common in Jupiter-like orbits}",
      journal = {Science},
         year = 2025,
        month = apr,
       volume = {388},
       number = {6745},
        pages = {400-404},
          doi = {10.1126/science.adn6088},
archivePrefix = {arXiv},
       eprint = {2504.20158},
 primaryClass = {astro-ph.EP},
       adsurl = {https://ui.adsabs.harvard.edu/abs/2025Sci...388..400Z}
}

@BOOK{perryman2018,
       author = {{Perryman}, Michael},
        title = "{The Exoplanet Handbook}",
         year = 2018,
       adsurl = {https://ui.adsabs.harvard.edu/abs/2018exha.book.....P}
}

@ARTICLE{gould2010,
       author = {{Gould}, A. and {Dong}, Subo and {Gaudi}, B.~S. and {Udalski}, A. and {Bond}, I.~A. and {Greenhill}, J. and {Street}, R.~A. and {Dominik}, M. and {Sumi}, T. and {Szyma{\'n}ski}, M.~K. and {Han}, C. and {Allen}, W. and {Bolt}, G. and {Bos}, M. and {Christie}, G.~W. and {DePoy}, D.~L. and {Drummond}, J. and {Eastman}, J.~D. and {Gal-Yam}, A. and {Higgins}, D. and {Janczak}, J. and {Kaspi}, S. and {Koz{\l}owski}, S. and {Lee}, C. -U. and {Mallia}, F. and {Maury}, A. and {Maoz}, D. and {McCormick}, J. and {Monard}, L.~A.~G. and {Moorhouse}, D. and {Morgan}, N. and {Natusch}, T. and {Ofek}, E.~O. and {Park}, B. -G. and {Pogge}, R.~W. and {Polishook}, D. and {Santallo}, R. and {Shporer}, A. and {Spector}, O. and {Thornley}, G. and {Yee}, J.~C. and {{\ensuremath{\mu}}FUN Collaboration} and {Kubiak}, M. and {Pietrzy{\'n}ski}, G. and {Soszy{\'n}ski}, I. and {Szewczyk}, O. and {Wyrzykowski}, {\L}. and {Ulaczyk}, K. and {Poleski}, R. and {OGLE Collaboration} and {Abe}, F. and {Bennett}, D.~P. and {Botzler}, C.~S. and {Douchin}, D. and {Freeman}, M. and {Fukui}, A. and {Furusawa}, K. and {Hearnshaw}, J.~B. and {Hosaka}, S. and {Itow}, Y. and {Kamiya}, K. and {Kilmartin}, P.~M. and {Korpela}, A. and {Lin}, W. and {Ling}, C.~H. and {Makita}, S. and {Masuda}, K. and {Matsubara}, Y. and {Miyake}, N. and {Muraki}, Y. and {Nagaya}, M. and {Nishimoto}, K. and {Ohnishi}, K. and {Okumura}, T. and {Perrott}, Y.~C. and {Philpott}, L. and {Rattenbury}, N. and {Saito}, To. and {Sako}, T. and {Sullivan}, D.~J. and {Sweatman}, W.~L. and {Tristram}, P.~J. and {von Seggern}, E. and {Yock}, P.~C.~M. and {MOA Collaboration} and {Albrow}, M. and {Batista}, V. and {Beaulieu}, J.~P. and {Brillant}, S. and {Caldwell}, J. and {Calitz}, J.~J. and {Cassan}, A. and {Cole}, A. and {Cook}, K. and {Coutures}, C. and {Dieters}, S. and {Dominis Prester}, D. and {Donatowicz}, J. and {Fouqu{\'e}}, P. and {Hill}, K. and {Hoffman}, M. and {Jablonski}, F. and {Kane}, S.~R. and {Kains}, N. and {Kubas}, D. and {Marquette}, J. -B. and {Martin}, R. and {Martioli}, E. and {Meintjes}, P. and {Menzies}, J. and {Pedretti}, E. and {Pollard}, K. and {Sahu}, K.~C. and {Vinter}, C. and {Wambsganss}, J. and {Watson}, R. and {Williams}, A. and {Zub}, M. and {PLANET Collaboration} and {Allan}, A. and {Bode}, M.~F. and {Bramich}, D.~M. and {Burgdorf}, M.~J. and {Clay}, N. and {Fraser}, S. and {Hawkins}, E. and {Horne}, K. and {Kerins}, E. and {Lister}, T.~A. and {Mottram}, C. and {Saunders}, E.~S. and {Snodgrass}, C. and {Steele}, I.~A. and {Tsapras}, Y. and {RoboNet Collaboration} and {J{\o}rgensen}, U.~G. and {Anguita}, T. and {Bozza}, V. and {Calchi Novati}, S. and {Harps{\o}e}, K. and {Hinse}, T.~C. and {Hundertmark}, M. and {Kj{\ae}rgaard}, P. and {Liebig}, C. and {Mancini}, L. and {Masi}, G. and {Mathiasen}, M. and {Rahvar}, S. and {Ricci}, D. and {Scarpetta}, G. and {Southworth}, J. and {Surdej}, J. and {Th{\"o}ne}, C.~C. and {MiNDSTEp Consortium}},
        title = "{Frequency of Solar-like Systems and of Ice and Gas Giants Beyond the Snow Line from High-magnification Microlensing Events in 2005-2008}",
      journal = {\apj},
         year = 2010,
        month = sep,
       volume = {720},
       number = {2},
        pages = {1073-1089},
          doi = {10.1088/0004-637X/720/2/1073},
archivePrefix = {arXiv},
       eprint = {1001.0572},
 primaryClass = {astro-ph.EP},
       adsurl = {https://ui.adsabs.harvard.edu/abs/2010ApJ...720.1073G}
}

@ARTICLE{wu2003,
       author = {{Wu}, Y. and {Murray}, N.},
        title = "{Planet Migration and Binary Companions: The Case of HD 80606b}",
      journal = {\apj},
         year = 2003,
        month = may,
       volume = {589},
       number = {1},
        pages = {605-614},
          doi = {10.1086/374598},
archivePrefix = {arXiv},
       eprint = {astro-ph/0303010},
 primaryClass = {astro-ph},
       adsurl = {https://ui.adsabs.harvard.edu/abs/2003ApJ...589..605W}
}

@ARTICLE{nagasawa2008,
       author = {{Nagasawa}, M. and {Ida}, S. and {Bessho}, T.},
        title = "{Formation of Hot Planets by a Combination of Planet Scattering, Tidal Circularization, and the Kozai Mechanism}",
      journal = {\apj},
         year = 2008,
        month = may,
       volume = {678},
       number = {1},
        pages = {498-508},
          doi = {10.1086/529369},
archivePrefix = {arXiv},
       eprint = {0801.1368},
 primaryClass = {astro-ph},
       adsurl = {https://ui.adsabs.harvard.edu/abs/2008ApJ...678..498N}
}

@ARTICLE{nagasawa2011,
       author = {{Nagasawa}, M. and {Ida}, S.},
        title = "{Orbital Distributions of Close-in Planets and Distant Planets Formed by Scattering and Dynamical Tides}",
      journal = {\apj},
         year = 2011,
        month = dec,
       volume = {742},
       number = {2},
          eid = {72},
        pages = {72},
          doi = {10.1088/0004-637X/742/2/72},
       adsurl = {https://ui.adsabs.harvard.edu/abs/2011ApJ...742...72N}
}

@ARTICLE{fabrycky2007,
       author = {{Fabrycky}, Daniel and {Tremaine}, Scott},
        title = "{Shrinking Binary and Planetary Orbits by Kozai Cycles with Tidal Friction}",
      journal = {\apj},
         year = 2007,
        month = nov,
       volume = {669},
       number = {2},
        pages = {1298-1315},
          doi = {10.1086/521702},
archivePrefix = {arXiv},
       eprint = {0705.4285},
 primaryClass = {astro-ph},
       adsurl = {https://ui.adsabs.harvard.edu/abs/2007ApJ...669.1298F}
}

@ARTICLE{1981A&A....99..126H,
       author = {{Hut}, P.},
        title = "{Tidal evolution in close binary systems.}",
      journal = {\aap},
         year = 1981,
        month = jun,
       volume = {99},
        pages = {126-140},
       adsurl = {https://ui.adsabs.harvard.edu/abs/1981A&A....99..126H}
}

@ARTICLE{ogilvie2004,
       author = {{Ogilvie}, G.~I. and {Lin}, D.~N.~C.},
        title = "{Tidal Dissipation in Rotating Giant Planets}",
      journal = {\apj},
         year = 2004,
        month = jul,
       volume = {610},
       number = {1},
        pages = {477-509},
          doi = {10.1086/421454},
archivePrefix = {arXiv},
       eprint = {astro-ph/0310218},
 primaryClass = {astro-ph},
       adsurl = {https://ui.adsabs.harvard.edu/abs/2004ApJ...610..477O}
}

@ARTICLE{ida2013,
       author = {{Ida}, S. and {Lin}, D.~N.~C. and {Nagasawa}, M.},
        title = "{Toward a Deterministic Model of Planetary Formation. VII. Eccentricity Distribution of Gas Giants}",
      journal = {\apj},
         year = 2013,
        month = sep,
       volume = {775},
       number = {1},
          eid = {42},
        pages = {42},
          doi = {10.1088/0004-637X/775/1/42},
archivePrefix = {arXiv},
       eprint = {1307.6450},
 primaryClass = {astro-ph.EP},
       adsurl = {https://ui.adsabs.harvard.edu/abs/2013ApJ...775...42I}
}

@ARTICLE{zhoulin2007,
       author = {{Zhou}, Ji-Lin and {Lin}, Douglas N.~C.},
        title = "{Planetesimal Accretion onto Growing Proto-Gas Giant Planets}",
      journal = {\apj},
         year = 2007,
        month = sep,
       volume = {666},
       number = {1},
        pages = {447-465},
          doi = {10.1086/520043},
archivePrefix = {arXiv},
       eprint = {0705.3105},
 primaryClass = {astro-ph},
       adsurl = {https://ui.adsabs.harvard.edu/abs/2007ApJ...666..447Z}
}

@ARTICLE{zhoulinsun2007,
       author = {{Zhou}, Ji-Lin and {Lin}, Douglas N.~C. and {Sun}, Yi-Sui},
        title = "{Post-oligarchic Evolution of Protoplanetary Embryos and the Stability of Planetary Systems}",
      journal = {\apj},
         year = 2007,
        month = sep,
       volume = {666},
       number = {1},
        pages = {423-435},
          doi = {10.1086/519918},
archivePrefix = {arXiv},
       eprint = {0705.2164},
 primaryClass = {astro-ph},
       adsurl = {https://ui.adsabs.harvard.edu/abs/2007ApJ...666..423Z}
}

@ARTICLE{2012A&A...541A.165R,
       author = {{Remus}, F. and {Mathis}, S. and {Zahn}, J. -P. and {Lainey}, V.},
        title = "{Anelastic tidal dissipation in multi-layer planets}",
      journal = {\aap},
         year = 2012,
        month = may,
       volume = {541},
          eid = {A165},
        pages = {A165},
          doi = {10.1051/0004-6361/201118595},
archivePrefix = {arXiv},
       eprint = {1204.1468},
 primaryClass = {astro-ph.EP},
       adsurl = {https://ui.adsabs.harvard.edu/abs/2012A&A...541A.165R}
}

@ARTICLE{liu2013,
       author = {{Liu}, Shang-Fei and {Guillochon}, James and {Lin}, Douglas N.~C. and {Ramirez-Ruiz}, Enrico},
        title = "{On the Survivability and Metamorphism of Tidally Disrupted Giant Planets: The Role of Dense Cores}",
      journal = {\apj},
         year = 2013,
        month = jan,
       volume = {762},
       number = {1},
          eid = {37},
        pages = {37},
          doi = {10.1088/0004-637X/762/1/37},
archivePrefix = {arXiv},
       eprint = {1211.1971},
 primaryClass = {astro-ph.EP},
       adsurl = {https://ui.adsabs.harvard.edu/abs/2013ApJ...762...37L}
}

@ARTICLE{liu2015,
       author = {{Liu}, Shang-Fei and {Agnor}, Craig B. and {Lin}, D.~N.~C. and {Li}, Shu-Lin},
        title = "{Embryo impacts and gas giant mergers - II. Diversity of hot Jupiters' internal structure}",
      journal = {\mnras},
         year = 2015,
        month = jan,
       volume = {446},
       number = {2},
        pages = {1685-1702},
          doi = {10.1093/mnras/stu2205},
archivePrefix = {arXiv},
       eprint = {1410.6815},
 primaryClass = {astro-ph.EP},
       adsurl = {https://ui.adsabs.harvard.edu/abs/2015MNRAS.446.1685L}
}

@ARTICLE{gu2003,
       author = {{Gu}, Pin-Gao and {Lin}, Douglas N.~C. and {Bodenheimer}, Peter H.},
        title = "{The Effect of Tidal Inflation Instability on the Mass and Dynamical Evolution of Extrasolar Planets with Ultrashort Periods}",
      journal = {\apj},
         year = 2003,
        month = may,
       volume = {588},
       number = {1},
        pages = {509-534},
          doi = {10.1086/373920},
archivePrefix = {arXiv},
       eprint = {astro-ph/0303362},
 primaryClass = {astro-ph},
       adsurl = {https://ui.adsabs.harvard.edu/abs/2003ApJ...588..509G}
}

@ARTICLE{sridhar1992,
       author = {{Sridhar}, S. and {Tremaine}, S.},
        title = "{Tidal disruption of viscous bodies}",
      journal = {\icarus},
         year = 1992,
        month = jan,
       volume = {95},
       number = {1},
        pages = {86-99},
          doi = {10.1016/0019-1035(92)90193-B},
       adsurl = {https://ui.adsabs.harvard.edu/abs/1992Icar...95...86S}
}

@ARTICLE{lin1996,
       author = {{Lin}, D.~N.~C. and {Bodenheimer}, P. and {Richardson}, D.~C.},
        title = "{Orbital migration of the planetary companion of 51 Pegasi to its present location}",
      journal = {\nat},
         year = 1996,
        month = apr,
       volume = {380},
       number = {6575},
        pages = {606-607},
          doi = {10.1038/380606a0},
       adsurl = {https://ui.adsabs.harvard.edu/abs/1996Natur.380..606L}
}

@ARTICLE{gonzales2013,
       author = {{Gonz{\'a}lez Hern{\'a}ndez}, J.~I. and {Delgado-Mena}, E. and {Sousa}, S.~G. and {Israelian}, G. and {Santos}, N.~C. and {Adibekyan}, V. Zh. and {Udry}, S.},
        title = "{Searching for the signatures of terrestrial planets in F-, G-type main-sequence stars}",
      journal = {\aap},
         year = 2013,
        month = apr,
       volume = {552},
          eid = {A6},
        pages = {A6},
          doi = {10.1051/0004-6361/201220165},
archivePrefix = {arXiv},
       eprint = {1301.2109},
 primaryClass = {astro-ph.EP},
       adsurl = {https://ui.adsabs.harvard.edu/abs/2013A&A...552A...6G}
}

@ARTICLE{liu2014,
       author = {{Liu}, F. and {Asplund}, M. and {Ramirez}, I. and {Yong}, D. and {Melendez}, J.},
        title = "{A high-precision chemical abundance analysis of the HAT-P-1 stellar binary: constraints on planet formation.}",
      journal = {\mnras},
         year = 2014,
        month = jul,
       volume = {442},
        pages = {L51-L55},
          doi = {10.1093/mnrasl/slu055},
archivePrefix = {arXiv},
       eprint = {1404.2112},
 primaryClass = {astro-ph.SR},
       adsurl = {https://ui.adsabs.harvard.edu/abs/2014MNRAS.442L..51L}
}

@ARTICLE{saffe2019,
       author = {{Saffe}, C. and {Jofr{\'e}}, E. and {Miquelarena}, P. and {Jaque Arancibia}, M. and {Flores}, M. and {L{\'o}pez}, F.~M. and {Collado}, A.},
        title = "{High-precision analysis of binary stars with planets. I. Searching for condensation temperature trends in the HD 106515 system}",
      journal = {\aap},
         year = 2019,
        month = may,
       volume = {625},
          eid = {A39},
        pages = {A39},
          doi = {10.1051/0004-6361/201935352},
archivePrefix = {arXiv},
       eprint = {1904.01955},
 primaryClass = {astro-ph.SR},
       adsurl = {https://ui.adsabs.harvard.edu/abs/2019A&A...625A..39S}
}

@ARTICLE{shen2005ApJ,
       author = {{Shen}, Z. -X. and {Jones}, B. and {Lin}, D.~N.~C. and {Liu}, X. -W. and {Li}, S. -L.},
        title = "{Spectroscopic Abundance Analysis of Dwarfs in the Young Open Cluster IC 4665}",
      journal = {\apj},
         year = 2005,
        month = dec,
       volume = {635},
       number = {1},
        pages = {608-624},
          doi = {10.1086/497264},
archivePrefix = {arXiv},
       eprint = {astro-ph/0508387},
 primaryClass = {astro-ph},
       adsurl = {https://ui.adsabs.harvard.edu/abs/2005ApJ...635..608S}
}

@ARTICLE{veras2005,
       author = {{Veras}, Dimitri and {Armitage}, Philip J.},
        title = "{The Influence of Massive Planet Scattering on Nascent Terrestrial Planets}",
      journal = {\apjl},
         year = 2005,
        month = feb,
       volume = {620},
       number = {2},
        pages = {L111-L114},
          doi = {10.1086/428831},
archivePrefix = {arXiv},
       eprint = {astro-ph/0501356},
 primaryClass = {astro-ph},
       adsurl = {https://ui.adsabs.harvard.edu/abs/2005ApJ...620L.111V}
}

@ARTICLE{spurzem2009,
       author = {{Spurzem}, R. and {Giersz}, M. and {Heggie}, D.~C. and {Lin}, D.~N.~C.},
        title = "{Dynamics of Planetary Systems in Star Clusters}",
      journal = {\apj},
         year = 2009,
        month = may,
       volume = {697},
       number = {1},
        pages = {458-482},
          doi = {10.1088/0004-637X/697/1/458},
archivePrefix = {arXiv},
       eprint = {astro-ph/0612757},
 primaryClass = {astro-ph},
       adsurl = {https://ui.adsabs.harvard.edu/abs/2009ApJ...697..458S}
}

@ARTICLE{zheng2015,
       author = {{Zheng}, Xiaochen and {Kouwenhoven}, M.~B.~N. and {Wang}, Long},
        title = "{The dynamical fate of planetary systems in young star clusters}",
      journal = {\mnras},
         year = 2015,
        month = nov,
       volume = {453},
       number = {3},
        pages = {2759-2770},
          doi = {10.1093/mnras/stv1832},
archivePrefix = {arXiv},
       eprint = {1508.01593},
 primaryClass = {astro-ph.EP},
       adsurl = {https://ui.adsabs.harvard.edu/abs/2015MNRAS.453.2759Z}
}

@ARTICLE{zhang2020,
       author = {{Zhang}, Yun and {Lin}, Douglas N.~C.},
        title = "{Tidal fragmentation as the origin of 1I/2017 U1 (`Oumuamua)}",
      journal = {Nature Astronomy},
         year = 2020,
        month = apr,
       volume = {4},
        pages = {852-860},
          doi = {10.1038/s41550-020-1065-8},
archivePrefix = {arXiv},
       eprint = {2004.07218},
 primaryClass = {astro-ph.EP},
       adsurl = {https://ui.adsabs.harvard.edu/abs/2020NatAs...4..852Z}
}

@ARTICLE{liu2019,
       author = {{Liu}, Shang-Fei and {Hori}, Yasunori and {M{\"u}ller}, Simon and {Zheng}, Xiaochen and {Helled}, Ravit and {Lin}, Doug and {Isella}, Andrea},
        title = "{The formation of Jupiter's diluted core by a giant impact}",
      journal = {\nat},
         year = 2019,
        month = aug,
       volume = {572},
       number = {7769},
        pages = {355-357},
          doi = {10.1038/s41586-019-1470-2},
archivePrefix = {arXiv},
       eprint = {2007.08338},
 primaryClass = {astro-ph.EP},
       adsurl = {https://ui.adsabs.harvard.edu/abs/2019Natur.572..355L}
}

@ARTICLE{Ochiai2014,
       author = {{Ochiai}, H. and {Nagasawa}, M. and {Ida}, S.},
        title = "{Extrasolar Binary Planets. I. Formation by Tidal Capture during Planet-Planet Scattering}",
      journal = {\apj},
         year = 2014,
        month = aug,
       volume = {790},
       number = {2},
          eid = {92},
        pages = {92},
          doi = {10.1088/0004-637X/790/2/92},
archivePrefix = {arXiv},
       eprint = {1406.6780},
 primaryClass = {astro-ph.EP},
       adsurl = {https://ui.adsabs.harvard.edu/abs/2014ApJ...790...92O}
}

@ARTICLE{Montet2014,
       author = {{Montet}, Benjamin T. and {Crepp}, Justin R. and {Johnson}, John Asher and {Howard}, Andrew W. and {Marcy}, Geoffrey W.},
        title = "{The TRENDS High-contrast Imaging Survey. IV. The Occurrence Rate of Giant Planets around M Dwarfs}",
      journal = {\apj},
         year = 2014,
        month = jan,
       volume = {781},
       number = {1},
          eid = {28},
        pages = {28},
          doi = {10.1088/0004-637X/781/1/28},
archivePrefix = {arXiv},
       eprint = {1307.5849},
 primaryClass = {astro-ph.EP},
       adsurl = {https://ui.adsabs.harvard.edu/abs/2014ApJ...781...28M}
}

@ARTICLE{Guo2025,
       author = {{Guo}, Kangrou and {Ida}, Shigeru and {Ogihara}, Masahiro},
        title = "{Formation of Free-Floating Planets via Ejection: Population Synthesis with a Realistic IMF and Comparison to Microlensing Observations}",
      journal = {arXiv e-prints},
         year = 2025,
        month = nov,
          eid = {arXiv:2511.03246},
        pages = {arXiv:2511.03246},
          doi = {10.48550/arXiv.2511.03246},
archivePrefix = {arXiv},
       eprint = {2511.03246},
 primaryClass = {astro-ph.EP},
       adsurl = {https://ui.adsabs.harvard.edu/abs/2025arXiv251103246G}
}

@ARTICLE{MMI2016,
       author = {{Ma}, Sizheng and {Mao}, Shude and {Ida}, Shigeru and {Zhu}, Wei and {Lin}, Douglas N.~C.},
        title = "{Free-floating planets from core accretion theory: microlensing predictions}",
      journal = {\mnras},
         year = 2016,
        month = sep,
       volume = {461},
       number = {1},
        pages = {L107-L111},
          doi = {10.1093/mnrasl/slw110},
archivePrefix = {arXiv},
       eprint = {1605.08556},
 primaryClass = {astro-ph.EP},
       adsurl = {https://ui.adsabs.harvard.edu/abs/2016MNRAS.461L.107M}
}

@ARTICLE{GanGuoMao2024,
       author = {{Gan}, Tianjun and {Guo}, Kangrou and {Liu}, Beibei and {Wang}, Sharon X. and {Mao}, Shude and {Buchner}, Johannes and {Fulton}, Benjamin J.},
        title = "{Relative Occurrence Rate between Hot and Cold Jupiters as an Indicator to Probe Planet Migration}",
      journal = {\apj},
         year = 2024,
        month = may,
       volume = {967},
       number = {1},
          eid = {74},
        pages = {74},
          doi = {10.3847/1538-4357/ad3deb},
archivePrefix = {arXiv},
       eprint = {2404.07033},
 primaryClass = {astro-ph.EP},
       adsurl = {https://ui.adsabs.harvard.edu/abs/2024ApJ...967...74G}
}
\bibliographystyle{aasjournal}



\end{CJK*}
\end{document}